\PassOptionsToPackage{final}{graphicx}
\documentclass[12pt,a4paper,twoside]{scrreprt}
\pdfoutput=1
\usepackage{diplomarbeitA-arXiv}

\begin{document}

	\pagestyle{empty}
	\maketitle
	\cleardoublepage

    \selectlanguage{english}
    \let\genericabstractname\englishabstractname
\begin{abstract}
\begin{center}
\abstractHeader
\parbox{.95\textwidth}
{ \setlength{\parindent}{3ex}
\setlength{\parskip}{0.3\baselineskip}

\noindent
{The embedded systems engineering industry faces increasing demands for more functionality, rapidly evolving components, and shrinking schedules. Abilities to quickly adapt to changes, develop products with safe design, minimize project costs, and deliver timely are needed. A response to the broader range of requirements and the problems brought along with system complexity are development methods based on models. Model-based development (MBD) follows a separation of concerns by abstracting systems with an appropriate intensity, such as the separation of functional requirements and specification from implementation details. MBD promises higher comprehension by modeling on several abstraction-levels, formal verification, automated code generation, and certification.}
\medskip
\newline
{This thesis demonstrates MBD with the Scicos/SynDEx framework on a distributed embedded system. Scicos is a modeling and simulation environment for hybrid systems. SynDEx is a rapid prototyping integrated development environment for distributed systems. Performed examples implement well-known control algorithms on a target system containing several networked microcontrollers, sensors, and actuators. Results of these demonstrations support the decision-making process of either preferring MBD or classical, hand-written approaches. The addressed research question tackles the feasibility of MBD for medium-sized embedded systems.}
\medskip
\newline
{In the case of single-processor applications experiments show that the comforts of tool-provided simulation, verification, and code-generation have to be weighed against an additional memory consumption in dynamic and static memory compared to a hand-written approach. Expenses for establishing a near-seamless modeling-framework with Scicos/SynDEx and an increased development effort indicate a high price for developing single applications, but might pay off for product families. A further drawback was that the distributed code generated with SynDEx could not be adapted to microcontrollers without a significant alteration of the scheduling tables. The Scicos/SynDEx framework forms a valuable tool set that, however, still needs many improvements. Therefore, its usage is only recommended for experimental purposes.\\}

}
\end{center}
\end{abstract}

\let\genericabstractname\germanabstractname
\begin{abstract}
\begin{center}
\abstractHeader

\parbox{.95\textwidth}
{ \setlength{\parindent}{3ex}
\setlength{\parskip}{0.3\baselineskip}

\noindent
{Die Erstellung eingebetteter Systeme wird mit erhöhten funktionellen Anforderungen, der raschen Weiterentwicklung von Komponenten, und verkürzten Lieferzeiten konfrontiert. Entwicklungsmethoden, die sowohl eine schnelle Anpassung an Veränderungen und fehlerfreies Design, als auch die Minimierung von Kosten und die Möglichkeit einer zeitgerechten Produktauslieferung bieten, werden benötigt. Eine Antwort auf erhöhte Anforderungen und Komplexität sind Entwicklungsmethoden basierend auf Modellen. Die modellbasierte Entwicklung trennt fachliche und technische Belange auf verschiedenen Ebenen, und unterstützt dabei die konzeptuelle Erfassung von Systemen, automatisierte Codegenerierung, formale Verifikation, und Zertifizierung.}
\medskip
\newline
{Diese Arbeit demonstriert den modellbasierten Entwurf von eingebetteten Systemen mit dem Scicos/SynDEx Framework. Scicos ist ein Modellierungs- und Simulationswerkzeug für hybride Systeme. SynDEx ist eine integrierte Entwicklungsumgebung für die Prototypenerstellung verteilter Systeme. Die Umsetzung des modellbasierten Ansatzes wird mit zwei konkreten Beispielen aus dem Bereich der Regelungstechnik auf einer Entwicklungsplattform bestehend aus vernetzten Mikrokontrollern, Sensoren, und Aktuatoren durchgeführt. Resultierende Ergebnisse helfen für mittelgroße eingebettete Systeme Vor- und Nachteile des modellbasierten Ansatzes gegenüber handgeschriebenen Ansätzen abzuwiegen.}
\medskip
\newline
{Einprozessoranwendungen zeigten einen erhöhten dynamischen und statischen Speicherverbrauch im Vergleich zu handgeschriebenen Ansätzen -- dagegen steht der von Werkzeugen gebotene Komfort durch Simulation, Verifikation und automatischer Codegenerierung. Der Einsatz des Frameworks ist für einzelne Applikationen unökonomisch, dies wurde durch den nötigen Aufwand für die Erstellung eines nahtlosen Modellierungs-Frameworks und den erhöhten Entwicklungsaufwand bezeugt. Ein weiterer Nachteil bestand darin, dass der mittels SynDEx generierte Code nicht an Mikrokontroller angepasst werden konnte ohne das Zeitverhalten der Applikation signifikant zu verändern. Das Scicos/SynDEx Framework bietet vielversprechende Ansätze, da jedoch noch viele Verbesserungen ausstehen, wird es derzeit nur für experimentelle Zwecke empfohlen.}

}
\end{center}
\end{abstract}
    \cleardoublepage

    \pagestyle{plain}
    \pagenumbering{roman}
    \setlength{\parskip}{5pt plus2pt minus2pt}

    \setlength{\parskip}{1mm}
    \linespread{0.0}

    \tableofcontents

	\newpage
	\phantomsection \label{phantom:listoffig}
    \addcontentsline{toc}{figure}{List of Figures}
    \listoffigures

	\newpage
	\phantomsection \label{phantom:listoftab}
	\addcontentsline{toc}{table}{List of Tables}
    \listoftables

    \linespread{1}
    \clearpage
    \cleardoublepage
    \setlength{\parskip}{5pt plus2pt minus2pt}

    \pagestyle{fancy}
    \fancyMain
	\renewcommand{\chaptermark}[1]{\markboth{\MakeUppercase{#1}}{}}
    \renewcommand{\sectionmark}[1]{\markright{\thesection\ #1}{}}
    \addtolength{\headheight}{2pt}

    \pagenumbering{arabic}
    \setcounter{page} {1}

	\chapter{Introduction} \label{sec:introduction}

Embedded systems engineering combines the fields of software, hardware and control engineering. The main characteristic which separates embedded development from plain software development is its intrinsic link to the physical world, often in a safety-critical environment. Embedded systems usually realize a set of requirements for precise control of electromechanical devices (e.g. \acp{MEM}).
\medskip
\newline
Advancements in microfabrication lead to cheaper and more powerful embedded components which come along with higher customer demands and tighter schedules. The increasing number of system functionalities implies higher complexity and diminished perceptibility especially when it comes to simultaneity: Systems that might appear as being simple at a first glance are in fact hard combinatorial problems, the limits of the conceptual landscape in the human mind are easily reached: the number of states in an automaton explodes with every additional degree of freedom.
\medskip
\newline
{Current development methods for embedded systems incorporate plan-driven and agile characteristics to a certain degree. Document-centric plan-driven methods provide less flexibility to requirement and specification changes than agile methods which focus on interaction with customers and adaptability. Development methods are responsible for the successes of IT-projects of which about $20\%$ failed according to studies between 2002 and 2006\cite{ElAmam2008}. Main reasons include over-budget, too many scope changes and the management being not sufficiently involved. These reasons are based on two common denominators, namely uncertainty and risk. At project start the possible solution space is blurred, estimations about costs and project time are vague.}
\medskip
\newline
A response to all these issues are model-based development (MBD) paradigms. In MBD concerns are separated by their importance for the current activity in the development process, such as the separation of implementation details from requirements specifications. Centering the development process around models allows a higher flexibility to change requests by automation, reduces risks by enabling simulation and verification in early project phases, and supports the management in planning project costs and time. Low costs, timely delivery and safe design are the driving forces of systems engineering. Introducing models has a price and many aspects have to be considered when shifting to a model-based design culture\cite{Smith07}. Educating stakeholders with new modeling techniques and tools might be expensive, requires time, and adopting already established processes or legacy systems is costly. These investments increase the development effort in the scope of short-term planning. Embedded systems are often restricted by computational power, memory and disk space. Are model-based solutions efficient? This thesis prepares information that helps to weigh costs against gain in the context of MBD.\\

Models act as a base for consistent, automatic artifact generation. Implementation (code generation) and verification can be carried out with trusted, standardized tools which are usually based on synchronous languages and are current research topics in computer science. Mathworks' MATLAB and Esterel Technologies' SCADE are well-established, commercial tool-sets on the market and specialized inside the domain of safety-critical embedded systems. An overall, generally applicable modeling theory does not exist, but modeling techniques and scientific tools provide research opportunities, such as the non-commercial Scicos/SynDEx framework. This framework combines hybrid systems modeling with Scicos and discrete temporal, distributed modeling, optimization and code-generation with SynDEx.\\

This thesis evaluated the benefits and costs of a MBD-based development process with the Scicos/SynDEx framework on a distributed target-platform. Capabilities of the framework were explored and MBD compared to classic, hand-written development. Development effort, executable code size, code structure (perceptibility), and memory consumption were compared. In literature there are no such examples of using Scicos/SynDEx on an distributed architecture with microcontrollers.\\

Objectives of this work are to exercise and analyze the outcomes of well-known control and data observation algorithms following a model-based development process. Examples are PID control algorithms, and a data observation application carried out on a scientific target-platform consisting of several microcontrollers and peripherals including sensors and actuators.\\

Algorithms in the discrete domain are modeled together with continuous environment peripherals. The demonstrations include a single- and a multi-processor example. In the single-processor application a PID algorithm controlling a cooling fan is realized. In the multi-processor example, temperature data is observed at one node, displayed on an LCD on another, and forwarded to the development workstation via an embedded gateway node.
\medskip
\newline
Implemented applications are hybrid systems which are designed, simulated and verified with Scicos. Temporal design, scheduling, optimization, distribution and code-generation of transformed Scicos models are performed with the help of SynDEx. A seamless modeling environment is approached by adapting the Scicos/SynDEx tool-chain.\\

The research in this thesis is limited to a multi-processor target-platform and the Scicos/SynDEx framework. Several other tools require research, but the results in this thesis can be used to compare this framework's performance to tool-sets incorporating other modeling techniques. The research shows strengths and weaknesses of the framework with respects to scheduling, modeled versus real behavior, code metrics and resource allocation. Estimations for the development effort of MBD and classic development are not quantitative, they are only based on a few examples.\\

Results of this thesis demonstrate how control algorithms can be modeled with this particular model-based development toolset. Problems and pressing issues were identified and contribute information for enhancing the Scicos/SynDEx framework. The data gained by comparing classical development to MBD eases the decision of migrating model-based paradigms into the development process. With model-based development the possible solution space is narrowed down in early stages by rapid prototyping and design faults are detected easier. However, all this comes with the price of extra development effort, program size and memory consumption.\\

\clearpage
\section{Outline}\label{sec:structure}
\vskip 1cm
The thesis is divided into \textbf{five chapters}: \textsl{\textbf{(1)} Introduction}, \textsl{\textbf{(2)} Concepts and Related Work}, \textsl{\textbf{(3)} Modeling Tools}, \textsl{\textbf{(4)} Demonstrations: MBD with Scicos/SynDEx}, and \textsl{\textbf{(5)} Conclusion}.
\vskip 1cm
{\noindent\textbf{Chapter 1} introduces the background, scope of research, applied methods, the objectives and relevance of this thesis.\\}
\medskip
\newline
{\textbf{Chapter 2} encapsulates important terms and definitions. Control theory relevant for the examples is introduced briefly. Basic concepts in modeling systems are presented, including model-based and model-driven development and approaches.\\}
\medskip
\newline
{\textbf{Chapter 3} explains the modeling tools Scicos and SynDEx in a detail necessary for understanding the examples. Syntax, semantics, underlying modeling techniques, and modeling with the used tools are surveyed. Strengths and weaknesses of temporal design with SynDEx is given special consideration to.\\}
\medskip
\newline
{\textbf{Chapter 4} shows the methods and realized experimental designs. The design of PID algorithms with Scicos and modeling hardware, communication, and software architectures with SynDEx is presented and discussed. Finally, this chapter summarizes the examples' results.\\}
\medskip
\newline
{\textbf{Chapter 5} reviews the results from the previous chapter. Advantages and drawbacks of systems design by modeling are discussed based on development effort and designs. Improvements are proposed for the SynDEx tool, as well as concepts for establishing a working modeling tool-chain.\\}

	\chapter{Concepts and Related Work}\label{chapter:concepts}

Model-based and model-driven development have been emerging terms in embedded systems development. Demonstrations in this thesis require the definition of concepts in control theory, modeling theory, and development methodologies. The following definitions resulted from a literature study to build a conceptual base for this thesis. Since the topic of this thesis derives from different disciplines such as computer science, control engineering, and embedded systems, the existing nomenclatures are often not clearly defined in literature, have ambiguous meaning, are used freely, and often depend strongly on their context.

\section{Basic Concepts in Systems Engineering}\label{sec:concepts:basics}

This section defines general methods and concepts needed in the following chapters of this thesis. The model-based development demonstrations require definitions in the fields of software, systems, models and control theory.\\

\paragraph{Software Engineering}$~~$\\
First defined in a \acf{NATO} conference in 1968\cite{nato1968}, \Keyword{Software Engineering} has been a term not easy to describe in a way that satisfies the whole IT - community. Even though there is no complete consensus, the \acf{IEEE} offers a standardized definition:
\begin{quote}
"(1) The application of a systematic, disciplined, quantifiable approach to the development, operation, and maintenance of software; that is, the application of engineering to software. (2) The study of approaches as in (1)."\cite{IEEE90}
\end{quote}

\paragraph{System}
\begin{quote}
"A collection of components organized to accomplish a specific function or set of functions." \cite{IEEE1471}
\end{quote}

\paragraph{Distributed System}$~~$\\
In a \Keyword{distributed system} the components are distributed in space and are contributing functionality to the whole system. A similar definition by IEEE:
\begin{quote}
"A computer system in which several inter-connected computers share the computing tasks assigned to the system."\cite{IEEE100-2000}
\end{quote}

\paragraph{Reactive System}$~~$\\
A \Keyword{reactive system} is a system which reacts to stimuli, outputs are set dependent on the input and the system function. Such a system is in a continuous interaction with its environment.

\paragraph{Hybrid System}$~~$\\
The combination of a system operating at discrete time instants and a corresponding environment located in a continuous time domain, can be called a \Keyword{hybrid system}.

\paragraph{Systems Engineering}
\begin{quote}
"Systems engineering is an interdisciplinary engineering management process that evolves and verifies an integrated, life-cycle balanced set of system solutions that satisfy customer needs."\cite{nla.cat-vn3847107}
\end{quote}

\paragraph{Architecture}
\begin{quote}
"The fundamental organization of a system embodied in its components, their relationships
to each other, and to the environment, and the principles guiding its design and evolution."\cite{IEEE1471}
\end{quote}

\paragraph{Life Cycle Model}
\begin{quote}
"A framework containing the processes, activities, and tasks involved in the development, operation, and maintenance of a software product, which spans the life of the system from the definition of its requirements to the termination of its use."\cite{IEEE1471}
\end{quote}

\paragraph{Software Process}$~~$\\
A set of activities and its following results building a software product is called a \Keyword{Software Process} (sometimes also called a Software Life Cycle or Software Development Process) which is an instance of a Software Development Methodology. Today, almost every software process contains following four basic process activities \cite{SommSoft2007}:\\
\begin{itemize}
\item{\textbf{Software Specification} - Customer and developer define how the software is developed within specified constraints.}
\item{\textbf{Software Development} - Design and development of the software.}
\item{\textbf{Software Validation} - Software test and check against the requirements/specification.}
\item{\textbf{Software Evolution} - Adaption of the software to fulfill new customer requirements.}
\end{itemize}

\paragraph{Development Process}$~~$\\
This term is often set in relation to a software process, but a brief stand-alone definition looks as follows. A \Keyword{Development Process} consists of \ldots
\begin{itemize}
\item[] \textbf{Phases} and corresponding activities.
\item[] \textbf{Activities} carried out by developers.
\item[] \textbf{Artifacts} as results of activities.
\end{itemize}

\paragraph{Software Process Models}$~~$\\
A \Keyword{Software Process Model} presents a simplified, abstracted view of a Software Process. Those models can contain activities as part of the software process, roles of persons, products and scheduling. Most of them build on following universal paradigms\cite{SommSoft2007}:\\
\begin{itemize}
\item{\textbf{Waterfall Model.} In this paradigm each software process activity is seen as its own development phase with an strictly defined start and end of the phase. The software process proceeds with a close of the preceding phase to the next phase. The concept of the Waterfall Model was initially created by Winston W. Royce in 1970.\cite{Royce1970}. This paradigm should be used if the requirements are fixed and no major changes are foreseen until the end of the project, otherwise, due to the inflexible partition of the phases, the development effort will rise tremendously.}

\item{\textbf{Evolutionary Development.} Specification, Development and Validation are proceeding parallel and with close cooperation with the customers. Information gained in each activity is shared to each other. The practice of building prototypes can help to get more knowledge about the customers' requirements.\\ This paradigm is good for a stepwise refinement of the specification but in the view of the management it is hard to measure the advancement of the project since it is not cost-effective to document each intermediate product. Another possible drawback is a bad structure of the system architecture due to the ongoing refining of the specification, which makes this paradigm not well-suited for very large projects ($>500\,000$ \ac{LOC}).}

\item{\textbf{Component Based Software Engineering.} This approach is focused on reusing and adapting software components.}
\end{itemize}

\paragraph{Software Life Cycle}
\begin{quote}
"The period of time that begins when a software product is conceived and ends when the software is no longer available for use. The life cycle typically includes a concept phase, requirements phase, design phase, implementation phase, test phase, installation and checkout phase, operation and maintenance phase, and sometimes, retirement phase. These phases may overlap or be performed iteratively, depending on the software development approach used."\cite{IEEE90}
\end{quote}

\paragraph{Process Iterations}$~~$\\
A way to handle requests for changes in the software product is to repeat the software process completely or partially. A \Keyword{Process Iteration} is either a repetition of the whole software process or single/several process activities. If an activity is repeated, all successors have also to be handled to avoid inconsistencies between specification, design, code and documentation.\\

\begin{itemize}
\item{\textbf{Incremental Development}\\
\Keyword{Incremental Development} is a combination of the Waterfall Model and Evolutionary Development: The specification and the design of the system and subsystems are defined once followed by the subsequent development, integration and validation of subsystems. Subsystems are delivered stepwise to the customers, starting with the most important functionalities.
}

\item{\textbf{Spiral Model}\\
Process activities are visualized using a spiral. Each loop of the spiral describes a process phase. Starting in the inner phase with a feasibility study, ensued by the definition of the specification, design, development etc. One main feature of this paradigm is a risk analysis embedded in each loop providing valuable information for the project management.\cite{BoehmSpiral88}}
\end{itemize}

\paragraph{Rational Unified Process (RUP)}$~~$\\
The \Keyword{Rational Unified Process} is a phase-oriented model spawned of the Unified Software Development Process. RUP applies \ac{UML} notations and combines parts of several software process models. In contrast to the other mentioned software processes, RUP supports three different views to the Software Process:\\
\begin{itemize}
\item{a \Keyword{dynamic perspective} showing all model phases timely ordered.}
\item{a \Keyword{static perspective} visualizing process activities.}
\item{a perspective proposing further procedures for the process.}
\end{itemize}

\paragraph{Software Development Methodology}$~~$\\
In literature the term \Keyword{Software Development Methodology} has ambiguous meanings and strongly depends on the context it is used in (e.g. some authors identify a methodology with a process). A software development methodology is a framework for solving a technical challenge. In a general point of view a software development methodology is an orchestration of:
\begin{itemize}
\item{Software process (software process model) which provides fragmentation and schedule of the software development in several phases.}
\item{Notations and techniques (e.g. object-oriented, prototyping) which are established for the documentation of products and intermediate products.}
\item{Management methods for the analysis and transformation of documents.}
\end{itemize}

\subsection{A Brief Glimpse on Control Theory}

Every technical process intriniscally comprises reaction delays and disturbances. Imagine a fan-cooling system of a computer-chip has to provide the correct amount of air-flow to ensure a certain operation-temperature that does not damage the chip. Just setting a reference speed will usually not result in the fan moving as desired due to disturbances. Reasons for this are characteristics of physical processes involved such as friction, acceleration time, inertia and many more. Not all fans will behave the same way because of production tolerances, and in the case of digital systems the references provided to the fan are of a discrete nature which is per se a reason for diverging behavior. Such dynamical systems require an automated control component that keeps it in a desired state such as following a pre-defined function. \Keyword{Control Theory} deals with mathematical approaches for the control of a system.\\

\begin{figure}[!htbp]
\begin{center}

\includegraphics[width=1.0\textwidth]{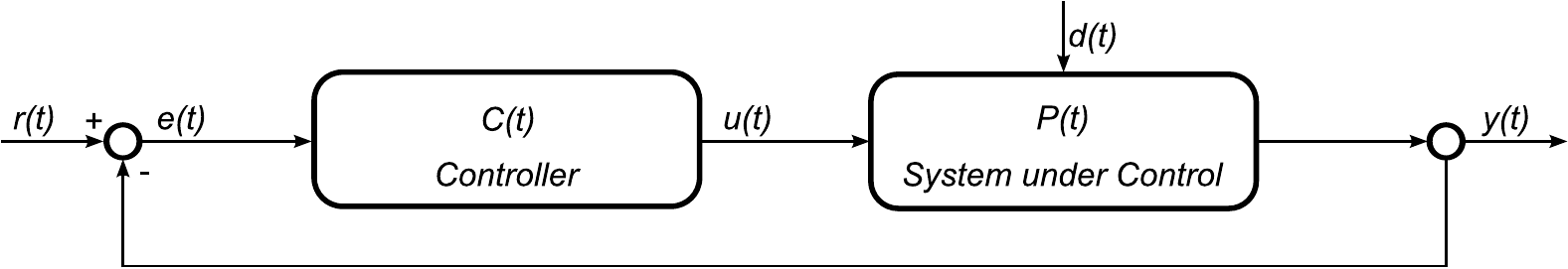}

\end{center}
\label{fig:controlTheoryBasics}
\caption{Single, closed-loop control structure.}
\end{figure}

A basic control structure is set up by a \Keyword{controller} $P$ providing \Keyword{inputs} $u(t)$ to the \Keyword{system under control (process)} $P$ that is interfered with by $d(t)$. The \Keyword{output} $y(t)$ of the system is fed back and the \Keyword{error} $e(t)$ between \Keyword{reference value} $r(t)$ and $y(t)$ calculated (figure~\ref{fig:controlTheoryBasics}). 

The behavior of the process $P(t)$ is determined with standardized test-functions and characterized by the resulting \Keyword{step responses}. This technique is often used to gain an approximate description of the process instead a complete mathematical model. A widely used control algorithm $C(t)$ is the PID (Proportional Integral Derivative) controller. Controllers are set-up and tuned with the goals of appropriate rise time, minimal overshoot and no steady-state errors. PID algorithms are tuned with several approaches such as the Ziegler-Nichols, Cohen-Coon, or the Chien-Hrones-Reswick formula. Fine-tuning of the algorithm might be required.

\subsubsection{PID Algorithms}

The output of a PID algorithm is composed by a proportional $K_P$, integral $K_I$ and derivative $K_D$ part, with the integral time $T_n$ and the derivative time $T_v$ in the \Keyword{standard form} :
\begin{eqnarray*}
u(t)=K_P * \left(e(t)+\frac{1}{T_n}\int_0^t e(\tau)d\tau+T_v \frac{\text{d}e(t)}{\text{d}t} \right)\\
\end{eqnarray*}
\ldots which is equivalent to the \Keyword{ideal parallel form}:
\begin{equation*}
u(t)=K_P\; e(t)+K_I \int_0^t e(\tau)\text{d}\tau + K_D \frac{\text{d}e(t)}{\text{d}t}
\end{equation*}
An implementation on a discrete controller requires a differential representation with sample period $T_S$:
\begin{equation*}
u(i)=K_P\; \left(e_i+\frac{1}{T_n}\sum_{v=0}^i e_v\; T_S+T_v\frac{e_i-e_{i-1}}{T_S}\right)
\end{equation*}

\subsubsection{Systems Under Control}

\begin{figure}[!htbp]
\begin{center}

\includegraphics[width=0.50\textwidth]{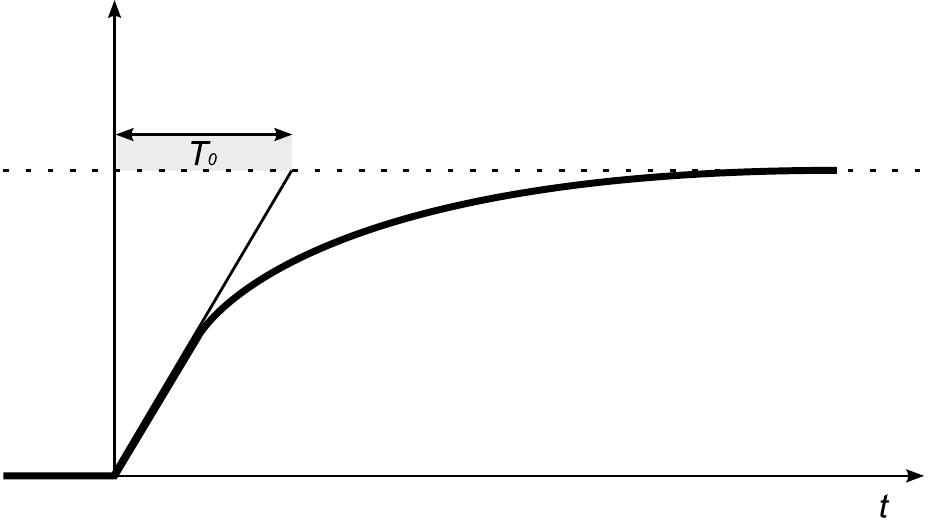}

\end{center}
\label{fig:controlTheoryPT1}
\caption{PT1: first-order delay element.}
\end{figure}

Dynamic system responses are mainly classified into $P$ (proportional), $I$ (integrational), first-order $PT1$ and second-order lag elements $PT2$. $PT1$ elements approximate direct-current motor behavior (figure~\ref{fig:controlTheoryPT1}):

\begin{equation*}
T_0\; y'(t)+y(t)=K_P\; u(t)
\end{equation*}
\ldots and the Laplace-transformed form:
\begin{equation*}
F_S(s)=\frac{Y(s)}{U(s)}=\frac{K_P}{1+s\;T_0}
\end{equation*}

Going into more detail would exceed the limits of this thesis -- further information is found in literature\cite{springer-prozessautomatisierung,elmenreich:09}.

\section{System Development Using Models}\label{sec:model}

Classical system development methodologies are mostly plan-driven where the results of each development phase are prerequisites for others. Changes in one phase affect the immediatly surrounding phases directly, they receive new development artifacts and pass them on to their subsequent following phase. In this way, deviations from the original design are handed on while every phase represents a potential source for additional errors in the solution.\\
The idea of introducing models into the development process counteracts error-chains by centering the process around a model. A model acts as a reference that is consistent to all phases in the cycle. 
The term \Keyword{model} was manifoldly defined in literature and in the following two definitions are presented:

\begin{quote}
"\textbf{Models} provide \textbf{abstractions} of a physical system that allow engineers to reason about that system by ignoring extraneous details while focusing on the relevant ones."\cite{Brown}
\end{quote}
\begin{quote}
"Engineering \textbf{models} aim to reduce risk by helping us better understand both a \textbf{complex problem} and its \textbf{potential solutions} before undertaking the expense and effort of a full implementation."\cite{Selic}
\end{quote}

These definitions emphasize:
\begin{itemize}
\item Abstraction that enhances understanding.
\item Representation of the problem domain on an appropriate level of abstraction that hides distracting details.
\item Reduction of risks by acquiring solution information early.
\item Minimize costs.
\end{itemize}

Introducing models in the development process requires extra effort spent on establishing a working infrastructure. Several tools already exist which provide a tool-chain reaching from requirements specification to implementation or at least assist in this process. Such tools are mostly proprietary and the question if the expenses are worth it is reasoned.\\

IT projects are very sensitive due to their per se complex nature, which makes it hard to imagine a possible solution for a given problem. Why IT projects fail has been studied for decades, for example in  the large-field studies/roadmaps of the Standish-Group\footnote{\url{http://www.standishgroup.com}}. Emam and Koru \cite{ElAmam2008} identify the reasons of IT project failures (table~\ref{fig:softwareCancellationsReasons}). The four largest entries have one common characteristic: They strongly depend on flexibility and system representation. Senior management is not sufficiently involved because projects could not be presented in an appropriate abstraction level, scope/requirement changes impact on the whole inflexible chain of development phases resulting in over budget.

\begin{table}[!Htbp]
\begin{tabular}{|l c|}
\hline
Reason for cancellation						&\vline	\xspace Percentage of respondents \\
\hline
Senior management not sufficiently involved	&	33 \\
Too many requirements and scope changes		&	33 \\
Lack of necessary management skills			&	28 \\
Over budget									&	28 \\
Lack of necessary technical skills			&	22 \\
No more need for the system to be developed	&	22 \\
Over schedule								&	17 \\
Technology too new; did not work as expected	&	17 \\
Insufficient staff							&	11 \\
Critical quality problems with software		&	11 \\
End users not sufficiently involved			&	6  \\
\hline
\end{tabular}
\caption[Reasons for project cancellations.]{Reasons for project cancellations 2007, source:\cite{ElAmam2008}}
\label{fig:softwareCancellationsReasons}
\end{table}

All the mentioned reasons for IT project failures are related to the combination of risk and uncertainty - the less information and the less clear the envisioned concept of the product is, the higher are the costs for necessary changes in later development stages. The concepts of uncertainty and risk in software engineering can have the following reasons:

\begin{itemize}
\item{Customers may not exactly know what they want.}
\item{Requirements can be ambiguous and interpreted diversely by customer and contractor.}
\item{Technical and management risks.}
\item{Development efforts are over- or underestimated.}
\item{Changes of methodologies and priorities during the project.}
\end{itemize}

\begin{figure}[!htbp]
\begin{center}
	\includegraphics[width=\textwidth]{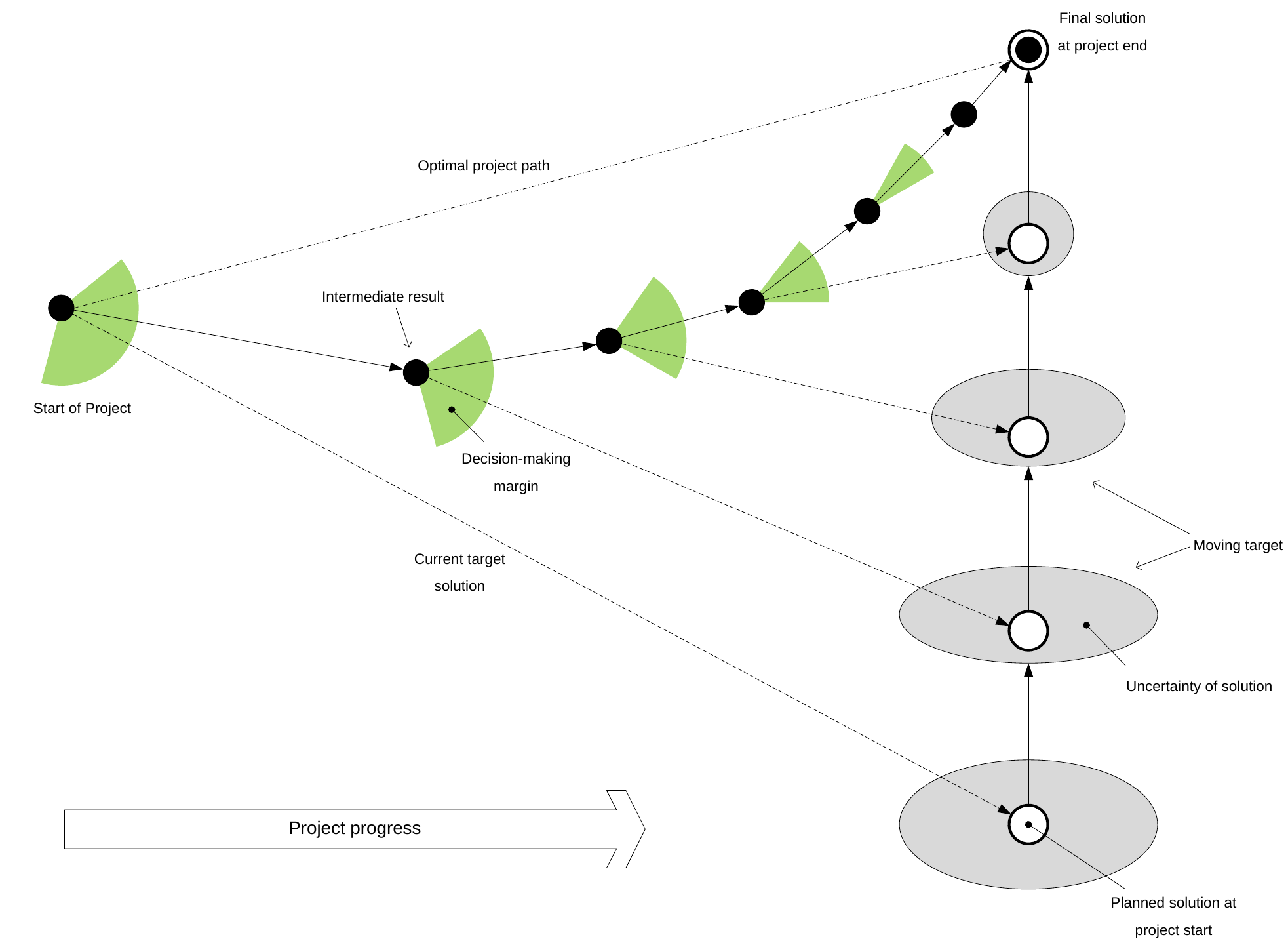}
   \caption[Uncertainty and risk.]{Uncertainty and Risk, source: \cite{Iteratec}, modified illustration}
   \label{fig:concepts:uncertainty}
\end{center}
\end{figure}

The relation between risk and uncertainty during an iterative software project are depicted in figure~\ref{fig:concepts:uncertainty}. The project starts with the definition of a the product accompanied by a high uncertainty about the real solution. The earlier the project time, the more unclear is the deviation from the defined end-product. Of course, requirements and specifications were defined under collaboration with the customer, however, there are always unknown characteristics (for example if a new technology could not perform as well as expected). These unknowns make a qualified prediction about the future outcomes hardly possible. With elapsing project time more information about development process and product is gathered - resulting in a reduction of uncertainty. According to the figure, an increased adaptability and flexibility to direction changes during the development is desirable. In this way the uncertainty is effectively narrowed and the development costs significantly decreased.

Efforts to master the system engineering task and the related pressing issues were usually pointing to the increase of manpower - which did not lead to satisfying results. Since then, a period of creating formalized system development methodologies was started: System development using models. Depending on the special use of models in the development cycle, the terms \Keyword{model-based development}, with a model as pure reference, and \Keyword{model-driven development}, with the focus on automatically artifact generation, are defined.\\
\clearpage
\subsubsection{Models and Systems}

Generally, it is desirable to describe systems with models. Integrating models can establish several advantages leading from improving the software development life cycle to estimations of project costs and formal verification of the system design.\\
A major concern in the development of embedded systems is to design systems and subsystems represented by different mathematical approaches: Controlled objects residing in a data and time domain with continuous nature have to be interfaced with controlling, discrete information systems. For example, a continuous systems might be described by an ordinary or partial differential equation, while discrete systems, such as a controller, could be described via a discrete event system. Shorty said, the composition of continuous and discrete dynamics is referred to as a \definition{hybrid system}. A formal definition of a \definition{hybrid system} follows. 

\subsubsection{Hybrid System}\label{sec:models:hybridSystem}

\begin{figure}[!htbp]
\begin{center}
	\includegraphics[width=0.25\textwidth]{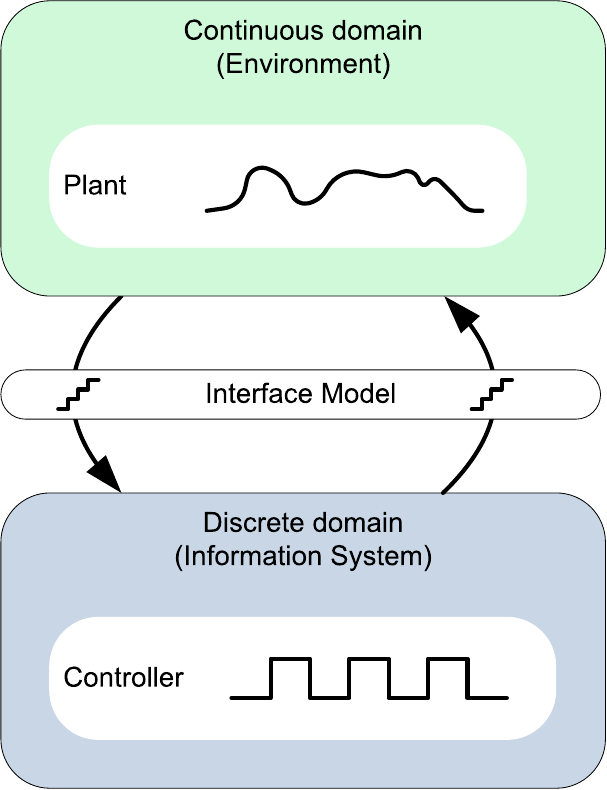}
   \caption{Scheme of a hybrid system.}
   \label{fig:concepts:hybridSystem}
\end{center}
\end{figure}

A \definition{hybrid system} has a continuously evolving nature including casual jumps. These jumps can either be caused by the continuously evolving system or a change of states in a controlling system. An example of a hybrid system would be a plant controlled by a discrete controller system interfaced by an interface-model (see figure~\ref{fig:concepts:hybridSystem}). Since hybrid systems are in the focus of this work, a formal definition of hybrid systems is being described.

\paragraph{Hybrid System - Formal Definition}$~~$\\
A literature research about the definition of a hybrid system ended with the work of Carloni, Passerone, Pinto and Sangiovanni-Vincentelli who gave a formal definition of a hybrid system\cite{LanguagesToolsForHybridSystemsDesign}. The following definition is essentially their work - recapitulations were made where it was possible without losing essential information.\\
Formally defining a hybrid system, requires some notations to be introduced beforehand. $X,U,V$ shall be vector fields over subclasses of continuous dynamical systems:\\
\begin{tabbing}
XX\= \kill
\>$X$\ldots Continuous state.\\
\>$U$\ldots Input.\\
\>$V$\ldots Disturbance.\\
\>$\mathcal{U_C}$\ldots Class of measurable input functions $u:\mathbb{R}\rightarrow U$.\\
\>$\mathcal{U}_d$\ldots Class of measurable disturbance functions $\delta:\mathbb{R}\rightarrow V$.\\
\>$\mathbf{S}_C(X,U,V)$\ldots is the class of continuous time dynamical systems which is defined by:\\
  \end{tabbing}
\begin{equation*}
\dot{x}(t)=f(x(t),u(t),\delta(t))\\
t \in \mathbb{R}\\
x(t) \in X\\
\end{equation*}
$f$ is a function such that for all $u \in \mathcal{U_C}$ and for all $\delta \in \mathcal{U_d}$, the solution $x(t)$ exists and is unique for a given initial condition.\\

\begin{defs}\label{sec:models:defHybridSystem}
\definition{Hybrid System.} A continuous time hybrid system is a tuple $\mathcal{H}=(\mathbf{Q},\mathbf{U}_D,E,X,U,V,S,Inv,R,G)$ where:
\begin{itemize}
\item $\mathbf{Q}$ \ldots set of states.
\item $\mathbf{U}_D$ \ldots set of discrete inputs.
\item $E \subset \mathbf{Q} \times \mathbf{U}_D \times \mathbf{Q}$ \ldots set of discrete transitions.
\item $X,U,V$ \ldots continuous state, input, disturbance.
\item $S:\mathbf{Q}\rightarrow \mathbf{S}_C(X,U,V)$ \ldots mapping associating to each discrete state a continuous time dynamical system (in terms of differential equations).
\item $Inv: \mathbf{Q} \rightarrow 2^{X \times \mathbf{U}_D \times U \times V}$\ldots mapping called invariant.
\item $R:E \times X \times U \times V \rightarrow 2^X$ \ldots reset mapping (the initial conditions upon entering a state).
\item $G:E \rightarrow 2^{X \times U \times V}$ \ldots guard mapping.
\end{itemize}
\end{defs}
A \Keyword{discrete time} hybrid system could similarly be defined by replacing $\mathbb{R}$ with $\mathbb{Z}$ for the independent variable, and by considering classes of discrete dynamical systems underlying each state.\\

The tuple $(\mathbf{Q}, \mathbf{U}_D, E)$ might be seen as the definition of an automaton characterizing a discrete transition structure:\\
\begin{tabbing}
XX\= \kill
\>$\mathbf{Q}$ \ldots state set.\\
\>$\mathbf{U}_D$ \ldots inputs.\\
\>$E$ \ldots transitions.\\
\end{tabbing}
This automaton can change the states if either a discrete input event occurs or the invariant in $Inv$ is not satisfied. 

The dynamical behavior of a hybrid system is described by \definition{executions}. An execution is a set of functions over time for the evolution of the continuous state as the system transitions through its discrete structure. The notation of a \definition{Hybrid Time Basis} is needed beforehand the definition of the a hybrid systems execution:\\

\begin{defs}\label{sec:models:defHybridSystemTimeBasis}
A \definition{Hybrid Time Basis} $\tau$ is a finite or an infinite sequence of intervals:\\
\begin{equation*}
	I_j = \{t \in \mathbb{R}:t_j \leq t\leq \acute{t_j}\},\: j \geq 0\newline
	\: where\: t_j \leq \acute{t_j}\: and\: \acute{t_j}=t_{j+1}.
\end{equation*}
\end{defs}

\begin{defs}\label{sec:models:defHybridSystemExecution}
\definition{Hybrid System Execution}. $T$ is considered to be the set of all hybrid time bases.
\begin{quotation}
An execution $\mathcal{X}$ of a hybrid system $\mathcal{H}$, with initial state $\hat{q} \in \mathbf{Q}$ and initial condition $x_0 \in \mathbf{X}$, is a collection $\mathcal{X}=(\hat{q},x_0,\tau,\sigma,q,u,\delta,\xi)$ where $\tau \in \mathcal{T},\sigma:\tau \rightarrow \mathbf{U}_D,q:\tau \rightarrow \mathbf{Q},u \in \mathcal{U_C}, \delta \in \mathcal{U}_d$ and $\xi: \mathbb{R} \times \mathbb{N}\rightarrow X$ satisfying:\\
\begin{enumerate}
\item Discrete evolution:
\begin{itemize}
\item $q(I_0)=\hat{q};$
\item for all $j,e_j=(q(I_j),\sigma(I_{j+1},q(I_{j+1})) \in E;$
\end{itemize}
\item Continuous evolution: the function $\xi$ satisfies the conditions
\begin{itemize}
\item $\xi(t_o,0)=x_0;$.
\item for all $j$ and for all $t \in I_j,$\\
\begin{center}
$\xi(t,j)=x(t)$\\
\end{center}
where $x(t)$ is the solution at time $t$ of the dynamical system $\mathbb{S}(q(I_j))$, with initial condition $x(t_j)=\xi(t_j,j)$, given the input function $u \in \mathcal{U}_C$ and disturbance function $\delta \in \mathcal{U}_d$.
\item for all $j,\xi(t_{j+1},j+1) \in R(e_j,\xi(\acute{t}_j,j),u(\acute{t}_j),v(\acute{t}_j))$.
\item for all $j$ and for all $t \in [t_j,\acute{t}_j]$,\\
\begin{center}
$(\xi(t,j),\sigma(I_j),u(t),v(t)) \in Inv(q(I_j))$.
\end{center}
\item if $\tau$ is a finite sequence of length $L+1$, and $\acute{t}_j \neq \acute{t}_L$,\\
then\\
\begin{equation*}
(\xi(\acute{t}_j,j),u(\acute{t}_j),v(\acute{t}_j)) \in G(e_j).
\end{equation*}
\end{itemize}

\end{enumerate}
\end{quotation}
\end{defs}

In their work they state, that the behavior of a hybrid system consists of all the executions that satisfy the Hybrid System Execution definition:
\begin{itemize}
\item \Keyword{Discrete Evolution Constraint}. The transitions (according to the transition relation $E$) of the system throughout its discrete states is constrained.
\item \Keyword{Continuous Evolution Constraint}. The execution must satisfy the system for each state and the invariant condition. If a invariant condition is violated, the system takes a transition to another state where the condition is satisfied. This constraint means, that a matching discrete input has to be supplied to the system.
\end{itemize}

An Hybrid System Execution is classified by its Hybrid Time Basis (definition \ref{sec:models:hybrid SystemExecutionClassification}).\\
\begin{defs}\label{sec:models:hybrid SystemExecutionClassification}
"A hybrid system execution is said to be (i) trivial if $\tau=\{I_o\}$ and $t_0 = \acute{t}_0$; (ii) finite if $\tau$ is a finite sequence; (iii) infinite if $\tau$ is an infinite sequence and $\sum_{j=0}^{\infty} \acute{t}_j-t_j=\infty$; (iv) Zeno, if $\tau$ is infinite but $\sum_{j=0}^{\infty}\acute{t}_j-t_j < \infty$."
\end{defs}

In this model of a hybrid system, an single input can lead to several valid executions - in that case, the system is non-deterministic (e.g. for incomplete systems or choice models). Defining priorities among the transitions can create a deterministic hybrid system.

\subsection[Model-Based Development]{Model-Based Development (MBD)}\label{sec:model:mbsd}

In \mbd the development process is centered around a model. This model has the purpose of pure documentation -- there is only a mental connection between model and system implementation. This model should in the best case be consistent in any state of the development process, but this might be an impossible task since every process phase may adapt the model without a process-wide modification doctrine. Advantages to the classical plan-driven methods without models exist, however errros may still be introduced if the models are not evolved by well-defined guidelines.\\

The term \mdd (MDD) is incorporated in the concept of \mbd. In this paradigm the model is not just pure documentation, it acts as specification and implemenation at the same time. The focus lies on the automatically generation of artifacts ensuring a system-wide model consistency: the generated code is always up-to-date. Several definitions of \mdd can be found in literature, a pertinent definition for this thesis is given by Stahl:
\begin{quote}
"Model-Driven Software Development is a generic name for techniques which automatically generate runnable software based on formal models."\cite{stahl}
\end{quote}
\noindent
A sub-variant of \mdd is \Keyword{Architecture-Centric \mdd}: Models are only used to generate the basic system infrastructure (e.g. skelleton components) which is enhanced manually by hand.

In this thesis the term \mbd is set equal to, or incorporates \mdd depending on the context the term is used in. Note that in literature, abbreviations relating to \mbd and \mdd exist, their meaning changes with the context. \Keyword{MBSD} can stand for model-based system development or model-based software development - same case for \Keyword{MDSD}. The terms \Keyword{MBE (model-based engineering)} and \Keyword{MDE (model-driven engineering)} are used when models meet system engineering issues (e.g. costs, risks, realization time). They relate to an approach of abstracting systems with the means of models and transform these systematically, with increasing concretization, until the level of executable models is reached. 

\subsubsection{Models in MDD}

\begin{figure}[!htbp]
\begin{center}
	\includegraphics[width=0.9\textwidth]{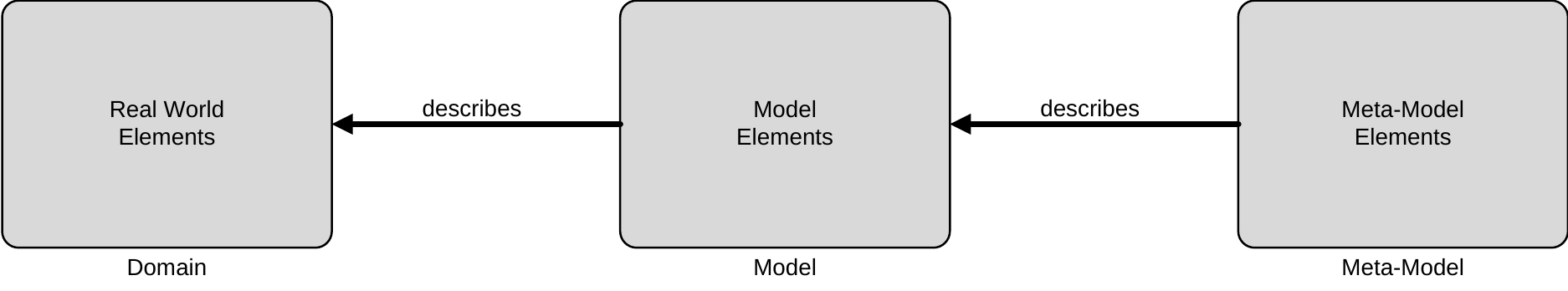}
   \caption[Model and meta-model relationships]{Model and meta-model relationships, source: \cite{stahl}, modified illustration.}
   \label{fig:concepts:models}
\end{center}
\end{figure}

In the context of MDD models describe contents of a domain. This description has to be done formalized, otherwise a partial automation of the software process is hardly possible. It is not enough to go straight forward into the design of the model, first of all, the structure of the domain has to be understood and on base of that an abstract \Keyword{meta-model} developed. Once a \Keyword{domain specific language} (DSL), a synonym for meta-model, is created, it is possible to build a formal model within all the characteristics given by the meta-model. Figure~\ref{fig:concepts:models} depicts a graphical overview of the relationship between formal models and meta-models.\\

\begin{figure}[!htbp]
\begin{center}
	\includegraphics[width=.80\textwidth]{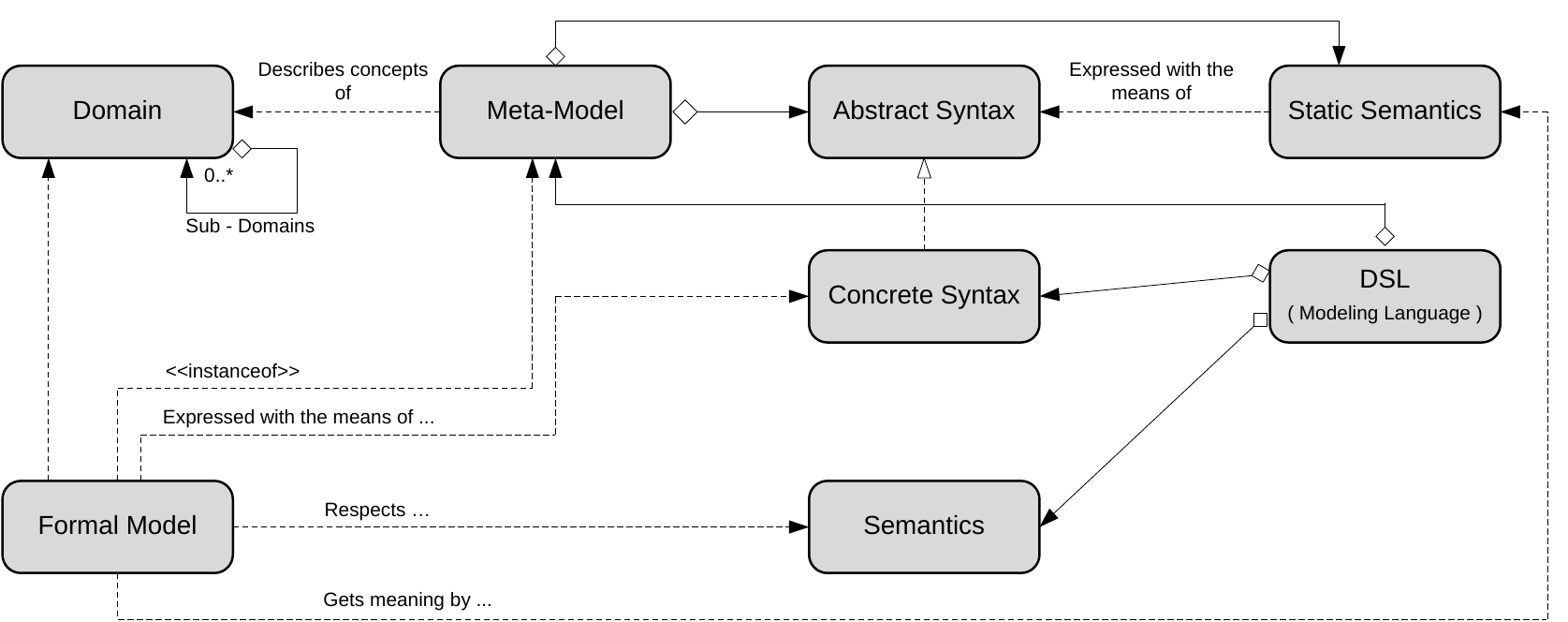}
   \caption[Models and DSLs]{Models and DSLs, source: \cite{stahl}, modified illustration.}
   \label{fig:concepts:modelspace}
\end{center}
\end{figure}

A meta-model describes how a model (or a modeling language) can be structured and defines constraints, validity and modeling rules. As it is the case in the \ac{MOF} 2.0, instances are described by a model, which itself is described by an \ac{UML} 2.0 model, which is in turn described by the \ac{MOF} 2.0 (this is a meta-meta-model for an instance). Automated actions, like generations of code or model-to-model transformations are based on the meta-model holding the abstract syntax and static semantics of a modeling language (see figure~\ref{fig:concepts:modelspace}). Formal models do not have to be graphical, a textual representation is also possible.\\

\subsubsection{MDD and Approaches to MDD}

This section gives an overview of some selected, already existing approaches to the \mdd paradigms. The concepts of the popular \ac{MDA} approach are focused.

\paragraph{Model Driven Architecture (MDA)}$~~$\\

\Keyword{MDA} is the specialized approach to \mdd by the \acf{OMG} which aims for a high portability (platform independence) and interoperability (neutral to manufacturers) of software systems. In contrary to GSD, MDA is not explicitly focused on system-families. \\
MDA is an approach to link various technologies together by using formal models (MOF, UML, etc.) describing their relations and configurations. Based on these formal descriptions models can be technology and platform independently used, allowing the building of functionalities and services by model design. \\
The three primary goals of MDA are portability, interoperability and re-usability through the architectural separation of concerns \cite{MDA}. \\

\begin{figure}[!htbp]
\begin{center}
	\includegraphics[width=.45\textwidth]{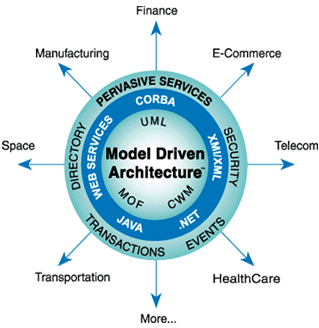}
   \caption[MDA - Overview.]{MDA - Overview, source: \cite{UML}}
   \label{fig:concepts:mdalogo}
\end{center}
\end{figure}

In MDA, three different default models emerged from three different corresponding viewpoints. \\
The first model is the \Keyword{Computation Independent Model} (CIM). The CIM describes the system in a less detailed manner, omitting details about internal  processes, behavior and dependencies within the system. The \Keyword{Platform Independent Model} (PIM) specifies the architecture of the underlying system without implementation (platform specific) details. Hence, the model should be suitable for different platforms that are described by a \Keyword{Platform Specific Model} (PSM). PSMs are platform dependent views that use the platform specifications described by the PIM. Lastly, working code (or at least pseudo-code) will be generated from the PSM. \\
A \Keyword{Platform Model} provides a set of technical concepts, representing the different kinds of parts that make up a platform and the services provided by that platform. It also provides concepts representing the different kinds of elements in specifying the use of the platform by an application.\cite{MDA}\\

\begin{figure}[!htbp]
\begin{center}
	\includegraphics[width=.65\textwidth]{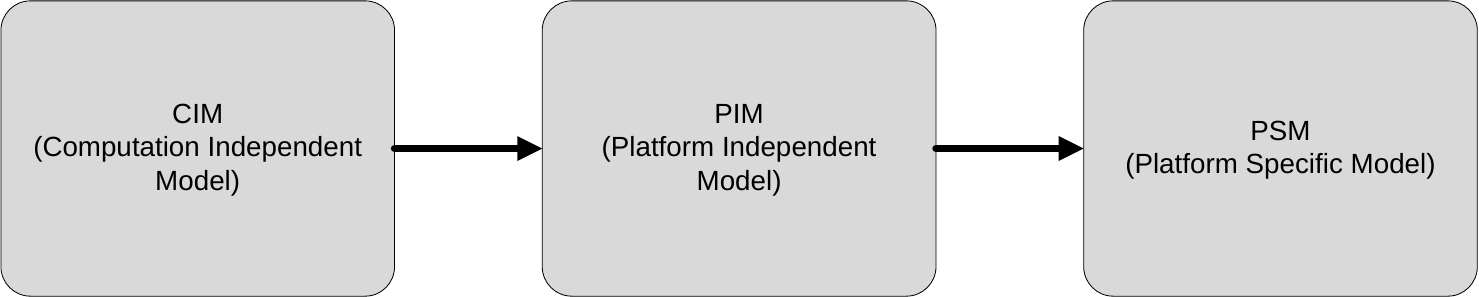}
   \caption{Dependencies in the Model Driven Architecture.}
   \label{fig:concepts:mdaDependencies}
\end{center}
\end{figure}

\begin{figure}[!htbp]
\begin{center}
	\includegraphics[width=.3\textwidth]{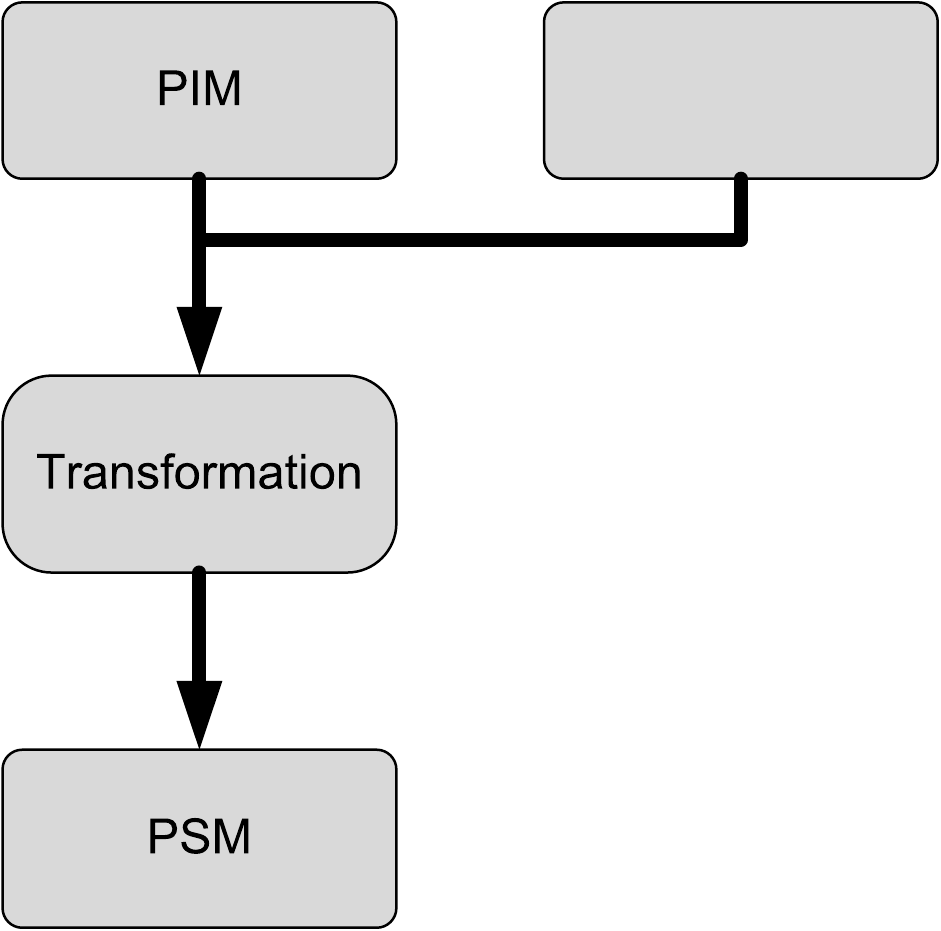}
   \caption[The MDA Pattern]{The MDA Pattern, source: \cite{MDA}, modified illustration}
   \label{fig:concepts:mdaPattern}
\end{center}
\end{figure}

The key concept of MDA is the \Keyword{Model Transformation}. In the Model Driven Architecture, additional information is added to a given PIM resulting in a PSM. These transformations are held as generic as possible. Any model can be transformed into another one in combination with a proper \Keyword{transformation specification}. This transformation is also called the \Keyword{MDA Pattern} (see \ref{fig:concepts:mdaPattern}). The types of transformations can be done manually, by using Profiles (UML), by using patterns or markings, or automatically. In the latter case, the design of PSMs becomes obsolete - all information for PIM to PSM transformations are already included in the PIM. The MDA pattern is not a one step process - it can be used in an incremental manner, meaning that a PIM can be transformed into various different PIMs before the generation of the PSM occurs. 

\paragraph{Other Approaches}$~~$\\

MDA might be the first thought when it comes to \mdd, but other approaches exist which are not of lesser significance.
\vspace{2cm}

\paragraph{Software Factories}$~~$\\

A \Keyword{software factory} pattern faciliates the creation of individual software by assembling components. Software factories are primarily used for interoperability while focusing more on productivity in contrary to \ac{MDA}.\cite{greenfield1}

\paragraph{Generative Programming}$~~$\\

\Keyword{Generative Programming (GP)} gained popularity with the release of the book "Generative Programming"\cite{eisenecker2000} which defines GP as follows:
\begin{quote}
"Generative Programming is a software engineering paradigm based on modeling software system families such that, given a particular requirements specification, a highly customized and optimized intermediate or end-product can be automatically manufactured on demand from elementary, reusable implementation components by means of configuration knowledge."
\end{quote}

\paragraph{Model-integrated Computing}$~~$\\

In embedded systems, it is crucial to incorporate methods for a fast, secure and reliable real-time environment. \Keyword{Model-integrated Computing} (MIC) sets its task to support development in this area, by providing an architecture for model-driven design - comparable to MDA. MIC is described on the ISIS page: 
\begin{quotation}
"MIC focuses on the formal representation, composition, analysis, and manipulation of models during the design process. It places models in the center of the entire life-cycle of systems, including specification, design, development, verification, integration, and maintenance."\footnote{\url{http://www.isis.vanderbilt.edu/research/MIC}}
\end{quotation}
	\cleardoublepage

	\chapter{Modeling Tools}\label{chapter:mbd}

Product requirements have been raising in parallel with cheaper embedded hardware. The complexity introduced by more demands and tight schedules require special techniques and tools to meet systems engineering requirements:
\begin{itemize}
\item \textbf{Minimized Costs}. The development effort should be kept as low as possible.
\item \textbf{Reduced Risks}. System complexity should be understood as early as possible, changes in late project phases and resulting costs avoided.
\item \textbf{Timely Delivery}. Time-to-market should be minimized and long overdues of deadlines prohibited.
\item \textbf{Safe Design}. Safety and reliability standards are of importance in safety-critical embedded systems.
\end{itemize}

Modeling Techniques, including domain-architectures, domain-specific languages and best practices, provide ways for designing embedded systems with models. These techniques are utilized by modeling tools such as the commercial tool-set SCADE and Mathworks MATLAB, or the scientific prototyping software SynDEx and the hybrid system simulation environment Scicos.\\

Formal techniques abstract the characteristics of embedded systems by models. Such models can be viewed in different ways to support the understanding of the system. Formal models have the advantage of being verifiable with formal verification techniques (theorem provers, model checkers, tests) and the possiblity of automatic code generation. In some cases it is possible to obtain qualification/certification for code-generators (for example the KCG code generator in the SCADE suite).\\

This thesis investigates the Scics/SynDEx framework capabilities for model-based development. Scicos is based on a Signal-like formalism. Signal is a synchronous language. SynDEx is a tool for the distribution and temporal design of distributed systems and implements the AAA-Methodology that optimizes/distributes a given algorithm onto a target-architecture.

\section{Scicos}\label{sec:mbd-tools:scicos}

\Keyword{Scicos} (Scilab Connected Object Simulator) is a simulator toolbox and graphical system modeler and focuses on the design and simulation of hybrid systems (section \ref{sec:models:hybridSystem}) and a variety of DAE hybrid systems. Furthermore, a C code generator is included. Scicos finds its purpose in several physical and biological domains like signal processing and systems control. The cornerstone for designing continuous dynamics is a Simulink-like language extending a synchronous language. The main features of Scicos (Version 4.2) are\footnote{Features are listed on \url{http://www.scicos.org}}:\\

\begin{itemize}
\item Graphical models, simulation and compilation of dynamical systems.
\item Combination of continuous and discrete-time behaviors in the same model.
\item Selection of model elements from palettes of standard blocks.
\item Programming of new blocks in C, Fortran, or Scilab language.
\item Run simulations in batch mode from the Scilab environment.
\item Generate C code from Scicos model using a code generator.
\item Run simulations in real-time with and real devices using Scicos-HIL.
\item Generate hard real-time control executables with Scicos-RTAI and Scicos-FLEX.
\item Simulate digital communications systems with Scicos-ModNum.
\item Use implicit blocks developed in the Modelica language.
\item Discover new Scicos capabilities using additional toolboxes.
\end{itemize}\label{scicos_featurelist}

Scicos is a toolbox of the scientific laboratory software \Keyword{Scilab}, a scientific open source software package for numerical computation which has been distributed freely since 1994. \ac{INRIA} and ENPC\footnote{ENPC - École nationale des ponts et chaussées, Engineering Institute University Paris-Est.} have been developing Scilab with commitments of the Scilab Consortium, which was established in 2003 (16.03.2003) and is holding 25 members\footnote{June, 2007. See http://www.scilab.org.}. Signs for the popularity of Scilab are some notable companies and organizations like Axs Ingénierie, Cril Technology, CEA, CNES, Dassault Aviation, EDF, ENPC, Esterel Technologies, INRIA, PSA Peugeot Citroën, Renault and Thales being members of the consortium as well as about $10.000$ downloads of the Scilab package per month \cite{scilabdownloads}.\\

Since this software is used for some research in following sections of this thesis, more pages were spent to introduce this software in a little more detail. In literature, the book "Modeling and Simulation in Scilab/Scicos"\cite{Campbell2006} is devoted to Scilab and Scicos - this section's work is leaned strongly towards it, equations and examples are taken from the book.

\subsubsection{Scilab/Scicos Models}

Scicos provides several ways of building models with different mathematical approaches, such as ordinary differential equations, boundary value problems, difference equations, and differential algebraic equations - they are described briefly in the following.

\paragraph{Ordinary Differential Equations.}
Scilab always assumes that an \Keyword{ordinary differential equation (ODE)} is written in first-order (\ref{equ:differential}). An ODE is is a differential equation with one independent variable (usually the time). ODEs of higher orders can be rewritten into ODEs of first-order by introducing new variables (see equation \ref{equ:differentialToFirst}).

\begin{equation}\label{equ:differential}
y=f(t,y)
\end{equation}

\begin{equation}\label{equ:differentialToFirst}
\begin{split}
\mbox{The second-order ODE:}\\
\ddot{y}_1=\dot{y}_1-y_2+\sin(t),\\
\ddot{y}_2=3\dot{y}_1+y_1-4y_2\\
\mbox{can be written as:}\\
\dot{y}_1=y_3,\\
\dot{y}_2=y_4,\\
\dot{y}_3=y_3-y_2+\sin(t),\\
\dot{y}_4=3y_3+y_1-4y_2.
\end{split}
\end{equation}

\subparagraph{Boundary Value Problems.}
A \Keyword{boundary value problem (BVP)} is a differential equation with boundary conditions. A two-point boundary value problem has the general form of equation \ref{equ:bvp}.
\begin{equation}\label{equ:bvp}
\dot{y}=f(t,y),\quad t_0 \leq t \leq t_f,\newline
0=B(y(t_0),y(t_f)).
\end{equation}

\paragraph{Difference Equations.}
Difference equations find their utilization in problems with discrete data - data values change only at discrete points in time.  Difference equations can often be approximated by numerical computations of differential equations. A solution solving a difference equation is a sequence $y(k)$ (equation \ref{equ:difference}).
\begin{equation}\label{equ:difference}
y(k+1)=f(k,y(k)),\quad y(k_0)=y_0.
\end{equation}

\paragraph{Differential Algebraic Equations.}
\Keyword{Differential Algebraic Equations (DAEs)} are a general form of differential equations for vector-valued functions composed of differential and algebraic equations (equation \ref{equ:dae}). A characteristic of DAEs is that they might only be solvable for certain initial conditions (so called consistent initial conditions). Scicos is able to solve index-one DAE systems. If a DAE model is given there are two ways: Rewrite the DAE model into a simpler DAE or an ODE.
\begin{equation}\label{equ:dae}
F(t,y,\dot{y})=0
\end{equation}

\subsubsection{Scicos Model Abstraction and Simulation}

\begin{figure}[!htbp]
\begin{center}
	\includegraphics[width=0.75\textwidth]{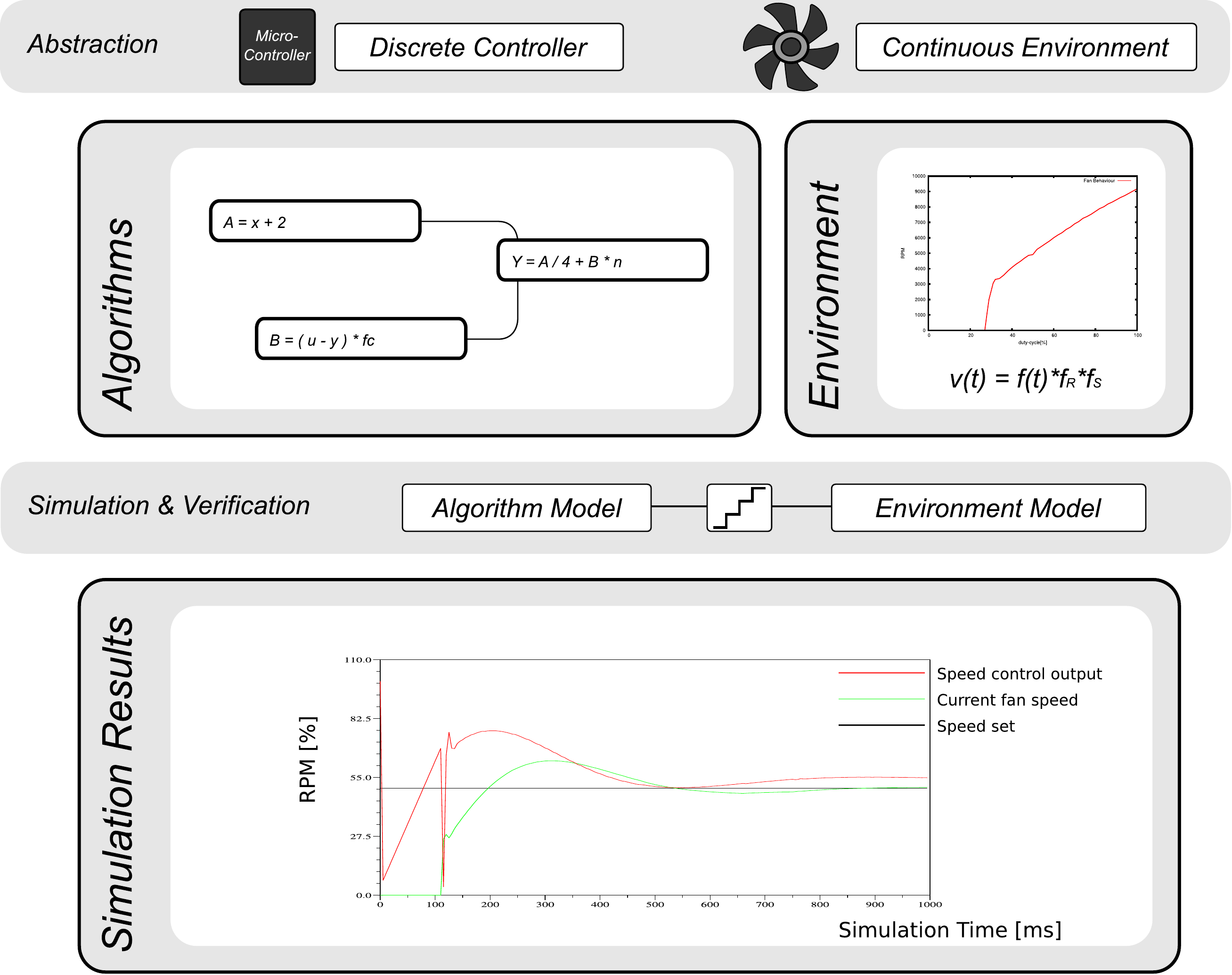}
   \caption[Scicos - Model abstraction and simulation]{Scicos - Model abstraction and simulation.}
   \label{fig:scicos:modelAbstraction}
\end{center}
\end{figure}

Subsystems are modeled for the control algorithms residing in the discrete domain and the environment located in the continuous domain. These two models are interfaced to each other, date and time discretized which enables the simulation of the complete system (see figure \ref{fig:scicos:modelAbstraction}).

\subsubsection{Scicos Formalism}

The formalisms used in Scicos are based on Signal and its extension to continuous-time systems \cite{NAN2003}. The simulator of Scicos uses two standard ODE/DAE numerical solvers which handle the continuous and discrete parts of the diagrams. A Scicos diagram is made by combining blocks connected via signals while single blocks offer a good opportunity to distribute the development to several teams.\\

\paragraph{Scicos Syntax}$~~$\\

\begin{figure}[!htbp]
\begin{center}
	\includegraphics[width=0.50\textwidth]{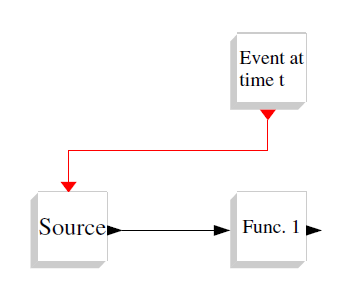}
   \caption[Simple Scicos diagram]{A simple Scicos diagram, source: \cite{Campbell2006}}
   \label{fig:scicos:scicosInheritance}
\end{center}
\end{figure}

\Keyword{Blocks} are connected via \Keyword{regular signals} and \Keyword{activation signals}. In that way a data precedence (regular signal) between computational functions (blocks) is achieved and the activation time of blocks (activation signal) is modeled.  Figure \ref{fig:scicos:scicosInheritance} depicts a simple Scicos diagram: An event generation block (clock block “Event at time t”) evolves the activation signal (signal between clock and "Source"). The activation signal triggers the functional block “Source”. The two regular blocks “Source” and “Func. 1” have a data signal (black) connecting them. The data signal models the passing of data between blocks and shows the “Source” block preceding the “Func. 1” block.

\subparagraph{Scicos Blocks}$~~$\\

A \Keyword{Scicos Block} is a graphical representation of a simulation function. In general, there are two types of representing blocks: \Keyword{basic blocks} and \Keyword{super blocks}. The purpose of \Keyword{super blocks} is to contain Scicos sub-diagrams and \Keyword{super blocks} in a single block. Thus, a design-hierarchy is created which supports the readability of low-level diagrams. Scicos is delivered with a library of several standard blocks for building algorithms.\\

If customized blocks are needed, the \Keyword{Scifunc} block can be used to define an algorithm by Scilab expressions, or new \Keyword{basic blocks} can be constructed by implementing an \Keyword{interfacing function} and a \Keyword{computational function}. The interfacing function handles the graphical behavior in Scicos, input/output port definitions and the initialization states of the block, while the computational function specifies the behavior of the block during the simulation. How an example \Keyword{basic block} can be constructed is briefly described in the appendix (\ref{sec:appendix:scicos:createNewBlocks}). Blocks may take parameters for defining initial and behavioral values. Parameters can be immediate values or symbolic parameters defined in the Scicos diagram context. 

\begin{figure}[!htbp]
\begin{center}
	\includegraphics[width=0.90\textwidth]{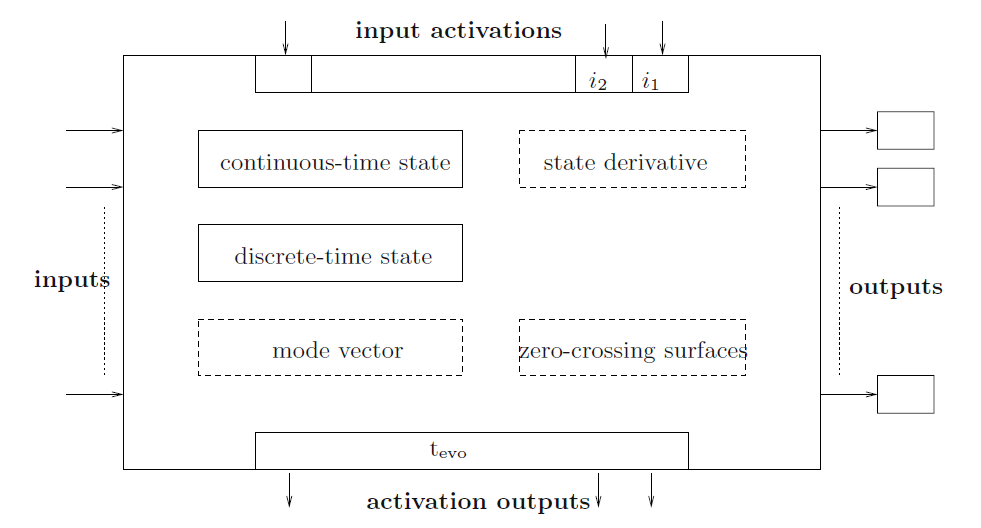}
   \caption[Scicos Block]{A Scicos Block.\cite{Campbell2006}}
   \label{fig:scicos:scicosBlock}
\end{center}
\end{figure}

A Scicos block comprises several components (figure~\ref{fig:scicos:scicosBlock}):

\begin{itemize}
\item \textbf{Regular inputs and outputs.} Vector input are passed through regular paths to the vector outputs.
\item \textbf{Input and output activations.} Activation paths link input event signals to output activation signals.
\item \textbf{Continuous-time vector state.} The continuous time state of the block.
\item \textbf{Discrete-time vector state.} The discrete time state of the block.
\item \textbf{Zero crossing surfaces and Mode vector.}  Inside Scicos the numerical integrator has problems with functions being not continuously differentiable. A mode can be defined to make sure that the numerical integrator never reaches such points of discontinuity inside an integration interval.
\item \textbf{State derivative.}
\end{itemize}

Depending on the combination of the simultaneously included features and how a block is activated, Scicos blocks can be classified into four different groups: \Keyword{Continuous Blocks}, \Keyword{Discrete Blocks}, \Keyword{Zero-Crossing Blocks} and \Keyword{Synchro-Blocks} \cite{LanguagesToolsForHybridSystemsDesign}.\\

A \Keyword{Continuous Basic Block (CBB)} continuously monitors its input ports and updates its output ports and states. A \Keyword{Discrete Basic Block (DBB)} is activated only if it receives an event on its activation input. A \Keyword{Zero-Crossing Basic Block (ZCBB)} is activated only if one of its regular inputs crosses a not differentiable point (e.g. crossing zero in an absolute value function). \Keyword{Synchro Basic Blocks (SBB)} can generate output activation signals that are synchronized with their input events - these blocks are the "event select block" and the "if-then-else block" represented in C code as "switch" and "if then else" respectively. These blocks can for example be used for a frequency division of an activation signal. A summary of the block types with their corresponding features is listed in table \ref{fig:scicos:blockTypes}.
\begin{table}[!Htbp]
\begin{center}
\begin{tabular}{|l|c c c c c c|}
\hline
& \begin{sideways}Regular Input\end{sideways} \vline \xspace & \begin{sideways}Regular Output\end{sideways} \vline \xspace	& \begin{sideways}Discrete State\end{sideways} \vline \xspace & \begin{sideways}Continuous State \xspace \end{sideways} \xspace \vline \xspace & \begin{sideways}Activation Output \xspace \end{sideways} \vline \xspace & \begin{sideways}Activation Input\end{sideways}\\
\hline
CBB & $\surd$ 	 & $\surd$ 		& $\surd$			 & $\surd$							& $\surd$	   & $\surd$ \\
DBB & $\surd$	& $\surd$ & $\surd$ & $\emptyset$ & $\surd$ & $\geq 1$ \\
ZCBB & $\surd$ & $\emptyset$ & $\emptyset$ & $\emptyset$ & $\surd$ & $\emptyset$ \\
SBB	& $1$ & $\emptyset$ & $\emptyset$ & $\emptyset$ & $\geq 2$ & $1$ \\
\hline
\end{tabular}
\end{center}
\caption{Scicos - Basic Block Interfaces.}
\label{fig:scicos:blockTypes}
\end{table}

\paragraph{Scicos Semantics}$~~$\\

In the following a summarized view about the behavior of blocks and signals in Scicos is presented.

\subparagraph{Scicos Signals}$~~$\\

\begin{figure}[!htbp]
\begin{center}
	\includegraphics[width=0.75\textwidth]{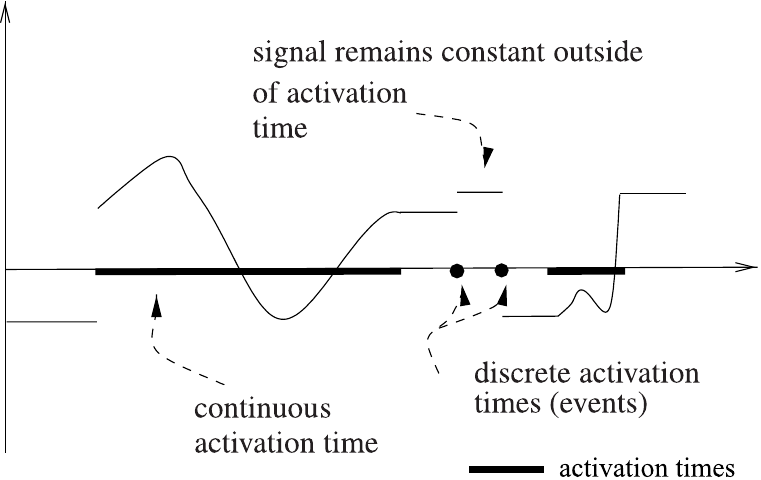}
   \caption[Scicos - Activation time dynamics]{Scicos - Activation time dynamics, source: \cite{Campbell2006}.}
   \label{fig:scicos:scicosTimeDynamics}
\end{center}
\end{figure}

Scicos activates blocks by activation signals. Such activations can make the block updating its respective in-/outputs, updating its states, and computing its state derivative. When such a block is activated depends on the activation type which can either be continuous or discrete. Continuous activation allows a signal to evolve permanently, while an \Keyword{event} triggers a block to update itself. Inbetween two events the regular signal remains unchanged (see figure~\ref{fig:scicos:scicosTimeDynamics}). These concepts provide a certain amount of time-control for the design of hybrid systems.

\begin{figure}[!htbp]
\begin{center}
	\includegraphics[width=0.75\textwidth]{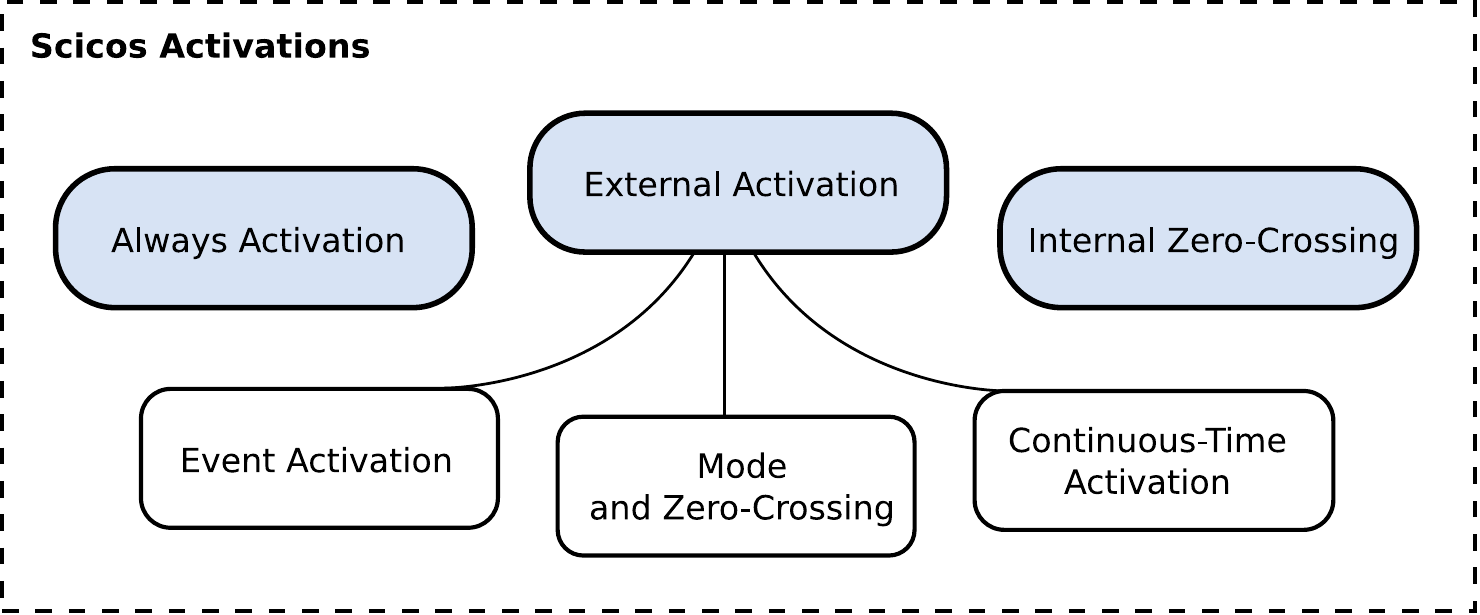}
   \caption{Scicos - Activations - Hierarchy.}
   \label{fig:scicos:scicosActivationsHierarchy}
\end{center}
\end{figure}

Scicos blocks are activated in three different ways - \Keyword{External Activation}, \Keyword{Always Activation} and \Keyword{Internal Zero-Crossing} (figure~\ref{fig:scicos:scicosActivationsHierarchy}).

\begin{itemize}
\item \textbf{External Activation}. A Block is activated when it either receives an activation signal on its activation input port, or inherits activation (\ref{sec:scicos:inheritanceAfter}). 

	\begin{itemize}
	\item \Keyword{Event Activation}. An event triggers the update of the block's output.
	\begin{equation*}
	\begin{split}
	y(t_e)&=f_1(t_e,x(t_{\bar{e}}),z(t_{\bar{e}}),u(t_e),\mu(t_e)).\\
	y(t_e)&\ldots \mbox{Vector-outputs.}\\
	u(t_e)&\ldots \mbox{Vector-inputs.}\\
	t_e&\ldots \mbox{Event time.}\\
	x(t_{\bar{e}})&\ldots \mbox{Continuous-time state.}\\
	z(t_{\bar{e}})&\ldots \mbox{Discrete-time state.}\\
	\end{split}
	\end{equation*}

A blocks can also implement its own activation outputs and provide thus a delay function:\\

	\begin{equation*}
	t_{evo}=f_3(t_e,x(t_{\bar{e}}),z(t_{\bar{e}}),u(t_e),\mu(t_e)).
	\end{equation*}

Updates of internal states are associated to following update-function:\\

	\begin{equation*}
	[z(t_e),x(t_e)]=f_2(t_e,x(t_{\bar{e}}),z(t_{\bar{e}}),u(t_e),\mu(t_e)).\\
	\end{equation*}

\item \Keyword{Continuous-Time Activation}. If a block is defined as \Keyword{always active} then it is activated at specified time intervals instead of events. The output depends on a continuous time activation period:\\
	\begin{equation*}
	y(t)=f_1(t,x(t),z(t),u(t),\mu(t)).
	\end{equation*}
		
\item \Keyword{Mode and Zero-Crossing}. If a function is not continuously differentiable (smooth) then the numerical integrator cannot post a solution. The \Keyword{mode} is used to avoid the integrator reaching such points inside an integration interval (the interval depends on the step size of the solver) by defining an integration start period. A \Keyword{Zero-Crossing Surface} is introduced to make sure the integration stops at points of discontinuity inside an integration interval.

   \end{itemize}

\item \textbf{Always Activation.} The block is defined to be always active (e.g. the Scicos sine-generator block). This is a special case of a continuous-time activation with a fictitious activation in put port.
\item \textbf{Internal Zero-Crossing.} The internal state of the block is updated if a zero-crossing occurs inside the block. Internal zero-crossing events are not predictable.
\end{itemize}

\subparagraph{Scicos Activation Inheritance}\label{sec:scicos:inheritanceAfter}$~~$\\

\begin{figure}[!htbp]
\centering
\begin{minipage}[b]{5cm}
	\includegraphics[width=0.9\textwidth]{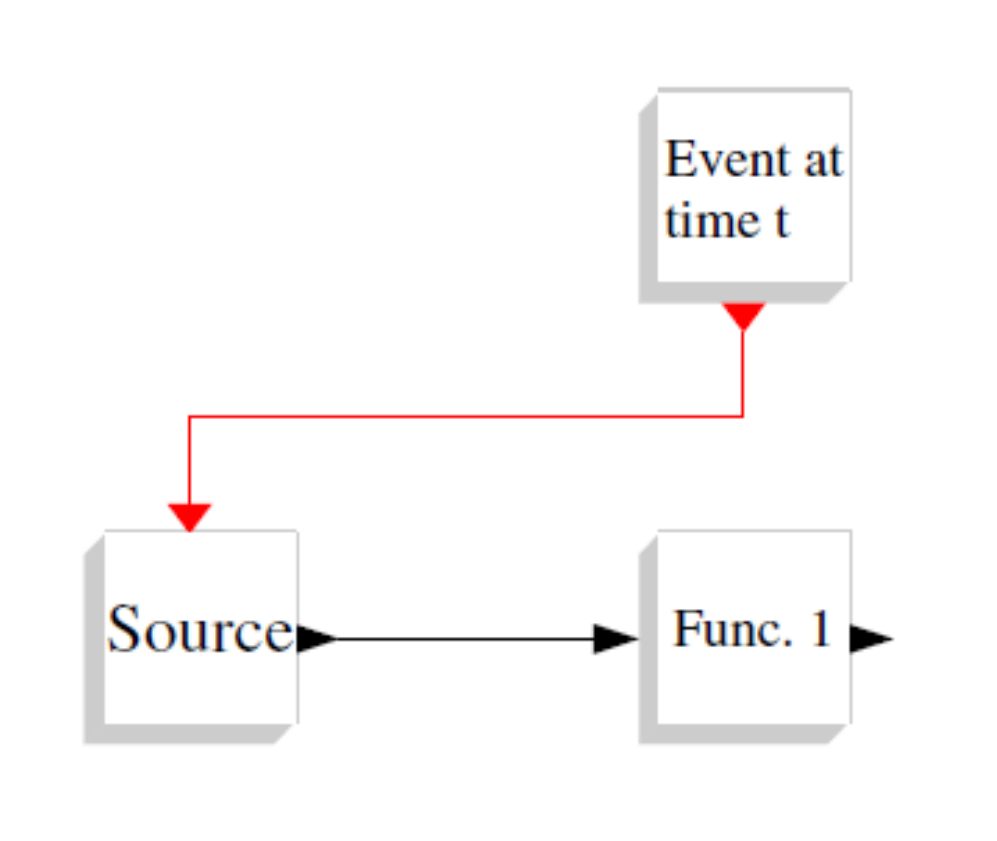}
\end{minipage}
\begin{minipage}[b]{5cm}
	\includegraphics[width=0.9\textwidth]{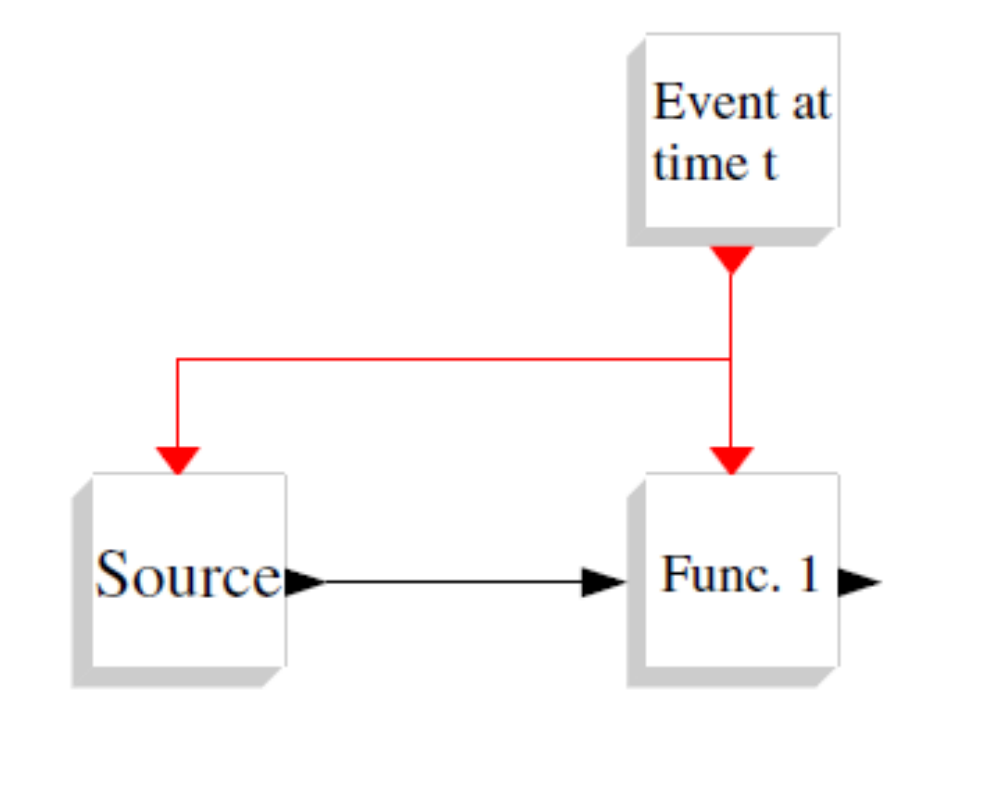}
\end{minipage}
   \caption[Scicos activation inheritance]{Scicos activation inheritance, before and after pre-compilation phase \cite{Campbell2006}.}
   \label{fig:scicos:scicosInheritanceConcept}
\end{figure}

Not all the Scicos blocks may have activation input ports. If a block does not, the activation is inherited through its regular input port. In figure~\ref{fig:scicos:scicosInheritanceConcept} a simple diagram is shown where block "Func. 1" inherits the activation from the block "Source" after the Scicos pre-compilation phase.

\subparagraph{Scicos Synchronism}\label{sec:scicos:synchronism}$~~$\\

\begin{figure}[!htbp]
\centering
\begin{minipage}[b]{6cm}
	\includegraphics[width=0.9\textwidth]{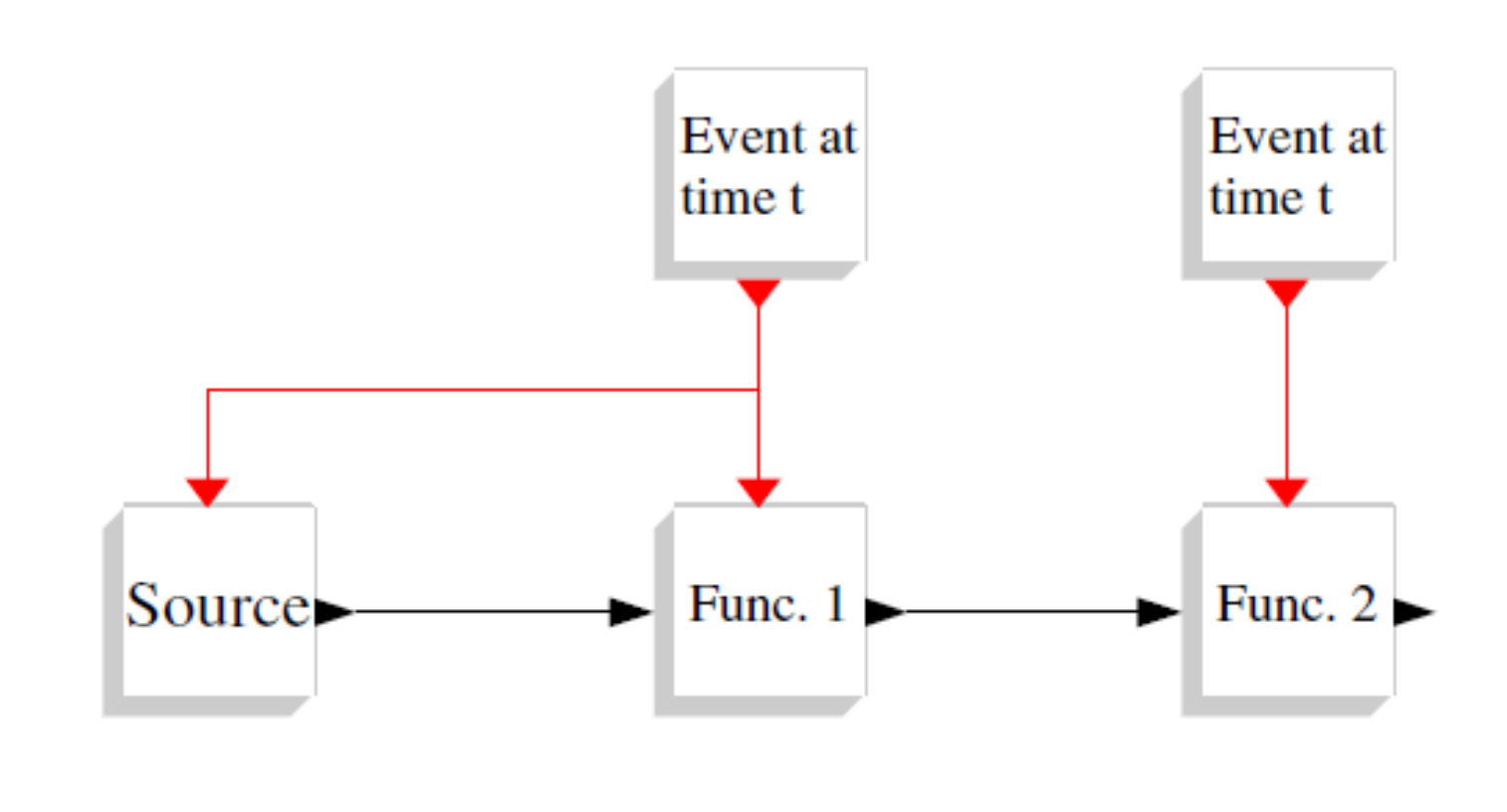}
\end{minipage}
\begin{minipage}[b]{6cm}
	\includegraphics[width=0.9\textwidth]{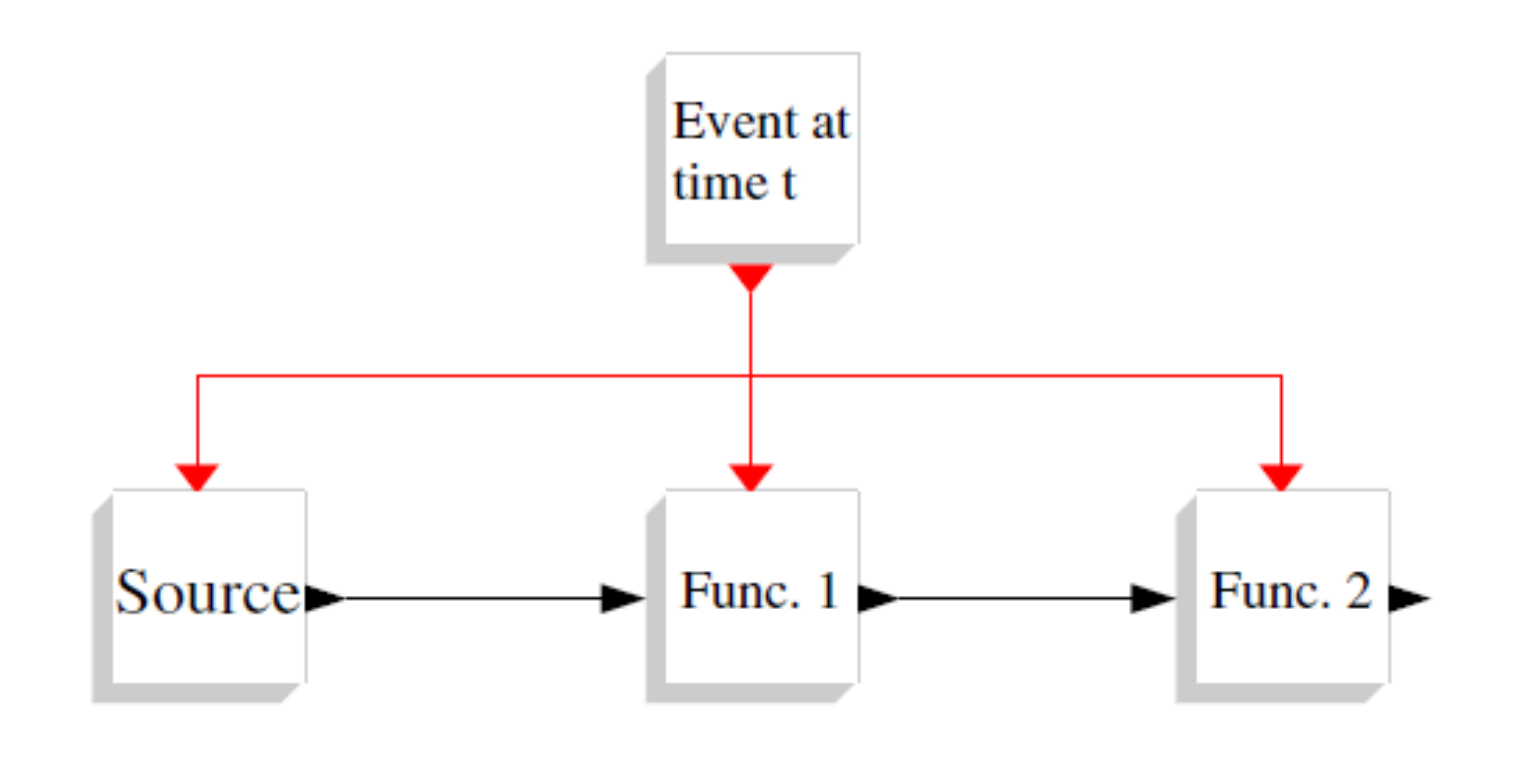}
\end{minipage}
   \caption[Scicos synchronism concepts.]{Scicos synchronism concepts. Asynchronous and synchronous diagrams, source: \cite{Campbell2006}.}
   \label{fig:scicos:scicosSynchronism}
\end{figure}

Scicos blocks are synchronized if they are activated by the same activation source (e.g. the \Keyword{Event Clock} block). If two blocks are connected together, Scicos executes them in the correct order. A design pitfall might be choosing two \Keyword{Event Clocks} with the same activation period to induce synchronism: This will not execute blocks synchronized - they might be activated by Scicos in any arbitrary order (see figure~\ref{fig:scicos:scicosSynchronism}).

\subparagraph{Memory Blocks}

\begin{figure}[!htbp]
\begin{center}
	\includegraphics[width=0.75\textwidth]{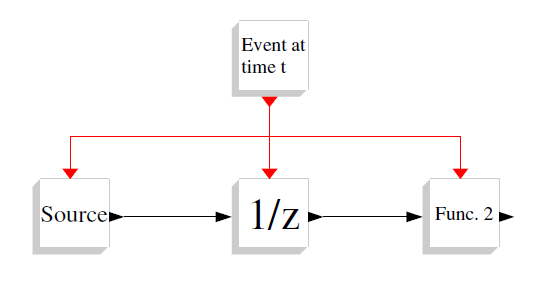}
   \caption[Scicos - Register block]{Scicos - Register block, source: \cite{Campbell2006}.}
   \label{fig:scicos:scicosRegister}
\end{center}
\end{figure}

Modeling a discrete delay in Scicos can be done with the \Keyword{Register} block: When it is activated it copies its internal state on the output and the input is copied into the internal state (see figure~\ref{fig:scicos:scicosRegister} for a simple Scicos diagram with a discrete delay).

\clearpage

\section{SynDEx}\label{sec:mbd-tools:syndex}

\Keyword{SynDEx (Synchronized Distributed Executive)} is a system level \acs{CAD} software based on the \Keyword{Algorithm Architecture Adequation (AAA)}\cite{mpcs94} and has been designed at \acs{INRIA} in the Rocquencourt Research Unit France, by the \acs{AOSTE} team. Its goals are to provide a tool for rapid prototyping and the optimization of distributed real-time embedded applications on multi-component architectures.

The key features of SynDEx are\footnote{as listed on \url{http://www.syndex.org/scicosSyndexGateway/index.htm}}:

\begin{itemize}
\item \textbf{Rapid prototyping} of complex distributed real-time embedded applications including automatic code generation.
\item \textbf{Automatic generation of safe and optimized distributed real-time code}. Formal verification of possible implementations can be done manually or automatically using the optimization heuristics based on multi-periodic distributed real-time scheduling analysis.
\item \textbf{Hardware/Software co-design through multi-component architecture} if some parts of the application must be implemented by software and run on processors, while others must be implemented by hardware and run on specific integrated circuits.
\item \textbf{Interface with domain specific languages} such as synchronous languages (\Keyword{Esterel}, \Keyword{SyncCharts}, \Keyword{Signal}) providing formal verifications, \acs{AIL}\cite{AIL2001} (a language for automobile), \Keyword{Scicos} a Simulink-like language, \Keyword{CamlFlow}\footnote{CamlFlow is a Caml to data-flow graph translator. Caml is a general-purpose programming language, designed with program safety and reliability in mind.} a functional data-flow language, \Keyword{UML2.0} with the \Keyword{\acs{MARTE}} profile, etc.
\item \textbf{System level CAD tool}. SynDEx offers a software environment reaching from the specification level (functional specifications, distributed hardware specifications, real-time and embedding constraints specifications) to the distributed real-time embedded code level, through (multiple) workstation functional and timing simulations.
\item \textbf{Interface with the integrated circuit level CAD tool} \Keyword{SynDEx-IC} which allows the implementation of a function (operation) onto a specific integrated circuit that can be used as a non-programmable component in the multi-component architecture.
\item \textbf{SynDEx is freeware} - free of charge for non-commercial applications. 
\end{itemize}

\subsubsection{SynDEx Models}

SynDEx leads the two fields of software and hardware development closer together by combining software algorithms (algorithm models) and highly abstracted target hardware (architecture model). The main focus of SynDEx lies in optimizing, scheduling and distributing algorithms under constraints (e.g.  given by the capabilities of the target hardware). Thus a hardware architecture model in SynDEx is a high level abstraction of the execution environment and is merely more than a pure description, while the algorithm models are abstracted at a level of the developer's desire, matched to constraints, computed and optimized.\\

In SynDEx models, information is transferred between the targets by a communication medium. A communication medium definition (e.g. a bus) contributes twofold: to the architecture model by defining the connections between the target processors, and the algorithm model, through the definition of durations for information transfers.\\

Summarizing, SynDEx combines three models covering several aspects in system development: An \Keyword{algorithm model} for the representation of computational functions, a \Keyword{medium model} defining the communication medium between the execution environments and an \Keyword{architecture model} describing the hardware structure and capabilities. With these models available, SynDEx performs an optimization of the computational function by algorithm adequation (\Keyword{adequation} is the seeking for an optimized implementation of an algorithm onto a distributed architecture by the execution of heuristics). The algorithm model is thus partitioned into several computational units, and these units are distributed onto the execution targets. Under consideration of hardware constraints (algorithm parts restraint to an execution environment) and timing properties of algorithm execution and communication media (durations), the optimization is performed by a greedy algorithm. This algorithm takes advantage of potential parallel structures, thus optimizes the execution time of the final executive. The result of the algorithm adequation is an algorithm distributed in portions over the whole hardware architecture, that has a total order for communication and algorithm actions and is guaranteed to be deadlock free.\\
This generated, distributed algorithm model can then be transformed into a \Keyword{macro code model (MCM)} based on a M4 implementation of the \acs{UNIX} macro processor. The target-independent MCM can then be translated into a target-dependent model (such as a programming language like C) with the use of customized M4 definitions. The resulting target-dependent model can then be translated into an executable model (e.g. by using a C compiler).\\

\paragraph{SynDEx Model Abstractions}$~~$\\

All SynDEx models reside in the discrete time and data domain. They are thought to be used for modeling discrete computation algorithms executed on a discrete hardware (e.g. microprocessors). Computational functions are interfaced to each other by: defined discrete data types, execution periods and durations. Algorithm portions, residing on different operators, communicate via the \Keyword{communication medium} which is an abstract representation of a synchronized communication channel.\\

\begin{figure}[!htbp]
\begin{center}
	\includegraphics[width=0.85\textwidth]{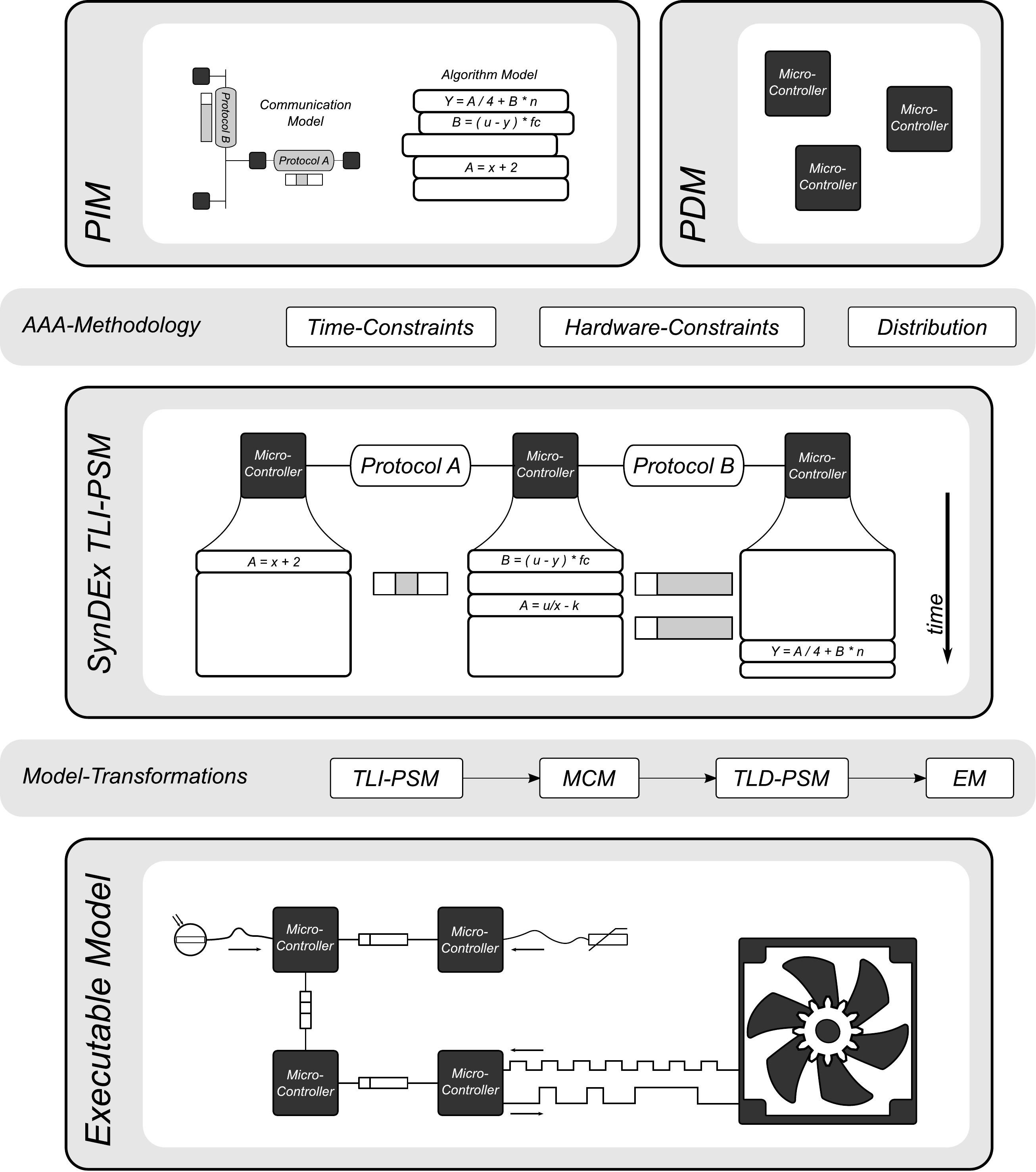}
   \caption{SynDEx: models and abstraction.}
   \label{fig:syndex:modelAbstractions}
\end{center}
\end{figure}

An algorithm model is conceptually a \Keyword{platform independent model (PIM)}, while the \Keyword{platform description model (PDM)} contains the hardware architecture information. Applying the algorithm adequation on these two models results in a partitioned algorithm that is dependent on the hardware in respect of timings and target placements. At this stage the algorithm represents a \Keyword{platform specific model (PSM)} - more precisely, a \Keyword{target-language independent PSM (TLI-PSM)}. The development-chain finishes with the transformation into a MCM, its expansion to a \Keyword{target-language dependent PSM (TLD-PSM)}, followed by a transformation into the \Keyword{executable (EM)} - see figure~\ref{fig:syndex:modelAbstractions}.
                                        
\subsubsection{SynDEx Formalism}

An \Keyword{algorithm model} is a (preferably) graphically specified directed, acyclic graph (DAG). The DAG consists of \Keyword{blocks} representing a sequence of operations, and \Keyword{signals} representing dependencies between operations.\\

\begin{figure}[!htbp]
\begin{center}
	\includegraphics[width=1.00\textwidth]{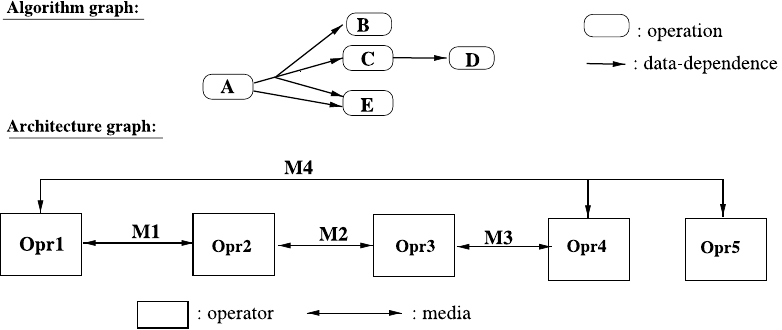}
   \caption[SynDEx formalism - Algorithm, architecture and communication models]{SynDEx formalism - Algorithm, architecture and communication models, source: \cite{codes99}.}
   \label{fig:syndex:AAAGraphs}
\end{center}
\end{figure}

An \Keyword{architecture model} describes the heterogeneous multiprocessor execution environment by a non-oriented hypergraph of operators. Operators in the architecture model are connected by a \Keyword{communication model}, also called a \Keyword{communication medium} (see figure~\ref{fig:syndex:AAAGraphs}).

SynDEx, as seen as a \acs{DSL}, is basically defined by a graphical syntax with semantics and corresponding meta-models defined by the \aaa. SynDEx's formal models are meant to describe a subset of the real-time embedded distributed systems domain.
\vskip 2cm

\paragraph{SynDEx Syntax}$~~$\\

In the DSL of SynDEx, graphically represented models (\Keyword{concrete syntax}) are instances of the AAA-Methodology's DAGs (\Keyword{abstract syntax}). Model corresponding semantics are defined by the AAA-Methodology. SynDEx-models can also be designed in a textual way, of course with the drawback of reduced cognitivity.\\

\begin{figure}[!htbp]
\begin{center}
	\includegraphics[width=1.00\textwidth]{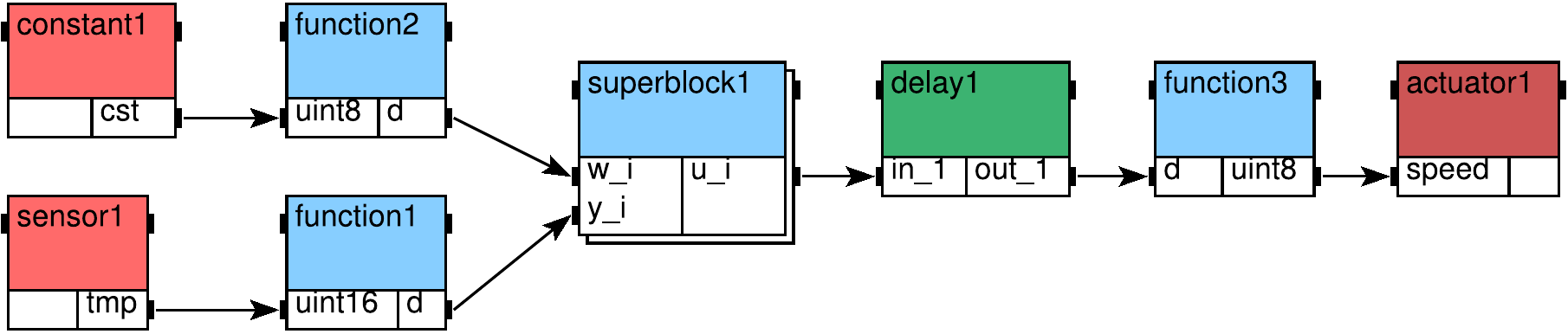}
   \caption[SynDEx syntax]{SynDEx syntax - graphically represented algorithm model containing all types of blocks.}
   \label{fig:syndex:demoAlgorithm}
\end{center}
\end{figure}

A simple, graphical overview of an algorithm model is depicted in figure~\ref{fig:syndex:demoAlgorithm}. A constant and a sensor block (e.g. temperature sensor) are providing data to the following functions \Keyword{function1} and \Keyword{function2} with the data types \Keyword{uint16} (16-bit unsigned integer) and \Keyword{uint8} respectively. After termination of function1 and function2 the data is forwarded to \Keyword{superblock1} (a block containing a sub-model). The results of this block are delayed by \Keyword{delay1}, computed in \Keyword{function3} (which takes a double data type as input, and provides an uint8 output) and finally passed to \Keyword{actuator1} (which could be e.g. an electric motor).

\begin{figure}[!htbp]
\begin{center}
	\includegraphics[width=0.75\textwidth]{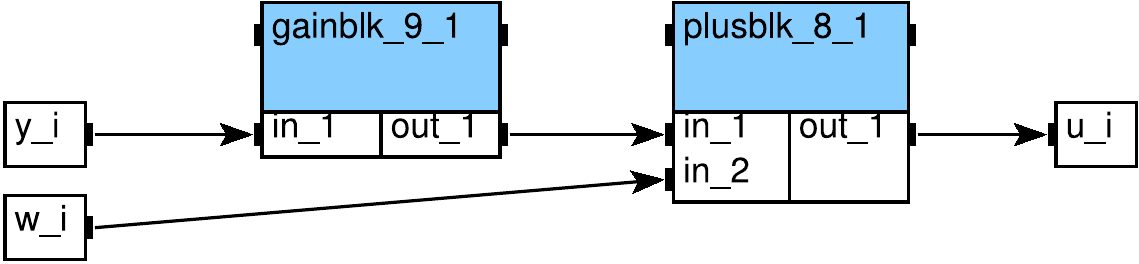}
   \caption[SynDEx syntax - Superblock internals]{SynDEx syntax - Internals of the superblock in figure~\ref{fig:syndex:demoAlgorithm}.}
   \label{fig:syndex:demoAlgorithmSuperblock}
\end{center}
\end{figure}

\subparagraph{SynDEx Blocks (Algorithm model)}$~~$\\

A SynDEx \Keyword{block} \Keyword{references} an \Keyword{algorithm definition}. A reference corresponds to exactly one algorithm definition, while a given algorithm definition may correspond to several references (a 1:n relationship). An algorithm definition is a platform-independent representation of a sequence of atomic operations, that means SynDEx considers all modeled blocks (except hierarchical blocks) to be not interruptable.\\

The characteristics of an \Keyword{algorithm definition} are:
\begin{itemize}
\item An algorithm definition \Keyword{name}.
\item \Keyword{Input / Output port definitions} with corresponding data-types.
\item \Keyword{Precedence dependencies} setting the logical execution orders between blocks.
\item \Keyword{Data dependencies} stipulating the order of the data flow between blocks.
\item \Keyword{Execution durations}. An integer value specifying the time this algorithm requires on a corresponding execution environment (e.g. the \ac{WCET} in processor clock cycles).
\item \Keyword{Operation Period} setting the periodically execution of the algorithm.
\item \Keyword{Allocation constraints} unchangeably pre-defining the allocation of the algorithm onto an execution environment.
\end{itemize}

Depending on the \Keyword{type} of the algorithm definition, appropriate input and output ports are declared (figure~\ref{fig:syndex:demoAlgorithm} holds an example of each type):
\begin{itemize}
\item \Keyword{Sensor}. A sensor block defines an input operation to the algorithm. In the algorithm model these are the first blocks called.
\item \Keyword{Actuator}. Actuators are outputs of the algorithm and thus the last operations called.
\item \Keyword{Function}. Function blocks represent a computational operation.
\item \Keyword{Constant}. These blocks act as a fixed value input to the algorithm and are also part of the first blocks called.
\item \Keyword{Delay}. Delay blocks act as discrete memory blocks.
\end{itemize}

\begin{figure}[!htbp]
\begin{center}
	\includegraphics[width=0.50\textwidth]{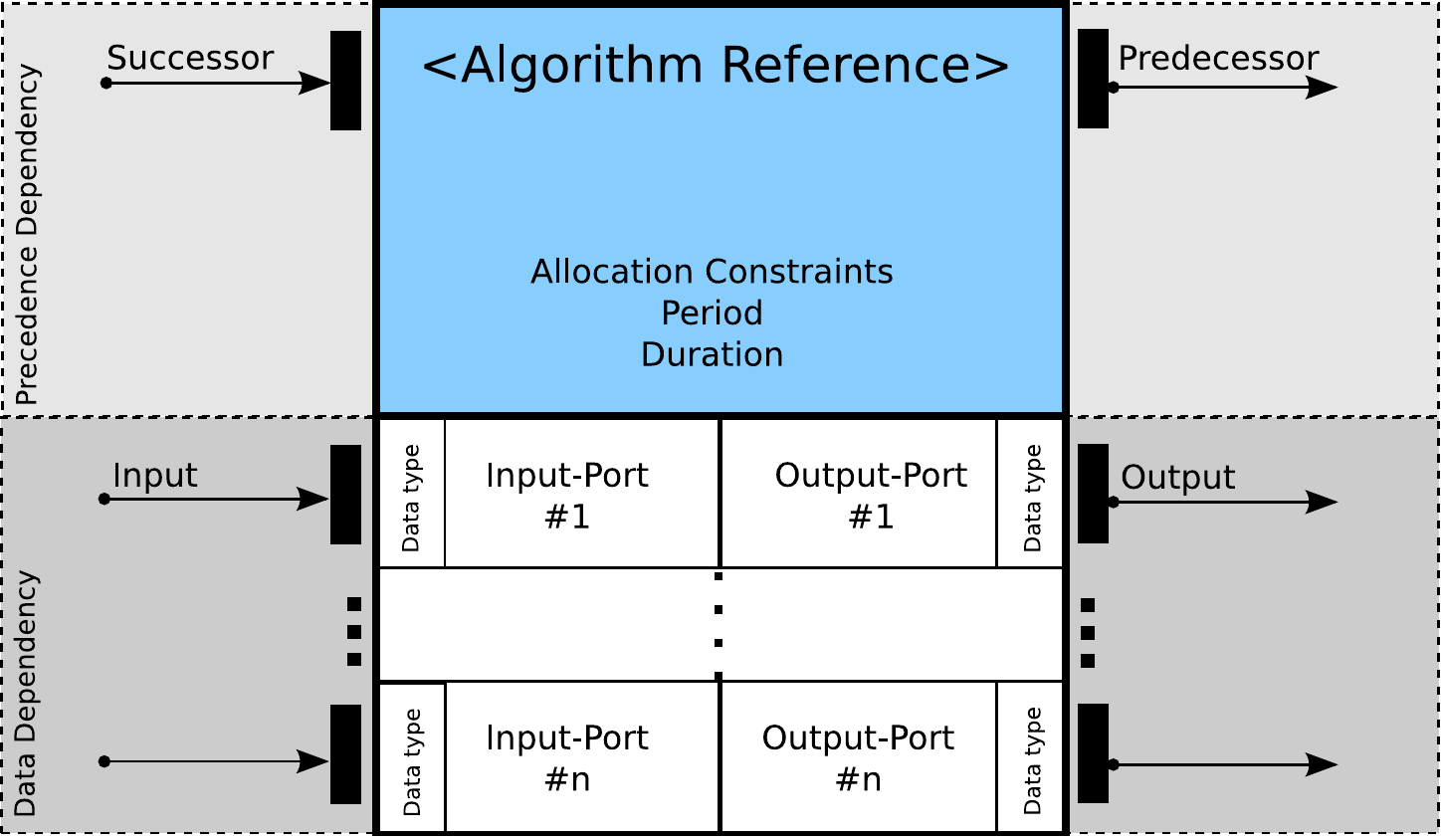}
   \caption{A SynDEx block.}
   \label{fig:syndex:algorithmDefinition}
\end{center}
\end{figure}

\begin{figure}[!htbp]
\begin{center}
	\includegraphics[width=0.10\textwidth]{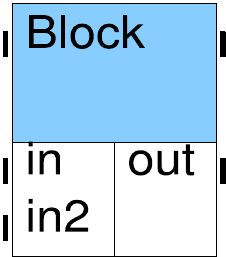}
   \caption[SynDEx block - editor view]{A SynDEx block as viewed in the graphical SynDEx editor with two inputs and one output port.}
   \label{fig:syndex:basicBlock}
\end{center}
\end{figure}

The scheme of a function-block can be obtained in figure~\ref{fig:syndex:algorithmDefinition}. \Keyword{Precedence dependencies} determine the order of block executions, thus they define if the block precedes or succeeds other blocks. \Keyword{Data dependencies} (in SynDEx so called \Keyword{strong precedence dependencies} assign the order of data-flows between blocks and by that also imply precedences. \Keyword{Input-ports} and \Keyword{output-ports} can be arbitrarily edited: Several ports can be declared with data types assigned to. \Keyword{Allocation constraints} are used to force SynDEx placing an algorithm on a desired target. For example this would be useful if special hardware functionalities are only available for one type of targets. If an algorithm shall be periodically called, the \Keyword{period} parameter of the block allows specifying execution periods. Before assigning periods or durations, a time-representation for the model has to be chosen: a relation between the time representation of the execution environment and the time representation in the SynDEx modeling environment. \Keyword{Durations} of algorithms can be defined manifold, for each operator (target) a duration has to be defined; SynDEx needs that information to perform the optimization and distribution of the algorithms. Figure \ref{fig:syndex:basicBlock} depicts the general scheme of a SynDEx blocks appearance in the graphical SynDEx editor.

\subparagraph{SynDEx Model Hierarchy and Conditions}$~~$\\

\Keyword{Superblocks} (the "superblock1", figure~\ref{fig:syndex:demoAlgorithm} and its contents in figure~\ref{fig:syndex:demoAlgorithmSuperblock}) can represent a set of blocks which lie on a lower level of the abstraction hierarchy. Finite automatons can be realized by superblocks with conditioned ports, that means an input port of the superblock realizes the state of the automatons like a switch statement (not only superblocks, but every block could be seen as a finite automaton). Additionally, the introduction of superblocks inside an algorithm model can increase the readability of the model. This kind of a block is not considered to be atomic, that means the set of connected blocks (a algorithm sub-model) beneath it might be partitioned by the \aaa.

\subparagraph{SynDEx Blocks (Architecture and Communication Model)}$~~$\\

An architecture model is built by \Keyword{operator blocks}, \Keyword{memory blocks} (with or without arbiter) and \Keyword{bus/mux/demux blocks} (with or without arbiter) connected by \Keyword{communication blocks}. Operators execute operations sequentially, communicators execute communications sequentially. 

\begin{figure}[!htbp]
\begin{center}
	\includegraphics[width=0.75\textwidth]{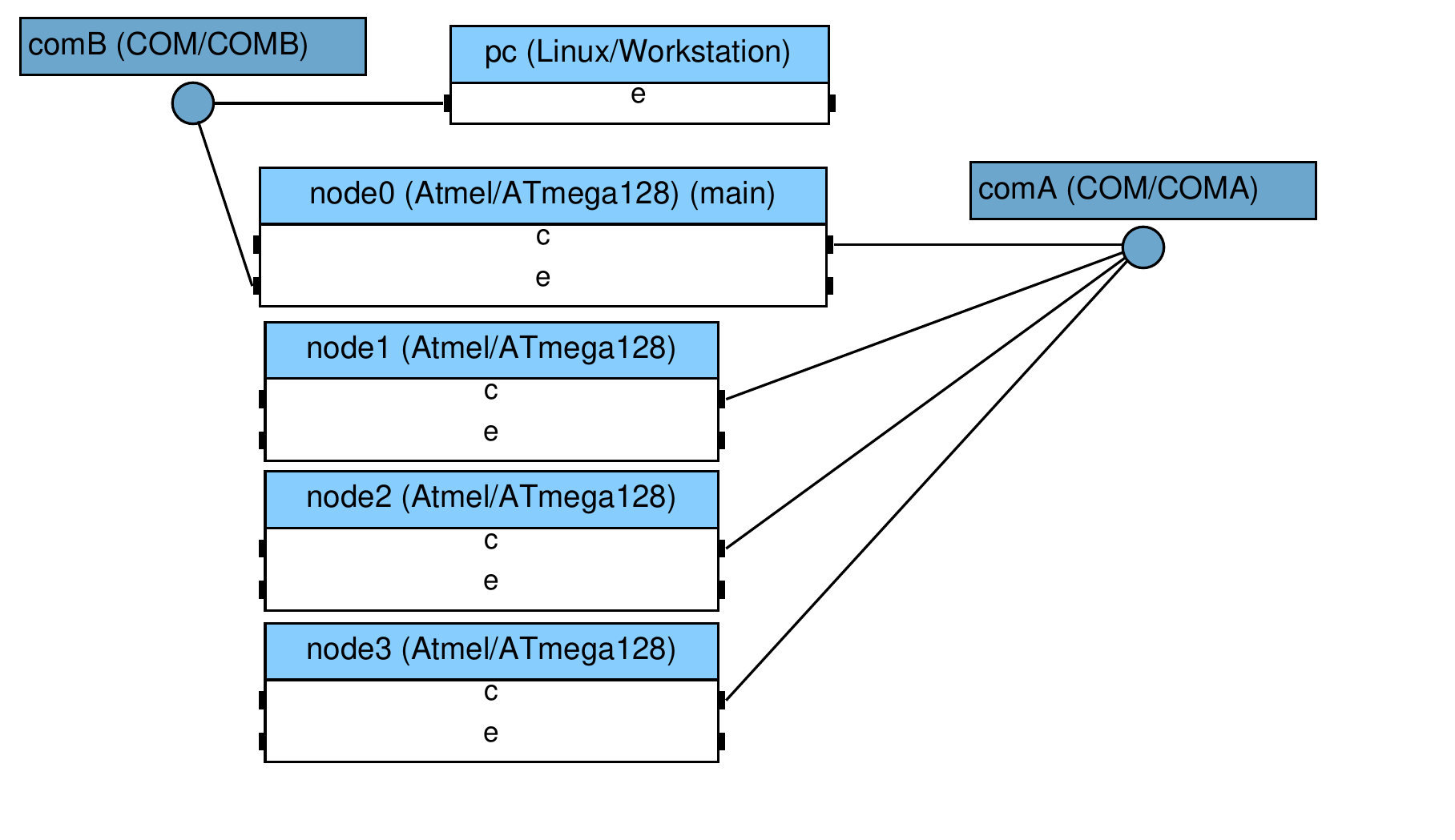}
   \caption[SynDEx architecture model]{SynDEx architecture model. Four nodes (ATmega128 microcontrollers) and a desktop PC (Linux workstation) are connected via communication gates "c" and "e".}
   \label{fig:syndex:architectureModeleBoard}
\end{center}
\end{figure}

The most important characteristics of an operator block (e.g. node0 in figure~\ref{fig:syndex:architectureModeleBoard}) are:
\begin{itemize}
\item \Keyword{Operator Name}. An identification.
\item \Keyword{Operator Type}. The type of the operator, e.g. an Atmel ATmega128 microprocessor.
\item \Keyword{Communication Gates}. They are used to specify full-duplex ports which may be used for the connection to communication blocks.
\end{itemize}

A \Keyword{communication block} is a model of a communication medium, which can be one of the two types: \Keyword{Sequential Access Memory (SAM)} and \Keyword{Random Access Memory (RAM)}. SAM can be defined as a \Keyword{point-to-point} medium where only two operators are connected to each other, or a \Keyword{multipoint} medium that allows several operators to be connected to each other via the communication medium.

An exemplary architecture model of a hardware setup containing four microcontrollers connected by a bus via the communication model \Keyword{comA}, and a desktop pc connected to target node0 via \Keyword{comB} is depicted in figure~\ref{fig:syndex:architectureModeleBoard}.

\paragraph{SynDEx Semantic (Algorithm  model)}$~~$\\

SynDEx blocks are connected via signals. Signals have the purpose of modeling data-flow and control-flow. Pure control-flow connections define rules for the ordering of the blocks in the final optimized SynDEx model while data-flow connections define data-flow (and imply control-flow) precedence dependencies. Blocks can only be connected, if all involved input and output ports are of the same data-type. This circumstance makes it clear how operations have to interface each other and therefore eliminates some issues of the error-prone process in data passing during the development process. \\

An algorithm model runs in a forever-loop, thus actuator blocks represent the last instances of a current iteration of the algorithm, while sensors and constant blocks stand for entry points at each iteration.

\subsubsection{SynDEx Modeling}

\begin{figure}[!htbp]
\begin{center}
	\includegraphics[width=0.90\textwidth]{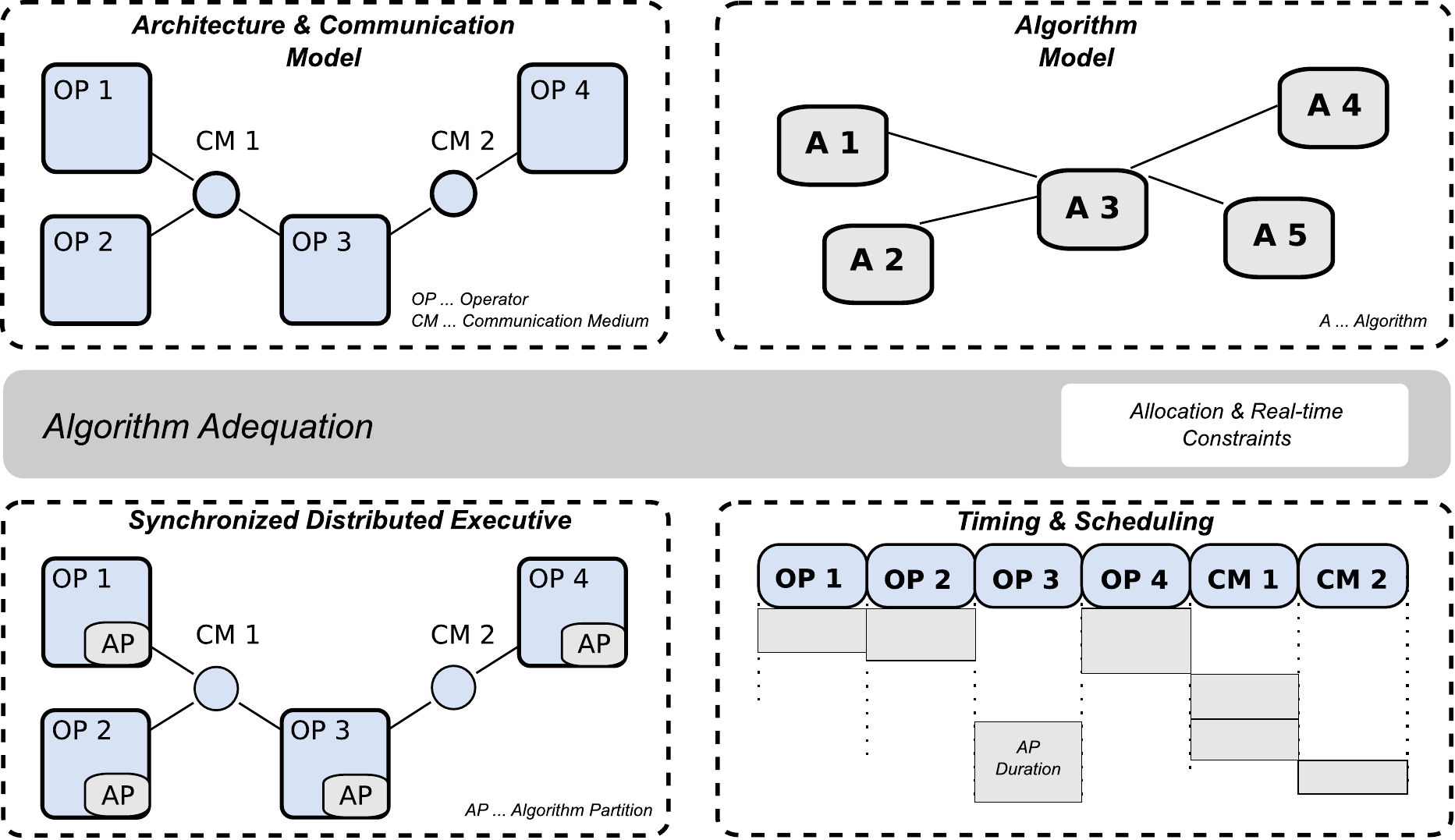}
   \caption{Design flow with SynDEx.}
   \label{fig:syndex:designFlow}
\end{center}
\end{figure}

Software development methodologies with SynDEx have strong similarities to rapid application development and prototyping. The design flow with SynDEx can be imagined as follows (figure~\ref{fig:syndex:designFlow}):

\begin{enumerate}
\item \Keyword{Algorithm model design}. The application is modeled with blocks and corresponding dependencies. For each block periods, durations (depending on the type of the target operator) and allocation constraints (the placement of a block onto a specific target) are defined.
\item \Keyword{Architecture and medium model design}. The architecture is modeled with operators connected by communication media.
\item \Keyword{Algorithm adequation execution}. SynDEx' algorithm adequation processes the models - the result is a synchronized executive having partitions of the algorithm distributed among the targets.
\item \Keyword{Assessment of the timing and scheduling model}. Blocks and their distribution onto the targets, timings and dependencies are graphically displayed. If the results are not satisfying the algorithm model has to be redesigned.
\end{enumerate}

If the resulting SynDEx model is schedulable and satisfying the requirements, a MCM can be transformed from SynDEx model. The macro code itself might then be transformed into an executive model by customized code generation definitions.

\paragraph{SynDEx Scheduling}$~~$\\

The following paragraphs start with a brief description of the SynDEx scheduling, especially the SynDEx scheduling tables. Later on a comparison of scheduling tables to real-world scheduling accompanied by a small discussion and proposals for the identified problems follows. Scheduling tables are results of the AAA methodology and presented graphically inside SynDEx.  In general, time representations of a scheduling table in SynDEx include:
\begin{itemize}
\item Medium Channels in multiprocessor architectures - Communication processes require channel resources, e.g. sending a data packet from one processor to another via an UART protocol.
\item Communication processes. Transfer of information from one processor to another.
\item Processors - Atomic tasks are executed on the processor with their periods and durations.
\item Tasks. Stateless, simple tasks with a duration and period. A simple task is a task that is executed until it's termination is reached without being interrupted \cite{Kopetz1997}.
\end{itemize}

Data is sent/received between targets by sender/receiver tasks connected by medium channels. In SynDEx every target is realized with its own communication thread running concurrently to the main routine, that means that the time consumption of a communication job includes two targets exchanging data (the sender and the receiver task). In fact there could be more targets involved that are simultaneously listening to the bus but discarding unintended packages.

\begin{figure}[!htbp]
\begin{center}
	\includegraphics[width=0.90\textwidth]{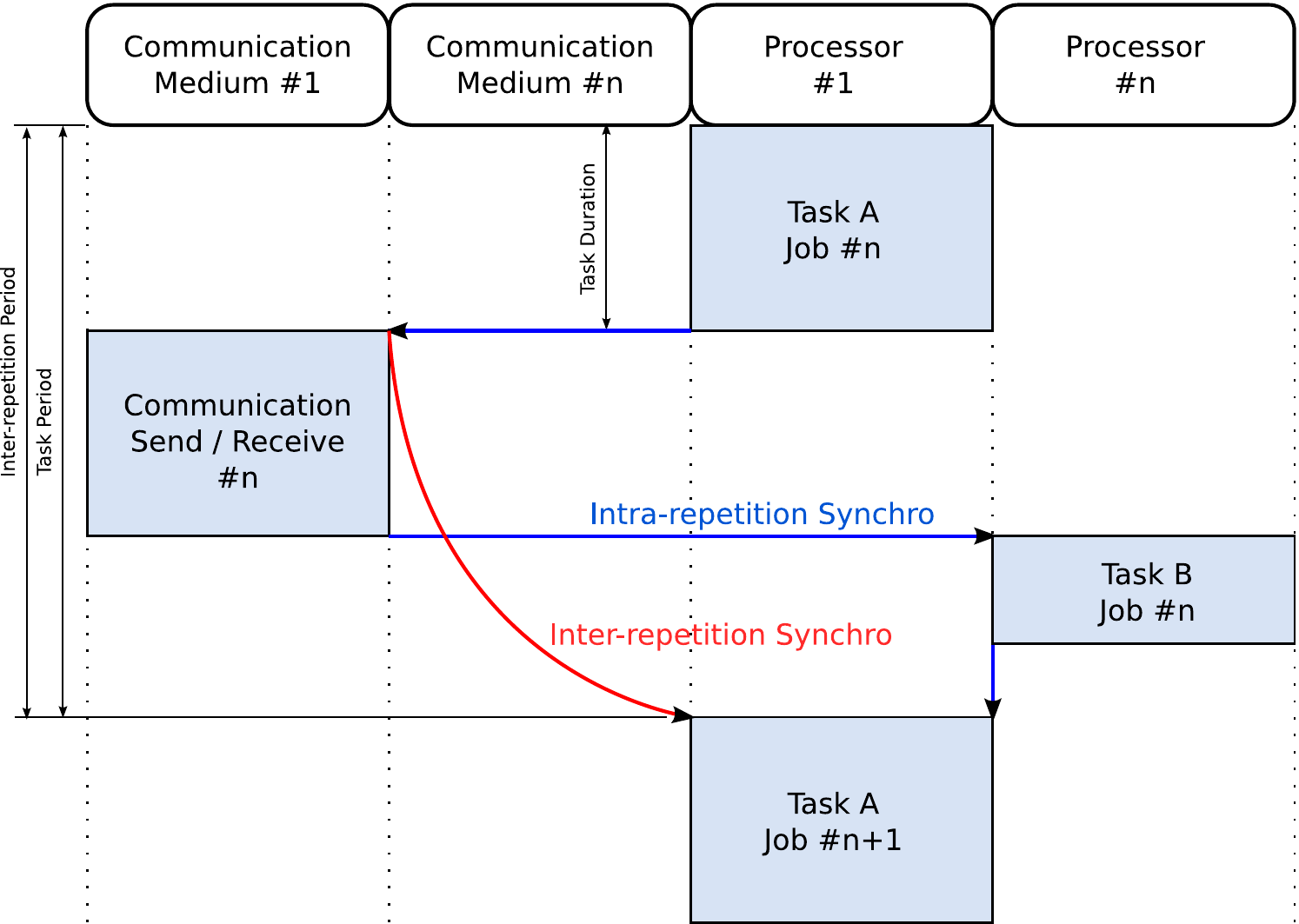}
   \caption{SynDEx - Scheduling table with two processors and two communication media.}
   \label{fig:syndex:schedulingTableBasics}
\end{center}
\end{figure}

The scheduling table does not hold the behavior of the algorithm during the init phase of the blocks. The scheduling table represents the timely behavior of the algorithm during the cyclic / periodic phases of the respective blocks, while init and end phases are not included. As depicted in figure~\ref{fig:syndex:schedulingTableBasics}, an algorithm is repeated after the \Keyword{Inter-repetition Period}. Every action in the distributed executive is synchronized. Synchronization can be thought of a token, or more tokens, being passed around from finished jobs to others. Preventing data-depending jobs being executed too early is done by creating communication threads and main loops on each target controlled by semaphores in such a way that tasks will be called in a correct execution order. \Keyword{Intra-repetition Synchros} order the execution of tasks within one whole execution of the algorithm, while \Keyword{Inter-repetition Synchros} delay the execution of tasks, in the scope of two successive repetitions of the whole algorithm, until their execution is needed.

\subparagraph{SynDEx Scheduling - Model versus Reality}$~~$\\ 

SynDEx creates a scheduling which resides conceptually in an ideal world. In the ideal world, execution durations of tasks are constant, therefore the same is valid for the periods of tasks. In a SynDEx multiprocessor architecture, all processors have a synchronized start-up and do not have clock drifts. All these characteristics have to be assumed to keep a scheduling model simple, and in cases of higher complexity even calculable.

\begin{figure}[!htbp]
\begin{center}
	\includegraphics[width=0.90\textwidth]{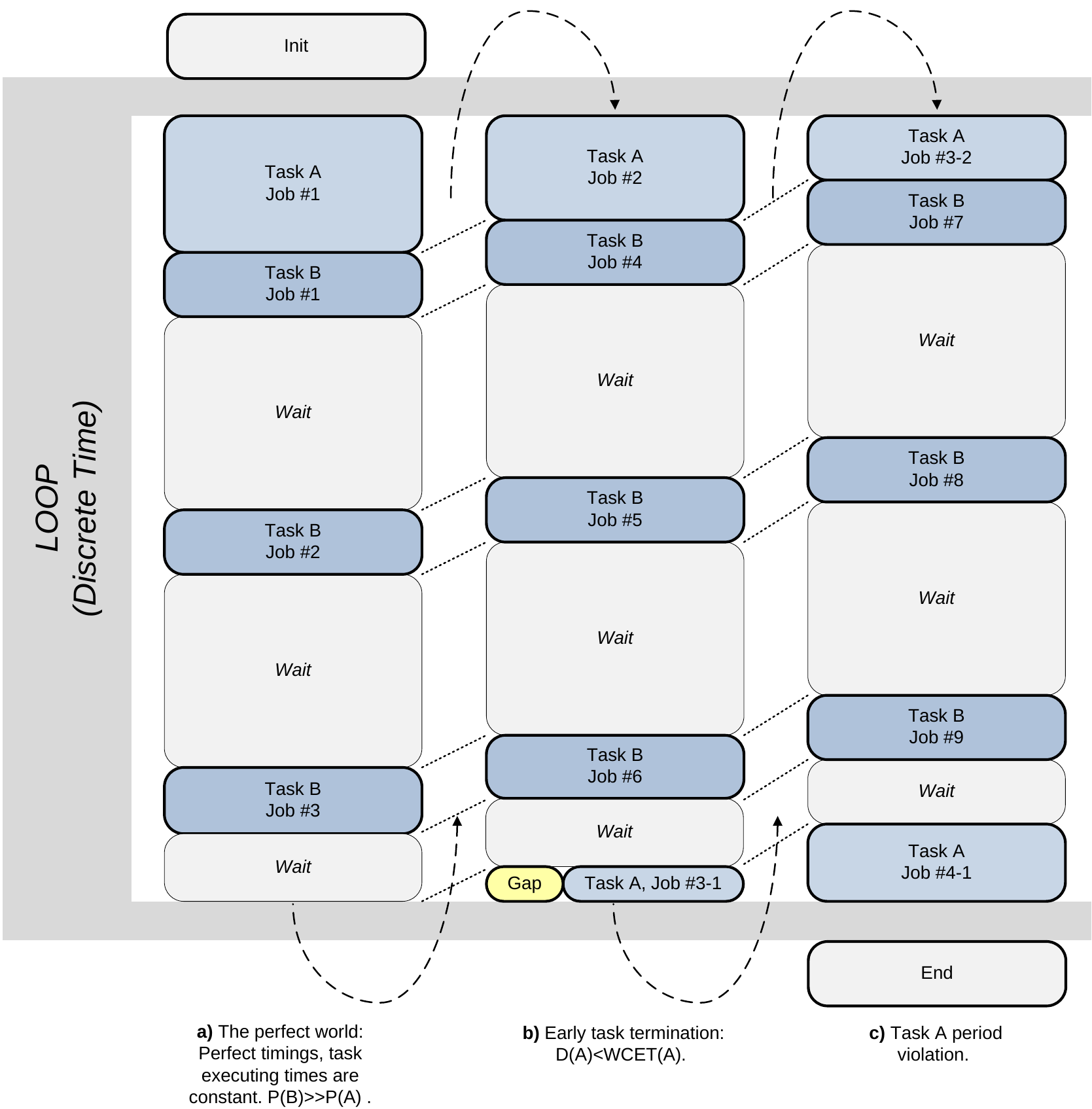}
   \caption[SynDEx - Scheduling with tasks of varying execution times]{SynDEx - Scheduling with tasks of varying execution times. a) In the perfect (modeled) world, all task durations and periods are exactly known and do not change during execution. The period of Task A $\mathbf{P(A)}$ is much smaller than the period of Task B $\mathbf{P(B)}$.
b) Executed on the target hardware, Task A terminates earlier than its WCET (duration of Task A $\mathbf{D(A)}\:<\: WCET(A)$). Therefore a gap in the scheduling occurs, which makes Task A being executed earlier in the next repetition of the loop.
c) Due to (b), the period of Task A is violated.
}
   \label{fig:syndex:schedulingGap}
\end{center}
\end{figure}

Every task requires the definition of durations (for the internal sequential calculations). In a general point of view, tasks can have various execution durations which are not preliminary known. Choosing the worst case execution time (WCET) for the task duration might be common practice because it is used to layout the system in the case of maximal load. Even though this sounds fairly reasonable, there are several issues arising when it comes to the realization of a scheduling model in the real world (see figure~\ref{fig:syndex:schedulingGap}). The tasks in this figure shall be seen as stateless, simple tasks which are just executed after another without any external synchronization (e.g. a timer unit that ensures the tasks exactly being executed at modeled times). Modeling task durations and fitting those to the real world, requires a stronger way of modeling which considers various possible execution times of tasks - in SynDEx this is for sure a deviation between the model and the real world. The problem of tasks terminating at unknown times is not just relating to SynDEx models, but rather a general issue.

Period violations occur if there are no synchronizing units responsible for the regular execution of the task. In  a pure event-driven task model there is no guarantee that the tasks will be executed exactly in their defined period, but might rather be executed earlier. Information are pushed in this kind of system. In a best-effort application these circumstances might be acceptable, but if an application with fault-tolerance is needed, e.g. taking sensor samples at fixed time periods, this kind of model might not be sufficient. Early execution of the tasks might even lead to some tasks being executed more times inside a time interval if there is no synchronization available.\\

\begin{figure}[!htbp]
\begin{center}
	\includegraphics[width=0.90\textwidth]{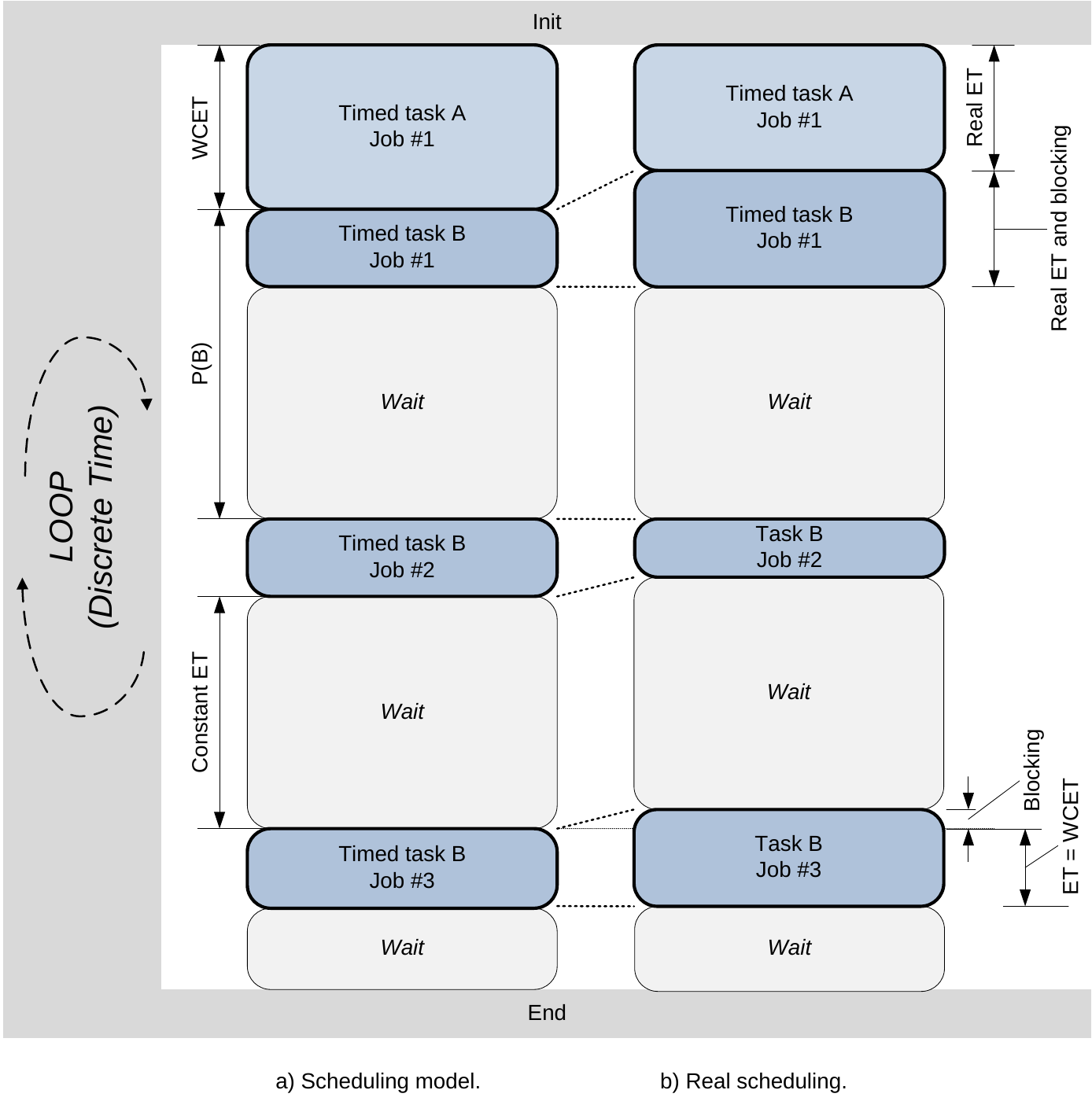}
   \caption[SynDEx - Model versus a real execution instance]{SynDEx - Model versus a real execution instance. a) In the SynDEx model execution times are constant, while b) at a real execution tasks may terminate earlier which is compensated by blocking timed tasks.}
   \label{fig:syndex:schedulingTimedTasks}
\end{center}
\end{figure}

If timer peripherals are provided on a target hardware, tasks with fixed periods can be realized, but still with the introduction of a deviation between model and reality. A fixed-period task has to include a hardware timer unit, which fires the task at desired time intervals and has to be blocking until the activation time is reached. Blocking is necessary to compensate in time for the early task terminations - this is not actually seen in the scheduling model, but is a method of realizing a fixed-period task (see figure~\ref{fig:syndex:schedulingTimedTasks}).

\begin{figure}[!htbp]
\begin{center}
	\includegraphics[width=0.90\textwidth]{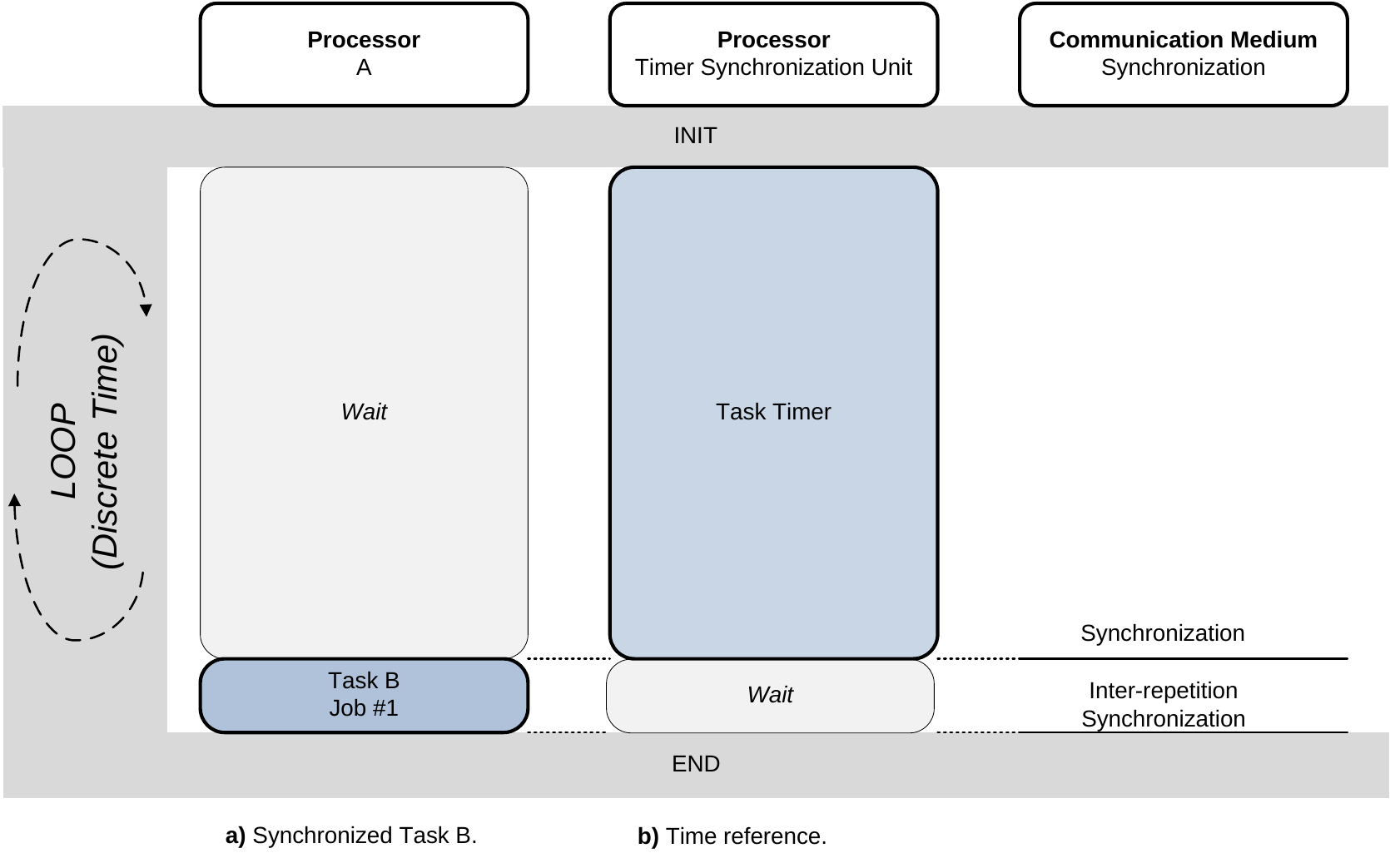}
   \caption{SynDEx - Synchronizing a task with a timing operator.}
   \label{fig:syndex:schedulingTimerSynchronization}
\end{center}
\end{figure}

Another way of synchronizing tasks is by introducing more operators into the hardware architecture model of SynDEx. These operators could model a timer unit on the processor itself and thus, by synchronization operations, maintain a fixed period of a task (see figure~\ref{fig:syndex:schedulingTimerSynchronization}). This is possible in the model, but the macro code generation part of SynDEx at the current version used in this thesis (7.0.0) is not fit for this issue - this solution would require additional customized model transformation and code generation programs. In the case of a multi-processor architecture, meaning several targets have to communicate and work collaboratively, the varying task durations will tend to average out since synchronization calls are done between the targets (synchronization and communication).\\

\clearpage

\section{Scicos-SynDEx Interface}\label{sec:mbd-tools:scicosSyndexInterface}

\begin{figure}[!htbp]
\begin{center}
	\includegraphics[width=0.80\textwidth]{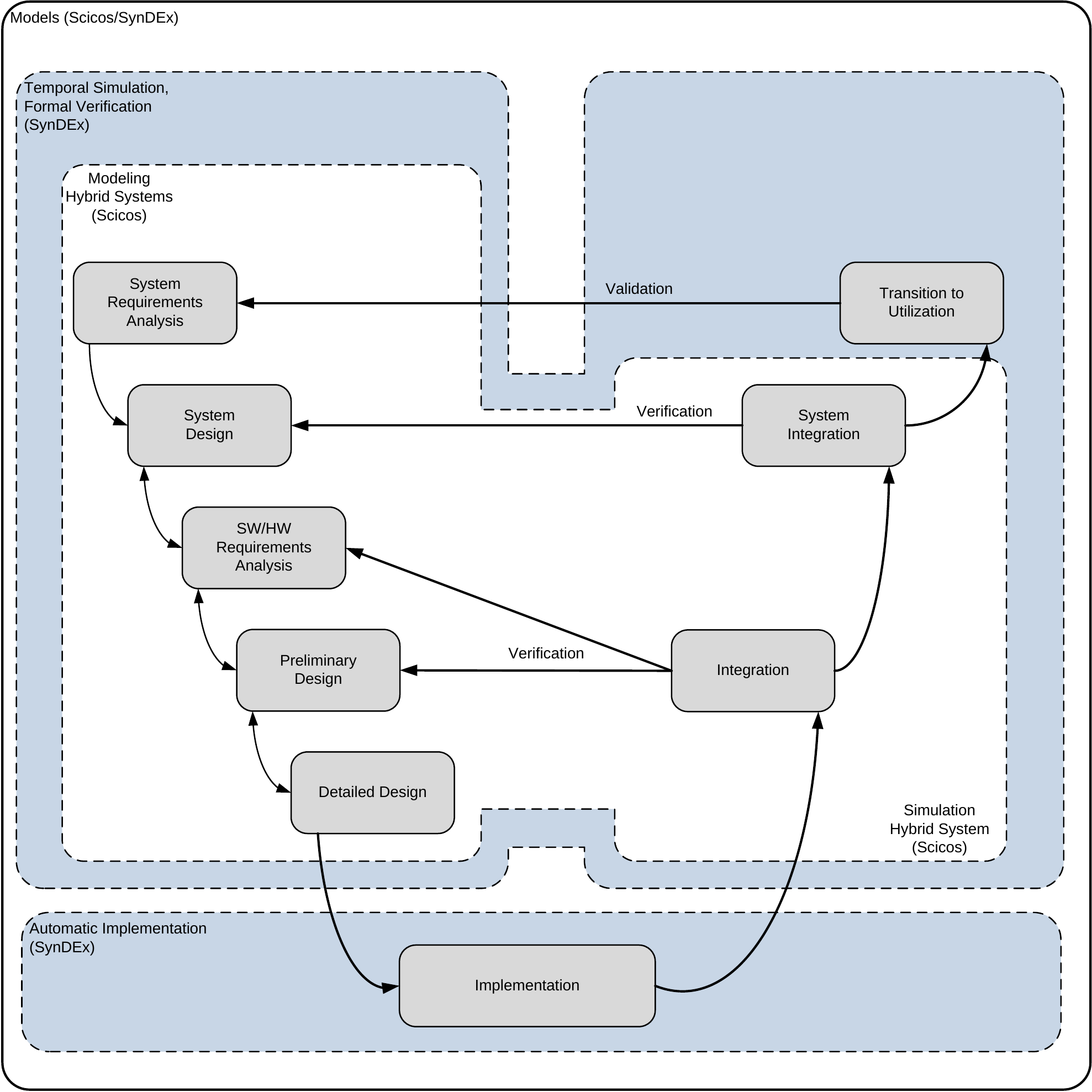}
   \caption[Scicos/SynDEx SDLC and V-Model '97]{A Software Development Lifecycle (V-Model '97) with the Scicos/SynDEx framework.}
   \label{fig:models:vmodelsdlc}
\end{center}
\end{figure}

The model-based character of the Scicos/SynDEx framework can improve the system development process. The Scicos/SynDEx framework System does not only enhance the interaction between several phases, but also reduces the number of phases in the process model in a sense of aggregating several phases in a single phase where a great deal is done by the used model-based framework. Consider the software development sub-model of the V-Model 97 (figure~\ref{fig:models:vmodelsdlc}) which provides a good example of showing the impact of introducing models in the system development process. This figure acts as an example how the model-based Scicos/SynDEx framework is applied to a document-centric, waterfall-model based software development model.

Scicos and SynDEx models are used as a formalized system representation in almost every phase of the process and thus providing a consistent view of the product throughout the whole software development life cycle. For example, a Scicos model might be used in meetings with project stakeholders and users in order to determine the requirements. The same model is also used in the system design phases and the verification phases in the development life cycle. As it might have been already noticed by the reader, a system process model like the one in figure~\ref{fig:models:vmodelsdlc} can not be executed as is with Scicos/SynDEx: A classical plan-driven, step-by-step proceeding is not what is happening when designing models, because during the design of a Scicos/SynDEx model, many phases are advanced at the same time, since the models contribute to all phases simultaneously: directly (model is the immediate representation of the system) or indirectly (Scicos provides the model for the SynDEx model which generates code in the implementation phase).

\begin{figure}[!htbp]
\begin{center}
	\includegraphics[width=0.70\textwidth]{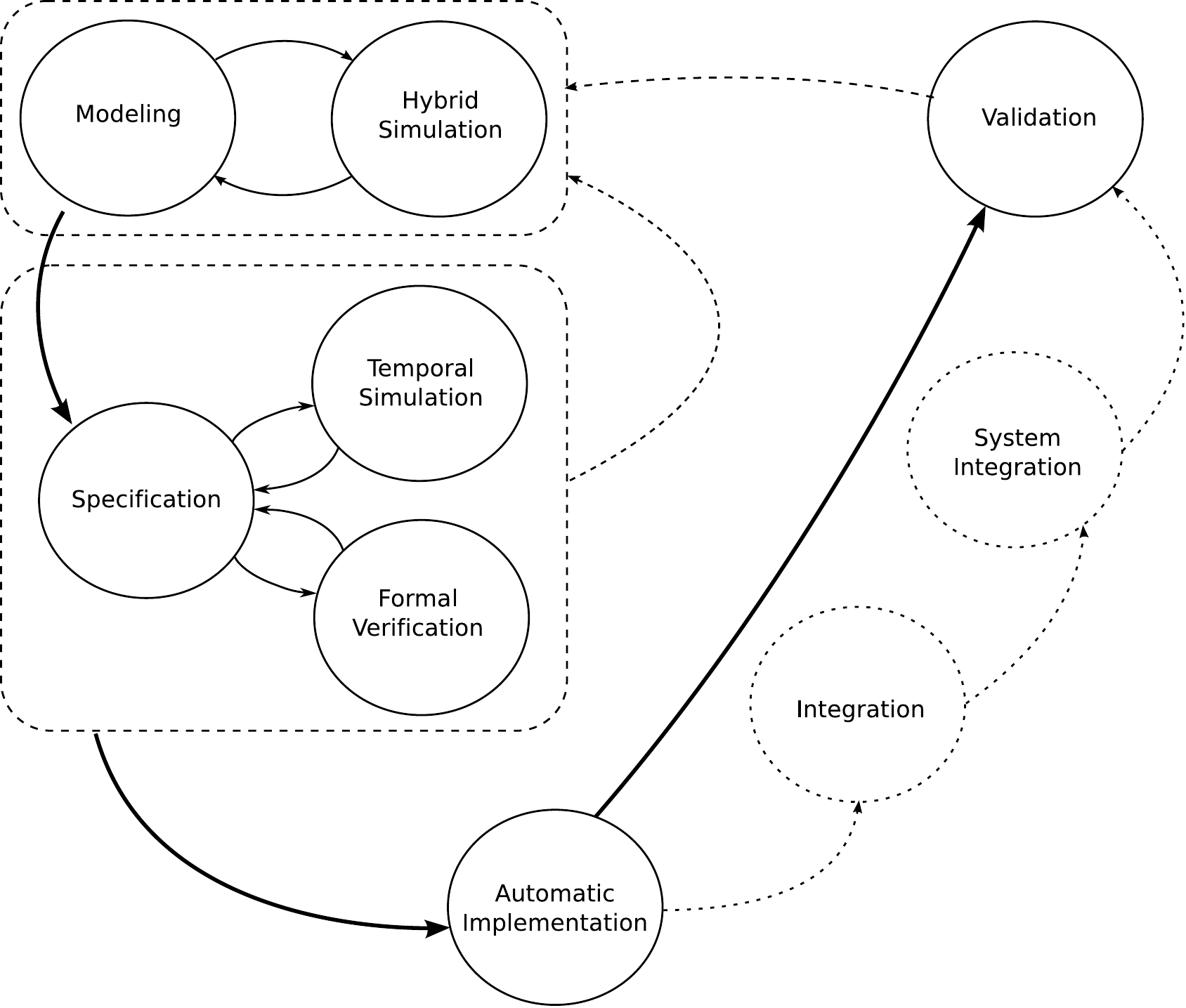}
   \caption[SDLC with Scicos/SynDEx]{The software development lifecycle with the Scicos/SynDEx framework, source: \cite{esm04} - modified illustration}
   \label{fig:models:sssdlc}
\end{center}
\end{figure}

These circumstances lead to a new software development life cycle with Scicos/SynDEx (see figure~\ref{fig:models:sssdlc}). At the start of this SDLC hybrid system models are defined in the Scicos context, simulated and models refined. If the simulation results are satisfactory, the model (specification part modeled in Scicos) can be model-to-model (Scicos-to-SynDEx) transformed and be part of the system specification represented by SynDEx models, which is, in the next phase of the SDLC, formal verified and treated with temporal simulation by the SynDEx software. In case of failing the requirements, the hybrid system part might have to be redesigned, that means the Scicos model has to be edited again, a model-to-model transformation done and the resulting model used again for the simulations in SynDEx. Automatic implementation, code generation, for different execution environments after successful simulation can be done by the SynDEx software and the results be validated afterwards. Instead of transcending from the executable code to integration and system integration verification phases, the validation phase is directly hit. The simulations and verifications done in the previous phases make it possible to skip, a now already integrated, additional verification phase (provided that there is enough trust in the software development tools). In case the results are positively validated the cycle can stop, otherwise the models have to be worked on again in the modeling phase.

\subsubsection{Scicos-SynDEx Gateway}\label{sec:mbd-tools:scicosSyndexGateway}

\begin{figure}[!htbp]
\begin{center}
	\includegraphics[width=0.90\textwidth]{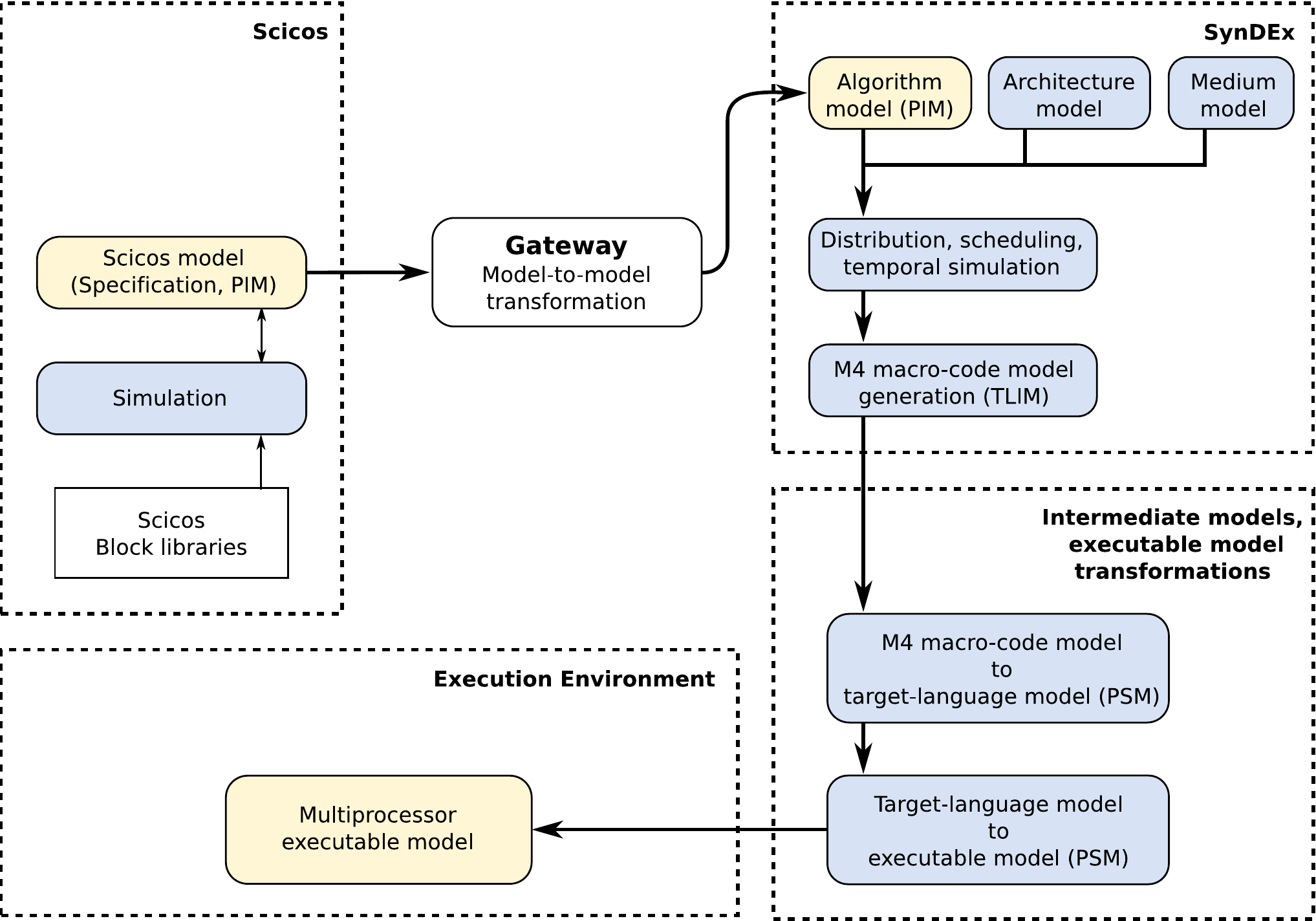}
   \caption[Scicos/SynDEx gateway]{Scicos/SynDEx gateway.}
   \label{fig:models:ssGateway}
\end{center}
\end{figure}

As mentioned above, the SDLC with the Scicos/SynDEx framework requires an intermediate step translating a Scicos model into a SynDEx model. This model-to-model transformation is done via the Scicos-SynDEx Gateway - an add-on tool-box of Scicos. With this gateway the hybrid system model designed in Scicos is transformed into a SynDEx model which can then be combined with other SynDEx models (see figure~\ref{fig:models:ssGateway}).

\clearpage
	\cleardoublepage
	\chapter[Demonstrations: MBD with Scicos/SynDEx]{Demonstrations of MBD with the Scicos/SynDEx Framework} \label{chapter:example}

Mapping the real world onto a model that retains all information of interest and represents models in a resource-friendly (memory, time) fashion should be the aim of any modeling software. The Scicos/SynDEx framework, Scicos for the modeling and simulation of discrete/continuous systems and SynDEx for the distribution and scheduling of models on a distributed system, might show a good way for modeling hybrid systems, simulation, optimization and scheduling (AAA Methodology) on a multi-processor system. How far and in which ways this framework can be applied to a distributed embedded system is shown by the implementation of two examples on a customized embedded system development board. One example implements a fan speed PID control circuit on a monoprocessor, using both, the Scicos and the SynDEx software tools, in combination. The second example implements a simple observer pattern on a multiprocessor system, where two observer nodes are notified about the temperature values from a sensor node. The architecture of the target hardware used for these examples is described in section \ref{sec:hw_arch}, information about the software architecture of the examples can be found in sections \ref{sec:exampleMono} and \ref{sec:exampleMulti} along with the corresponding Scicos and SynDEx models.\\

Measuring the capabilities of a model-based framework is not an easy task. Sufficiently understanding of the modeling software is a prerequisite for assessments. What kinds of requirements and specification can successfully be modeled with this specific model-based development framework? In which ways should the application be distributed among all the resources? What is an efficient solution? How is the scheduling of the algorithms done? How can one visualize the resource allocation (time and data) of the possible solution? What code metrics can be applied to measure the quality of the produced code? These are just a couple of questions which will at least partially be answered by the implementation of the two examples following in this section.
	\section{Hardware Architecture}\label{sec:hw_arch}

\begin{figure}[!htbp]
\begin{center}
	\includegraphics[width=0.75\textwidth]{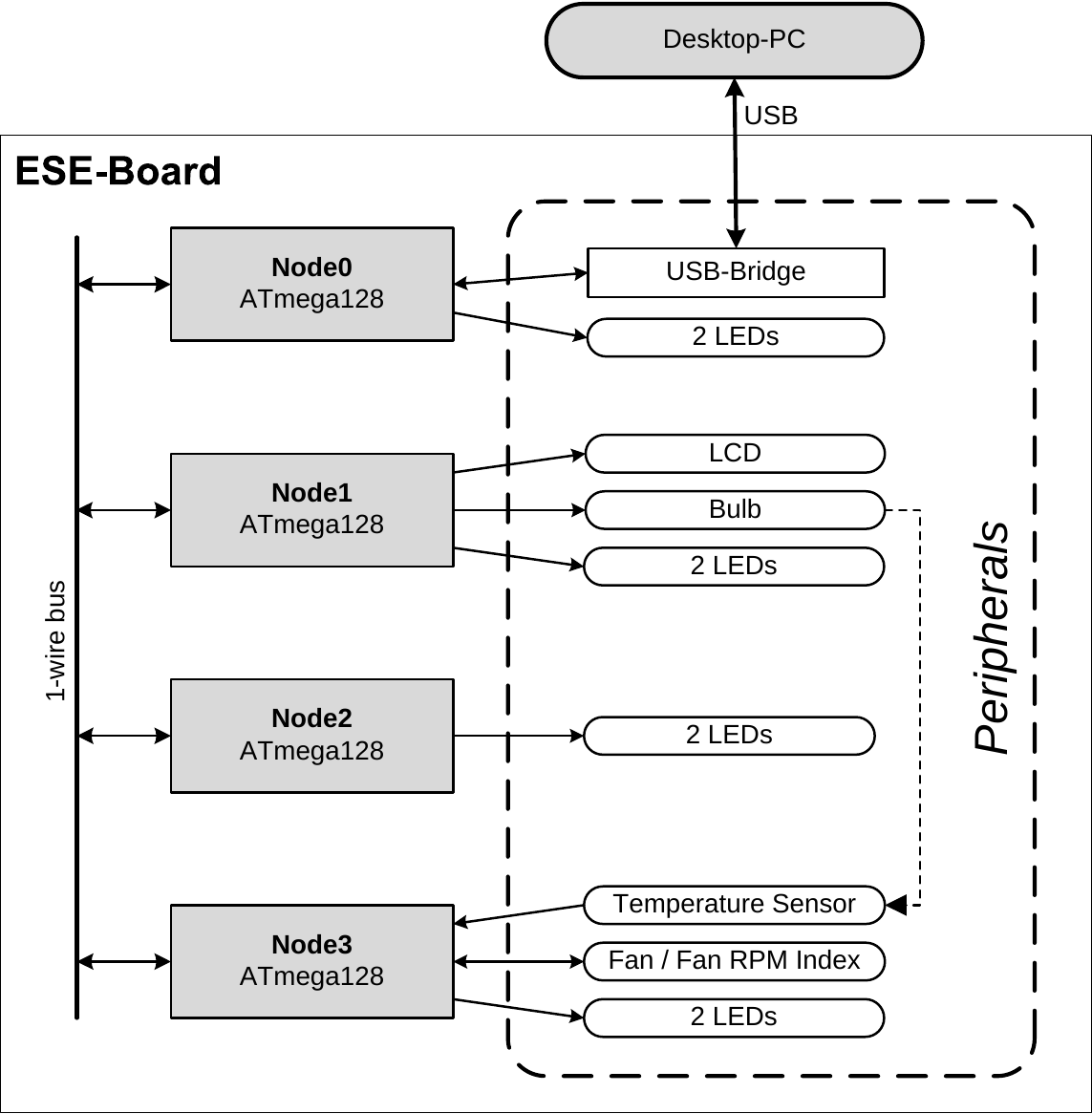}
   \caption{ESE-Board - Simplified layout.}
   \label{fig:HWArch:eBoardSchematics}
\end{center}
\end{figure}

\begin{figure}[!htbp]
\begin{center}
	\includegraphics[width=1.0\textwidth]{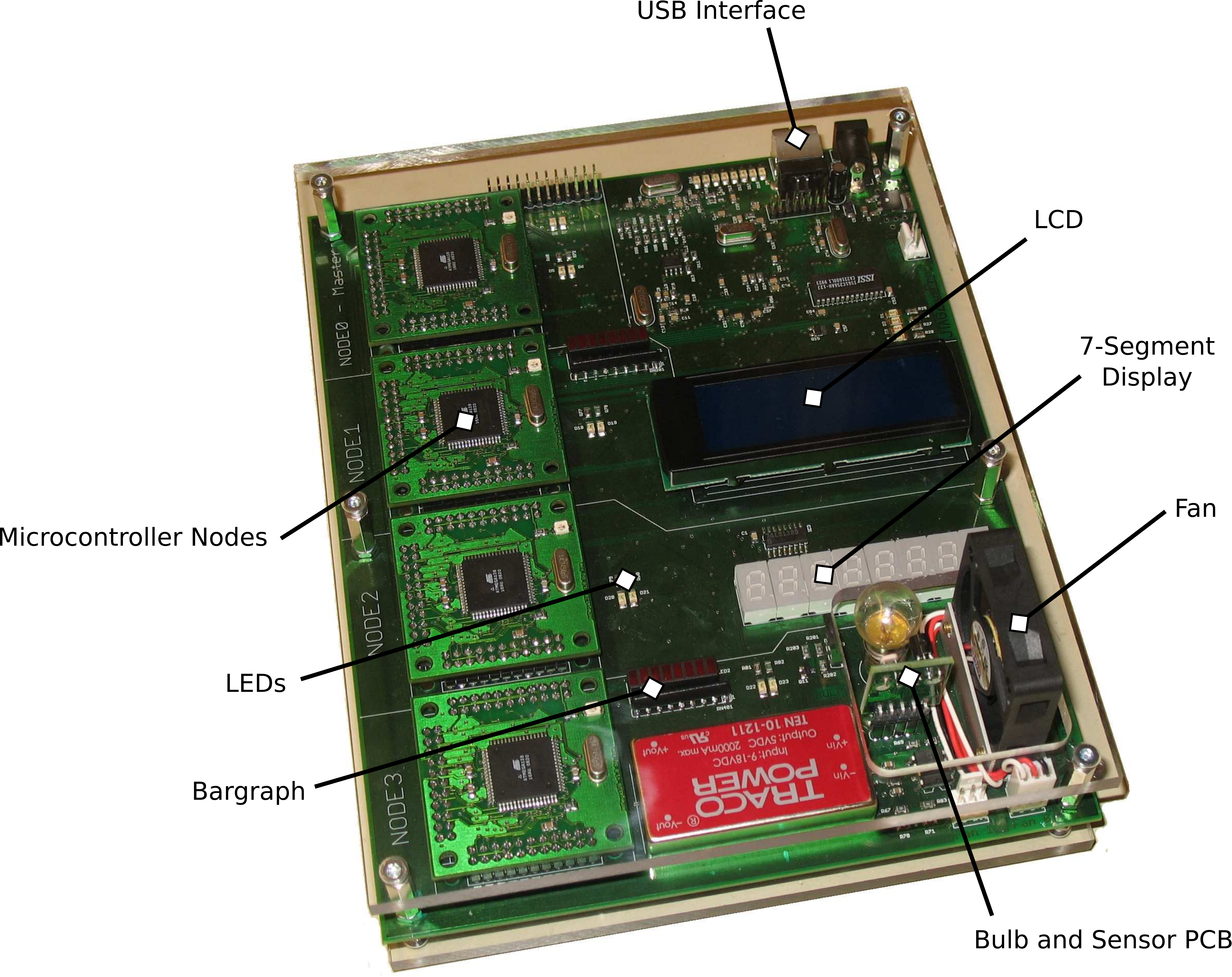}
   \caption[ESE-Board - Photography]{ESE-Board - Photography, source: \cite{Koessler2009}, modified illustration.}
   \label{fig:HWArch:eBoardPhoto}
\end{center}
\end{figure}

The hardware used for the evaluation of the Scicos/SynDEx modeling capabilities is a development and educational printing circuit board with sensor/actuator peripherals (from now on to be referred as the "ESE-Board" - Embedded Systems Engineering Board) created by K\"ossler at the Technical University of Vienna (complete information regarding the board can be found in the creator's master thesis \cite{Koessler2009}).
Basically, this board consists of four 8-bit ATmega128 microcontrollers (Node0 - Node3) where each controller node is interconnected to the others by a bus. Furthermore, there are various Input/Output peripherals assigned to each microcontroller, such as light intensity and temperature sensors, as well as actuators like a fan, liquid crystal displays, LEDs and LED bar-graphs. The programming of the microcontrollers is done by an USB interface via an USB-to-Serial circuit. Thus, this PCB is a good evaluation board for hybrid systems. The microcontrollers realize the control part of the system, while the controlled peripherals implement a real-world plant (dense time, continuous data values).  A very simplified layout of the board is shown in figure~\ref{fig:HWArch:eBoardSchematics} and a real-world photography in figure~\ref{fig:HWArch:eBoardPhoto}. Note that only for the examples' most relevant ESE-Board parts are depicted.

\subsection[Target Platform Components]{The components of the ESE-Board}\label{sec:hw_arch:eBoard}

The nodes with their attached peripherals form a distributed embedded system, which can be accessed by a PC workstation through an USB interface. Several peripherals are connected to the microcontrollers, that includes two LEDs for each node. An LCD, a light bulb, a temperature sensor and a fan are connected to one specific microcontroller. Every node (Node0-Node3) is an ATmega128\footnote{The Atmel Corporation, \url{http://www.atmel.com}} microcontroller holding following basic characteristics (only the most important ones are listed):\\

\begin{itemize}
\item 8-bit microcontroller.
\item 32 x 8 General Purpose Working Registers + Peripheral Control Registers.
\item Fully Static Operation.
\item Advanced RISC Architecture.
\item 128KB In-System Programmable Flash Memory.
\item 4KB EEPROM.
\item 4KB Internal SRAM.
\item 133 Instructions, most of them take a single clock cycle for the execution.
\item JTAG (IEEE std. 1149.1 Compliant) Interface.
\item Two 8-bit Timer/Counters with separate prescalers and Compare Modes.
\item Two Expanded 16-bit Timer/Counters with separate prescaler, Compare Mode and Capture Mode.
\item Two 8-bit PWM Channels.
\item 8-channel, 10-bit ADC.
\item Dual Programmable Serial USARTs.
\item Master/Slave SPI Serial Interface.
\item 53 Programmable I/O Lines.
\end{itemize}

The time-source for the microcontrollers is an external clock block running with 14.745600 MHz. Every node is connected to the bus via a HW- or SW-USART (Hardware- or Software- \Keyword{Universal Synchronous and Asynchronous Receiver Transmitter}). These USARTs are fully compatible with the AVR\cite{AVR} UART regarding baud rate generation, transmitter operation, transmit buffer functionality and receiver operation. The examples in this thesis, realized with the AVR tools, will only use the asynchronous mode of the USART, therefore they are depicted as UART. Following paragraphs describe only the parts of the nodes used for the examples in this thesis.

\paragraph{Node0}$~~$\\

\Keyword{Node0} acts as the connector between the ESE-Board and an external device (usually an USB-wired desktop PC) and is connected to two LEDs. The UART1 block of the microcontroller is connected to the "USB to Serial" IC interfacing a possible external device.
\clearpage
\begin{figure}[!htbp]
\begin{center}
	\includegraphics[width=0.5\textwidth]{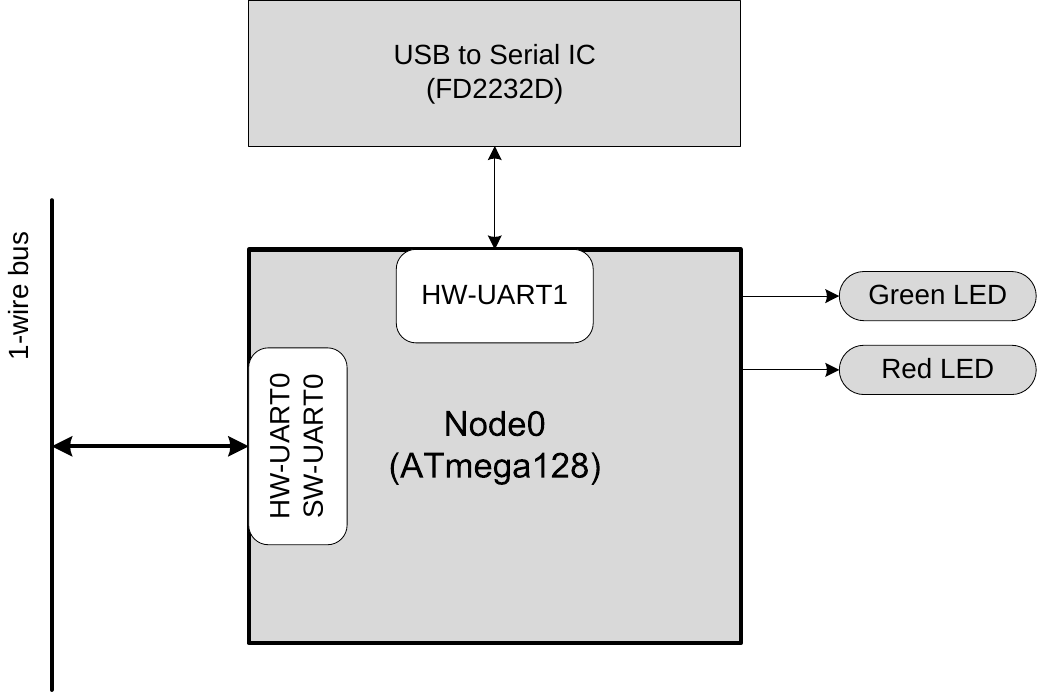}
   \caption{ESE-Board - Node0.}
   \label{fig:HWArch:eBoardNode0}
\end{center}
\end{figure}

\paragraph{Node1}$~~$\\

\Keyword{Node1} is responsible for controlling an \acf{LCD} device and a light bulb. The background light of the LCD is adjusted by \acf{PWM}, the data I/O handled by \acf{SPI}.

\begin{figure}[!htbp]
\begin{center}
	\includegraphics[width=0.5\textwidth]{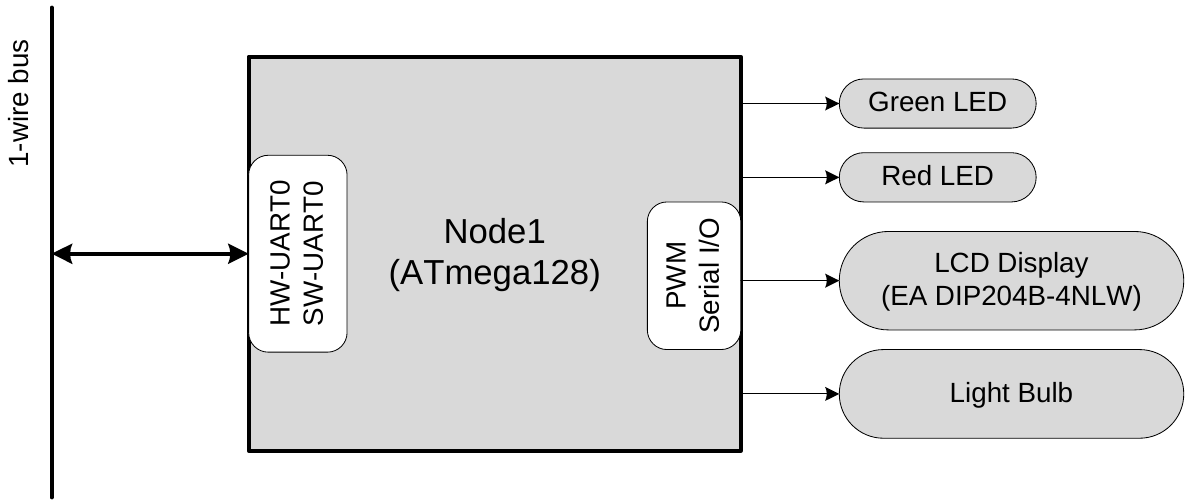}
   \caption{ESE-Board - Node1.}
   \label{fig:HWArch:eBoardNode1}
\end{center}
\end{figure}

\paragraph{Node2}$~~$\\

\Keyword{Node2} is connected to two LEDs.

\begin{figure}[!htbp]
\begin{center}
	\includegraphics[width=0.5\textwidth]{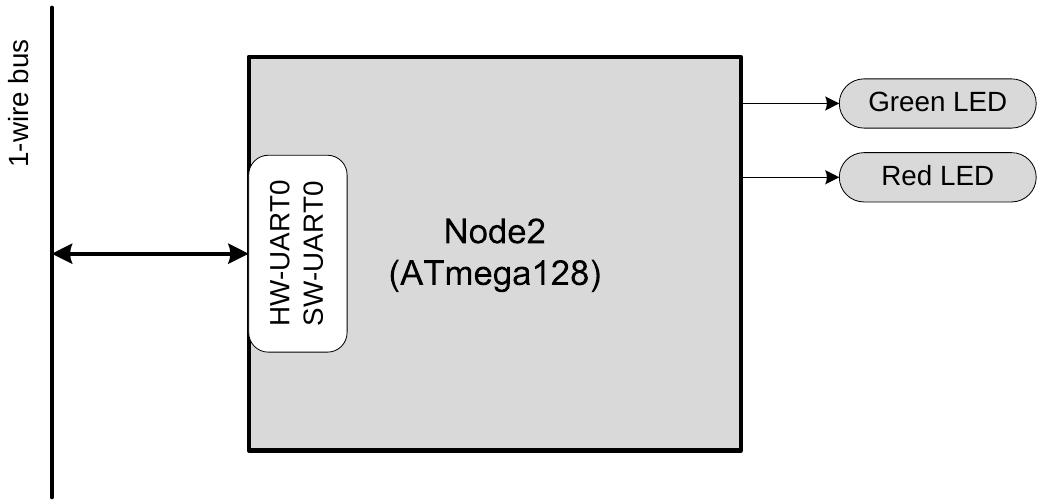}
   \caption{ESE-Board - Node2.}
   \label{fig:HWArch:eBoardNode2}
\end{center}
\end{figure}

\paragraph{Node3}\label{sec:hw_arch:node3}$~~$\\

\Keyword{Node3} controls a \acf{DC} fan (DA04010B12S-017), two LEDs, and acquires data from a temperature sensor (temperature sensitive Zener diode, National Semiconductor, LM135). The speed of the fan is controlled using PWM, the current \acf{RPM} are measured by observing the frequency generator output supplied by the fan (emitting two pulses per round). Temperature data values are converted via the \acf{ADC} unit of the microcontroller.

\begin{figure}[!htbp]
\begin{center}
	\includegraphics[width=0.5\textwidth]{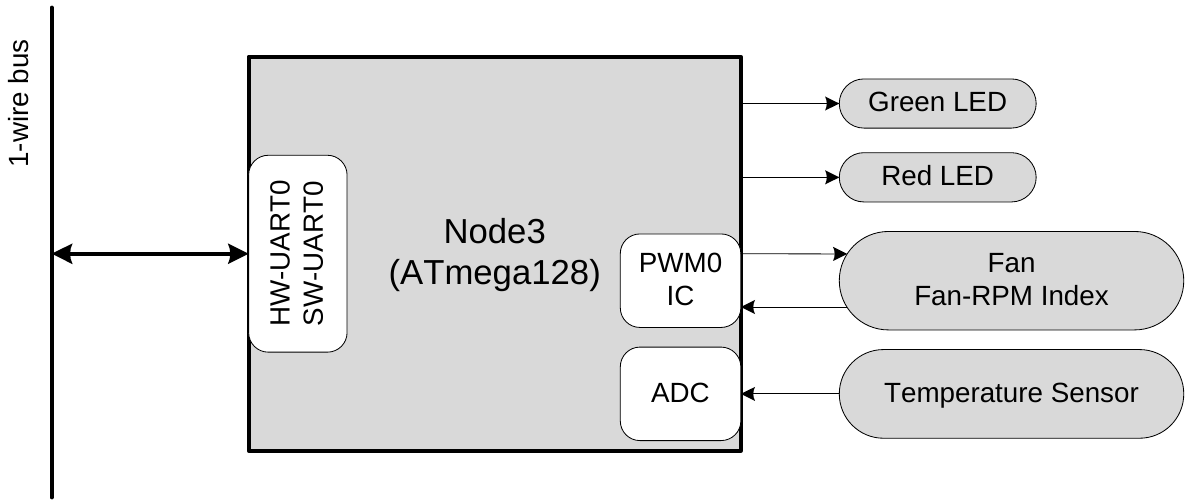}
   \caption{ESE-Board - Node3.}
   \label{fig:HWArch:eBoardNode3}
\end{center}
\end{figure}
	\section[Example Monoprocessor]{Example Monoprocessor - A PID Controller with Scicos and SynDEx}\label{sec:exampleMono}

\begin{figure}[!htbp]
\begin{center}
	\includegraphics[width=0.75\textwidth]{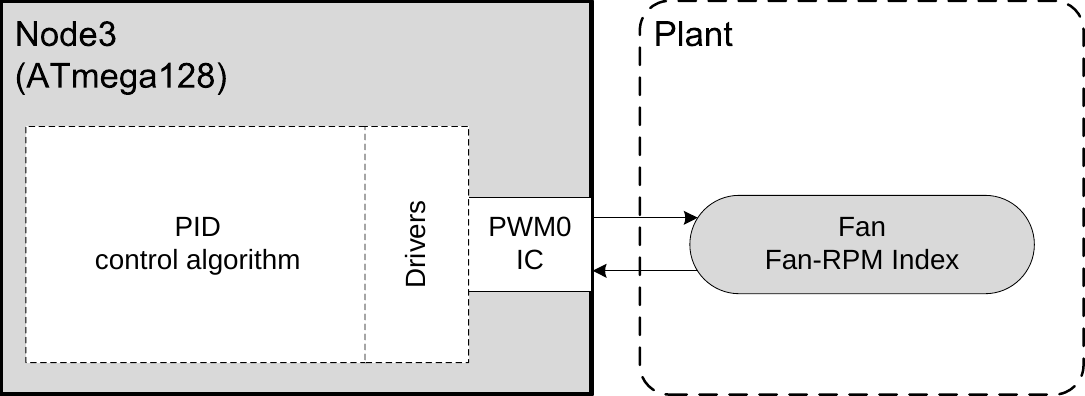}
   \caption[PID-Example. A PID algorithm placed on Node3] {PID-Example: A PID algorithm placed on Node3 controlling the fan.}
   \label{fig:ExampleMono:Principle}
\end{center}
\end{figure}

This example exercises a development process with the combination of Scicos and SynDEx. A simple application is designed and run on a monoprocessor target. Node3 of the ESE-board will act as the execution environment of a digital PID control algorithm with anti-windup (confer to \cite[section 5.4.2]{elmenreich:09}) that regulates the connected peripheral air fan. The fan (the plant) resides in the continuous domain (figure~\ref{fig:ExampleMono:Principle}). The PID algorithm and the fan will be modeled and simulated with Scicos. After simulations, we demonstrate the temporal design of the PID algorithm and model it in SynDEx in order to support code generation for the hardware target.\\

The peripherals connected to Node3 (an ATmega128 microcontroller) provide an excellent target to evaluate a PID control algorithm: The speed of the fan can be controlled by the PID algorithm through a customized fan driver, while speed data is obtained by a fan speed measurement driver.

The topic of designing a digital PID algorithm was chosen for several reasons: PID algorithms are very widely used control algorithms, have basic mathematics operations included, enforce the use of memory blocks for calculations, touch the topic of discrete PID implementation and their relation to continuous domain PID algorithms. Furthermore, the PID algorithm requires reading data from a sensor, the calculation of depending data, and writing of control values to an actuator. Through simulations of algorithm and architecture designs possible solution candidates for software algorithms and target hardware designs can be found. Latter is already realized by the ESE-Board hence the focus of this work lies mostly on software design.\\
Details about the target hardware architecture is found in section \ref{sec:exampleMono:hw_arch}, while the software architecture, including a short requirements description and the design of the models with their simulation and model transformations, is located in section \ref{sec:exampleMono:sw_arch}. These sections include several comments and considerations which were posed during the development process.\\

The results of the example are helpful to reasonably answer following questions:
\begin{itemize}
\item How well is the Scicos/SynDEx framework suited to contribute to the development of a simple control algorithm?
\item How much memory will the generated application use on the embedded target?
\item How much effort is necessary (e.g. person hours) to develop such an application with the framework? Is the time spent reasonable? Does it need considerable time for a developer to understand/setup the framework?
\end{itemize}
The presented experiment will provide data allowing a reasonable statement about code size and quality for a typical embedded application.

\subsection[Hardware Architecture]{Hardware Architecture (Execution Environment)}\label{sec:exampleMono:hw_arch}

The chosen execution environment - plant and controller- consists of several ESE-Board components with following parts playing major roles:
\begin{itemize}
\item \textbf{Controller device:} Node3 (depicted in \ref{fig:ExampleMono:Principle} and section \ref{sec:hw_arch:node3}).
\item \textbf{Controlled peripherals:} The air fan connected to Node3 (actuator) takes the set fan speed as argument, and the fan sensor provides fan speed data. The air fan's input line is connected to the PWM0 output of the microcontroller, the speed index line to an input pin (IC - interrupt capture) of the microcontroller.
\end{itemize}
Note that this section contains the physical description of a target execution environment (hardware architecture), while the formally modeled hardware architecture is designed with SynDEx (following in this example).
The hardware description could be done in more detail on the physical level, for example by modeling the hardware based on the blueprints of the board, but this is not intended for the design of the models at the chosen design abstraction level (keep the models simple). Simple models, dislodged from a very low level representation, are easier to understand and design, but at the same time they might hide details which could be crucial for designing a model representing the real world in adequate details.

\subsection{Software Architecture}\label{sec:exampleMono:sw_arch}

A digital PID control algorithm with anti-windup is designed in a generic way. This makes it possible to compare the results to other frameworks. For instance, the same algorithm could be implemented with the SCADE software in order to compare the results to this example with Scicos and SynDEx. Focusing on a simple example with only a PID algorithm leads to the idea of following rough requirements that just demand an air fan to be operated at a given speed.

\paragraph{Requirements}$~~$\\

A fan (with a maximum of $10\,000\:RPM$) is regulated by a digital PID control algorithm on a single processor. At system start, the speed of the fan is supplied with a constant percent value, for example a value of $100\%$ results in $10\,000\:RPM$, $50\:\%$ in $5\,000\:RPM$, $0\:\%$ in $0\:RPM$. A speed tolerance of $15\:\%$ is acceptable, thus a speed input value of $50\:\%$ may result in $5\,000\pm 750\:RPM$.

\paragraph{Considerations}$~~$\\

Before going into solving the algorithmic problems, some economic factors have to be considered. Imagine oneself being in the role of the software company reading these requirements. Following thoughts could occur:
There is a large variety of fans and control electronics available on the market. Which electronics equipment is the cheapest? How can such electronic components be evaluated in a fast way - is it possible to create a solution using these components? If necessary, would it be easily possible to replace a component by another model (e.g. the fan of choice could not be delivered anymore)?

A top-down model design approach using Scicos and SynDEx might lead to a working prototype. Scicos can be used to simulate the PID algorithm with a model of some different hardware fans in question. If the fan models used in the evaluation phase are considered to be accurate, the PID algorithm can be designed and simulated. Thus, the feasibility of a prototype can be confirmed or denied. Furthermore, information about necessary hardware components and algorithm designs can be gathered by including SynDEx in the design process. SynDEx can handle the temporal design of the application and automatically generate code for an electronic control logic of choice (e.g. an embedded microcontroller in this case).
\vspace{2cm}
\paragraph{Hybrid System Design and Verification}$~~$\\

Obviously, the requirements describe a hybrid system: The fan is located in the real world with its continuous time and data characteristics. The design of the functional PID model and the fan model are implemented with Scicos. If the model of the real-world fan is implemented with sufficient accuracy, the parameters of the PID algorithm (that are located in a discrete data and time domain) can be calculated.\\

Scicos is a domain specific language and is used for modeling during the development and simulation phase. Scicos, as an internal DSL (which is based on Scilab), provides the means for hybrid systems modeling. Scicos models are meant to be platform-independent, thus their generic nature makes it possible to transform them into other artifacts or even executable models. Platform independence pertains mostly the controller algorithm part of the Scicos models - they are a part of the development which should be portable to different target hardware architectures. Whereas for the model of the plant (air fan) hardware independence is an issue which might not be realized easily. Every plant has different characteristics which need to be modeled in a sufficient level of detail - if only a very abstract representation of the plant is required, then the plant model could be considered to have some degree of platform independence - a real hardware fan could then be replaced by another model (or just a fan of the same model) without changing the plant model. The term \Keyword{platform independence} might be seen not adequate for a plant model where no code will be executed on. Nevertheless, such models are needed for the simulation and verification of the controller part and moreover, their reuse in combination with various other plant instantiations would be of advantage.

\subsubsection{PID Algorithm - Scicos Model}\label{sec:exampleMono:sw_arch:Scicos}

\begin{figure}[!htbp]
\begin{center}
	\includegraphics[width=0.35\textwidth]{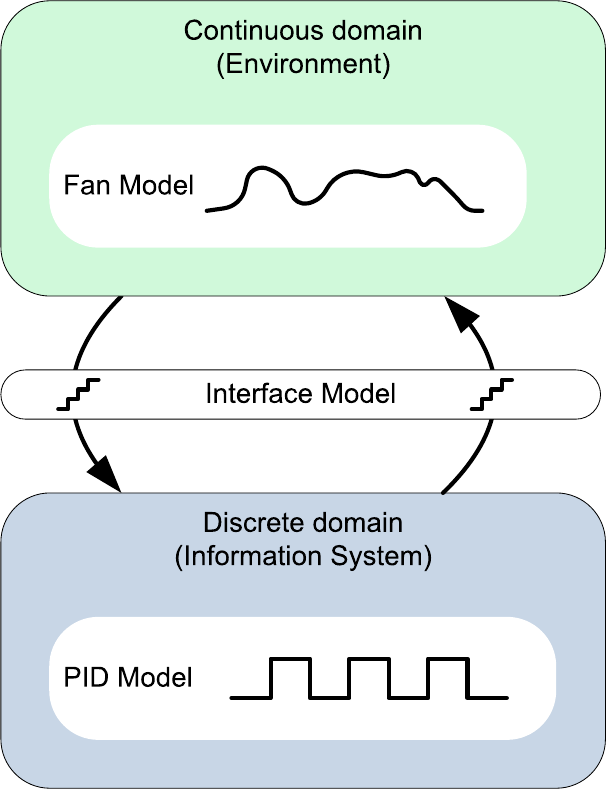}
   \caption[PID-Example. A hybrid system]{PID-Example - a hybrid system.}
   \label{fig:Scicos:HybridSystem}
\end{center}
\end{figure}

The Scicos model consists basically of two main components, the fan (plant) and the PID control algorithm (discrete controller). Both domains are connected via interface components (figure~\ref{fig:Scicos:HybridSystem}). Interface components are necessary since the drivers implemented on the microcontroller take integer values as arguments.

Before going towards the modeling of the plant and controller, some fundamental assumptions have to be made. In which detail should the fan be modeled? An ideal solution would be a model that represents real-world behavior. It might be common sense that an ideal solution is hardly possible, for several reasons, such as the fact that memory space and computation power are limited. This example does not focus on a sophisticated model, but on the methods and concepts in designing models with the Scicos/SynDEx framework.

\paragraph{Plant Modeling}$~~$\\

\begin{figure}[!htbp]
\begin{center}
\subfigure[Step response.]{\label{fig:fanStepResponse}\includegraphics[width=0.49\textwidth]{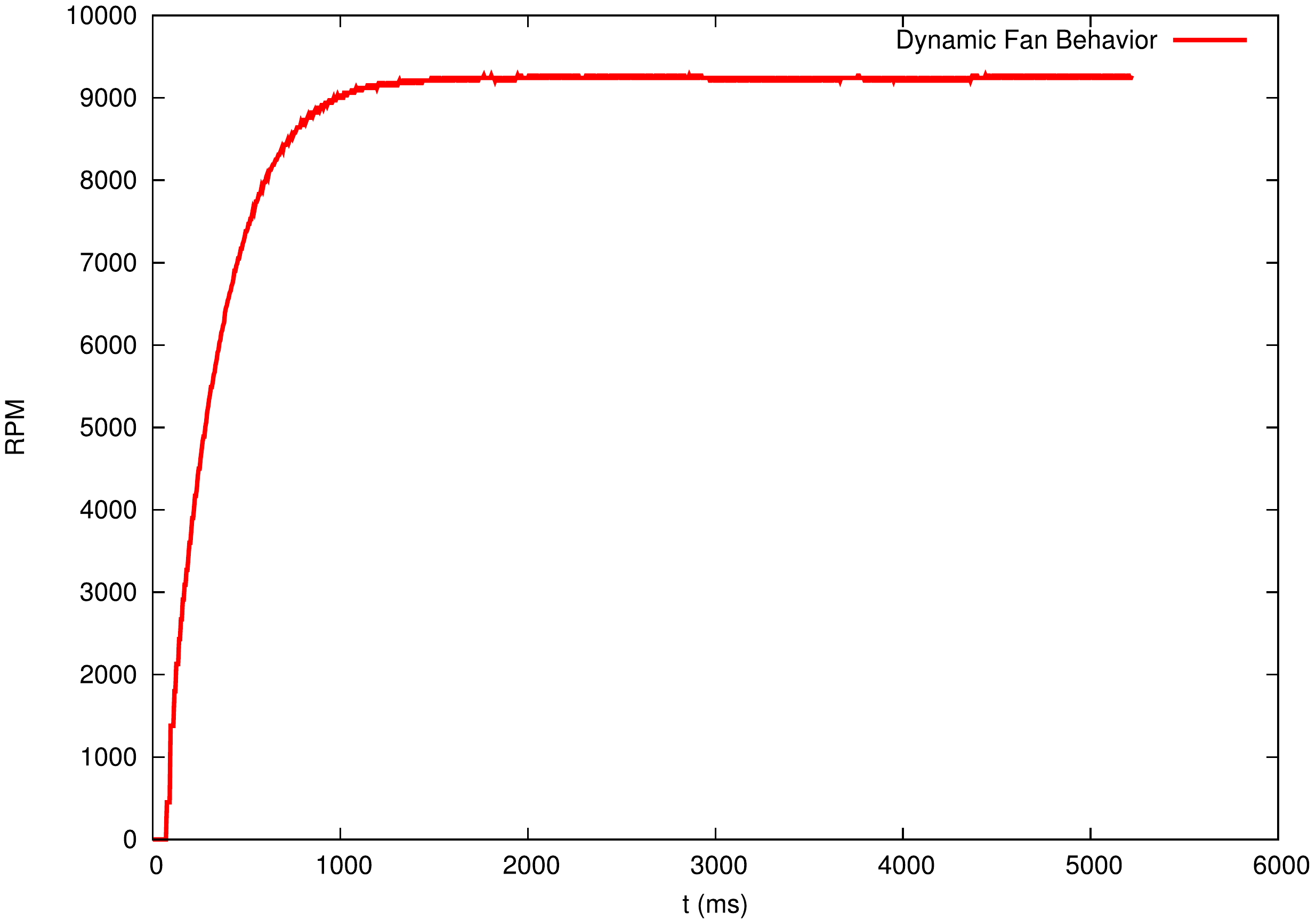}}
\subfigure[Static behavior.]{\label{fig:fanStatic}\includegraphics[width=0.49\textwidth]{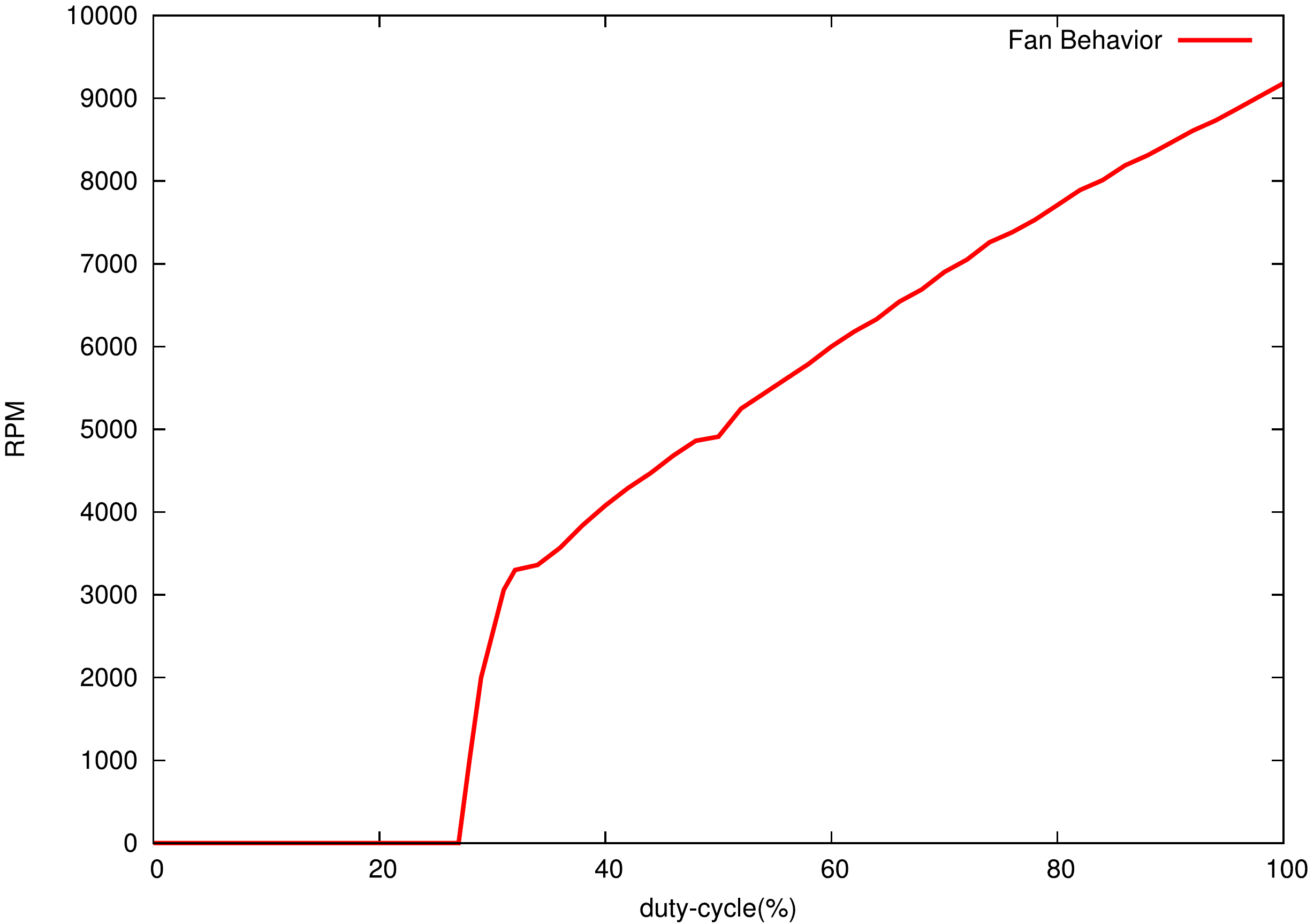}}
\end{center}
\caption{Cooling fan behavior.}
\label{fig:fanMeasuredBehavior}
\end{figure}

The behavior of the fan is modeled based on the dynamic and static behavior of the real hardware. The step response function (figure~\ref{fig:fanStepResponse}) has to be combined with the \ac{PWM}-\ac{RPM} curve function (figure~\ref{fig:fanStatic}).

\begin{figure}[!htbp]
\begin{center}
	\includegraphics[width=0.65\textwidth]{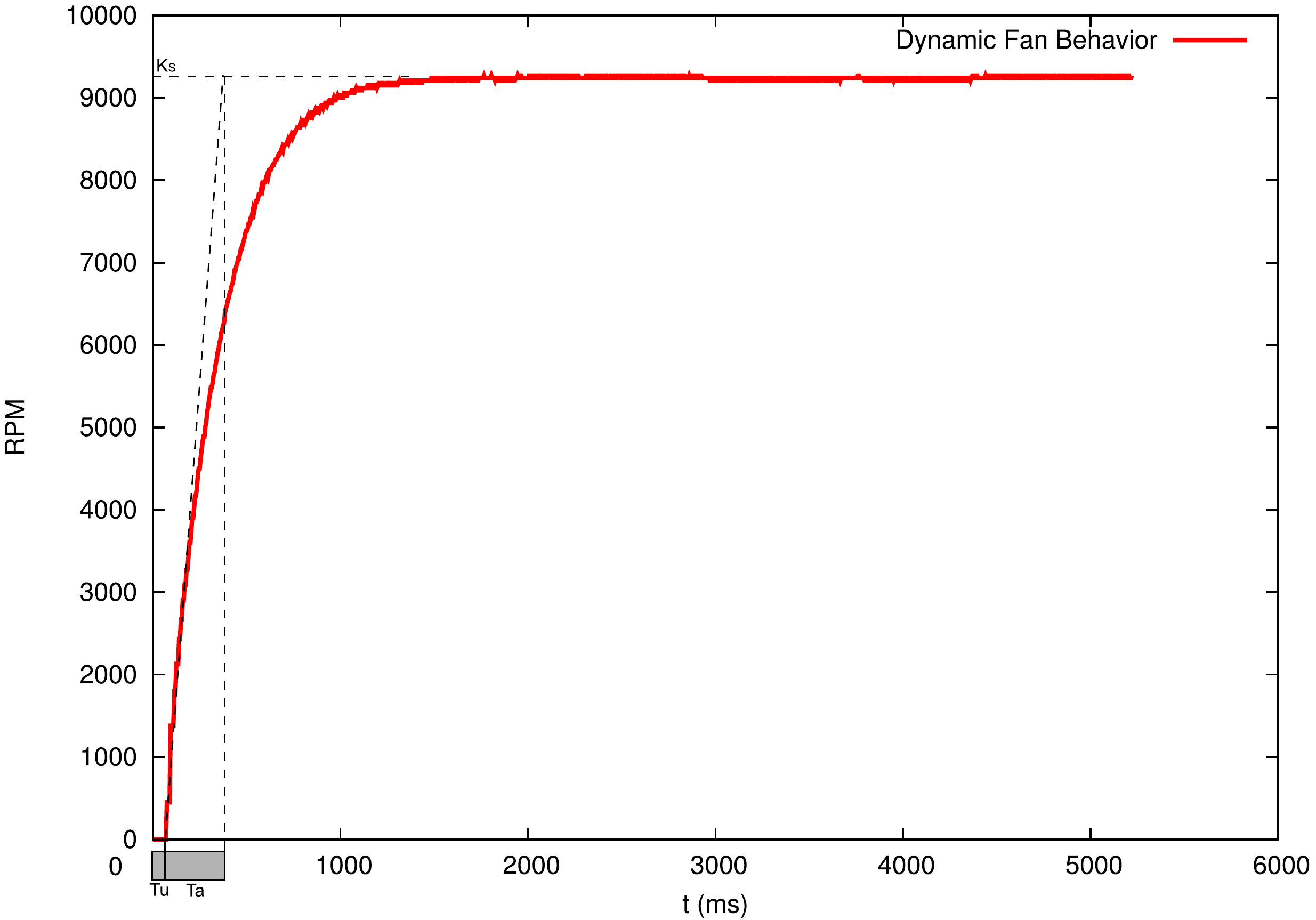}
   \caption[PID-Example. Fan, function approximation]{PID-Example - Approximation of the fan's PT1 and dead-time behavior.}
   \label{fig:fanModelParameterApproximation}
\end{center}
\end{figure}

\begin{figure}[!htbp]
\begin{center}
	\includegraphics[angle=270,width=0.9\textwidth]{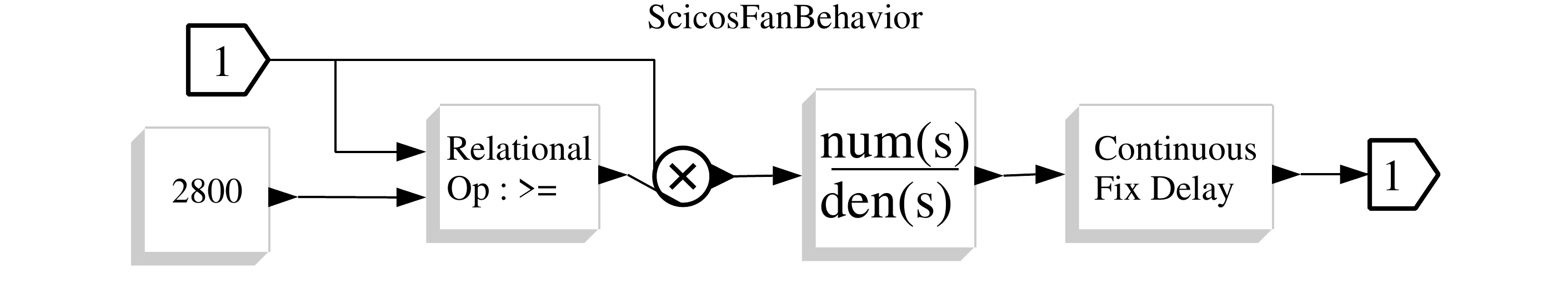}
   \caption[PID-Example. Fan, Scicos behavior model]{PID-Example - fan behavior Scicos model.}
   \label{fig:fanModelScicosAndPT1}
\end{center}
\end{figure}

The PT1 behavior is approximately characterized by $x(s)=\frac{K_S}{1+s \cdot T_a}$ with $T_a=330\:ms$ and $K_S=0.925\:\frac{\%}{\%}$ (figure~\ref{fig:fanModelParameterApproximation}). The controller input is modeled as percentage of the maximum \ac{PWM} input to the fan motor. The controller output is modeled as a percentage of the nominal maximum fan speed ($10\,000\: RPM$). Additionally, there is a deadtime of about $T_u=75\:ms$, modeled with $x(s)=\frac{1}{1+s\cdot T_u}$. The static behavior is a function correlating the set \ac{PWM} duty-cycle (dc) and the fan's \ac{RPM} (the signal dampening of about $7.75\:\%$ is already considered in the dynamic behavior):\\

\[
f(x_{dc})\, RPM = \left\{
	\begin{array}{l l}
	0 & \quad \text{if $x_{dc}<28\%$}\\
	PWM \cdot 9\,250 & \quad \text{if $x_{dc}\geq 28\%$}\\
	\end{array} \right.
\]
 The corresponding Scicos model is depicted in (figure~\ref{fig:fanModelScicosAndPT1}). \\

\begin{figure}[!htbp]
\begin{center}
\includegraphics[angle=0,width=0.75\textwidth]{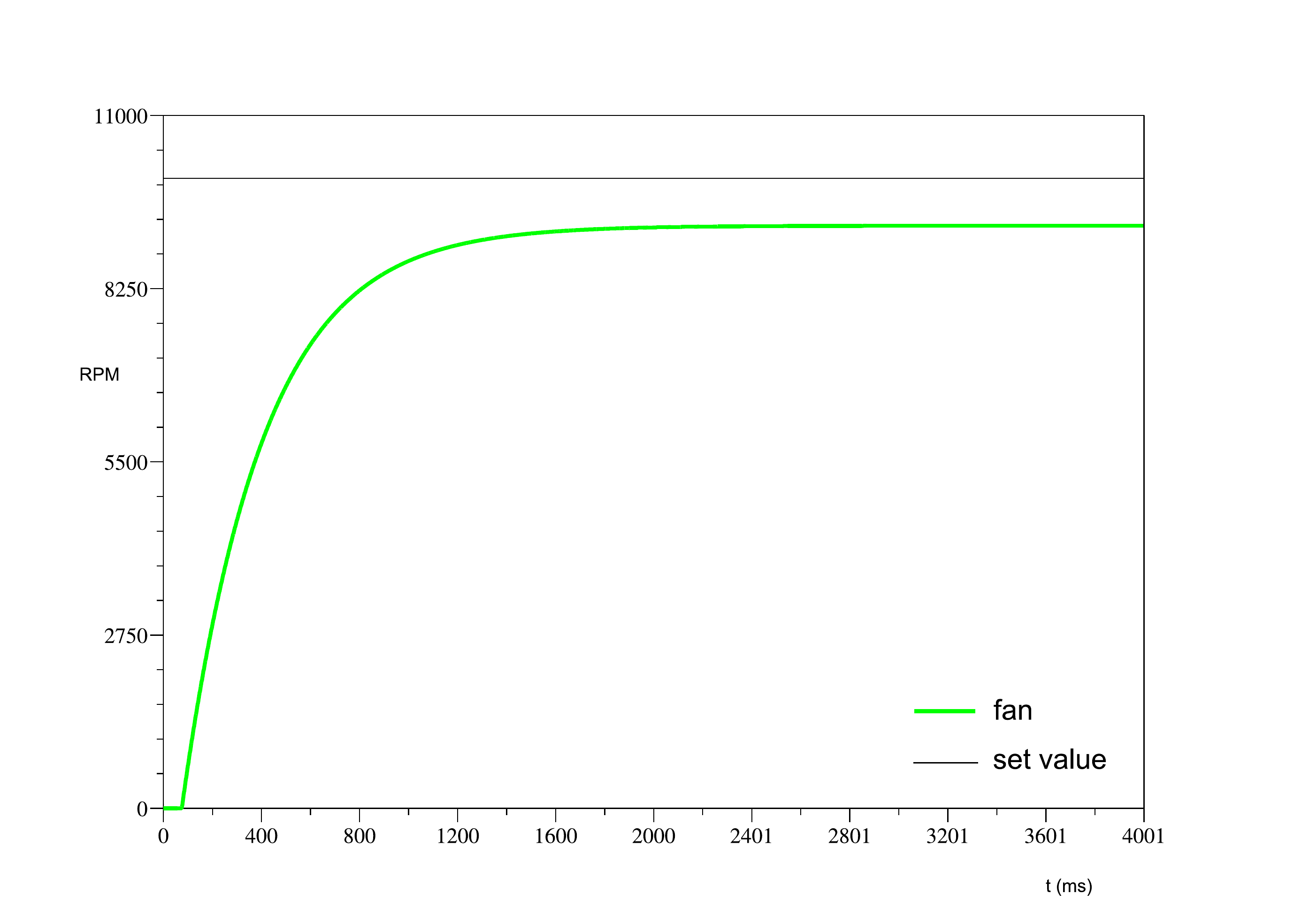}
\caption[Fan, Scicos model simulation]{Fan, simulated Scicos model, $T_a=330\: ms,\: T_u=78\: ms,\: K_S=0.925\frac{\%}{\%}$.}
\label{fig:fanBehaviorScicosSimulatedBeauty}
\end{center}
\end{figure}

These characteristics are combined into a single Scicos model. The simulation results of the path are shown in figure~\ref{fig:fanBehaviorScicosSimulatedBeauty}.

\paragraph{Controller Modeling}$~~$\\

\begin{figure}[!htbp]
\begin{center}
\includegraphics[angle=270,width=0.95\textwidth]{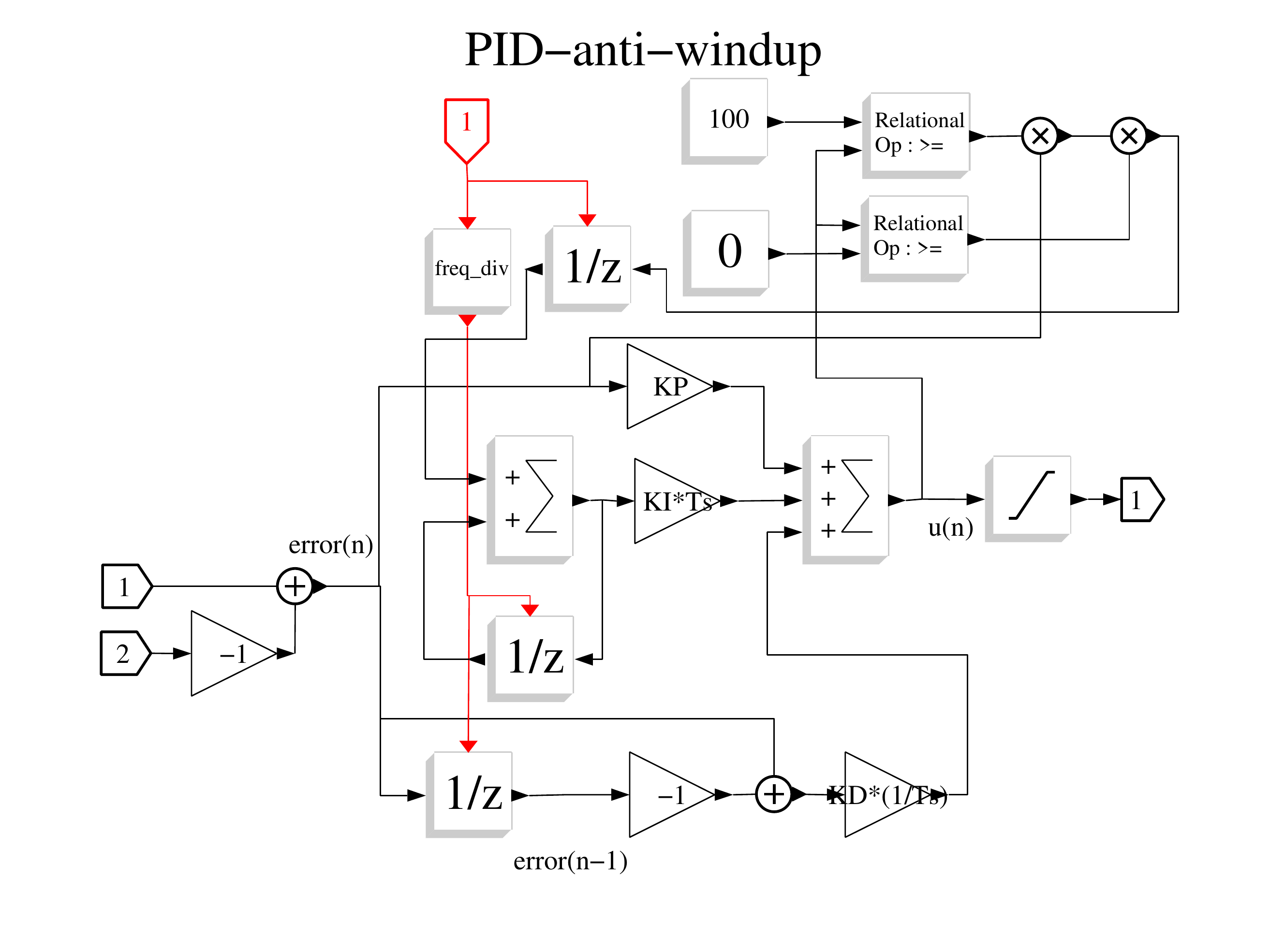}
\caption[PID controller, Scicos model]{PID controller, Scicos model.}
\label{fig:pidControllerScicos}
\end{center}
\end{figure}

The modeled controller is a typical discrete PID controller with anti-windup (figure~\ref{fig:pidControllerScicos}). The anti-windup feature is necessary in order to avoid negative effects from the integrator component when the control output from the PID algorithm is pruned to the maximum output. The additional 1/Z block (memory) is inserted to avoid algebraic loops - it is triggered with a higher frequency than the rest of the algorithm.

\paragraph{Hybrid System Simulation}$~~$\\

\begin{figure}[!htbp]
\begin{center}
\includegraphics[angle=270,width=0.85\textwidth]{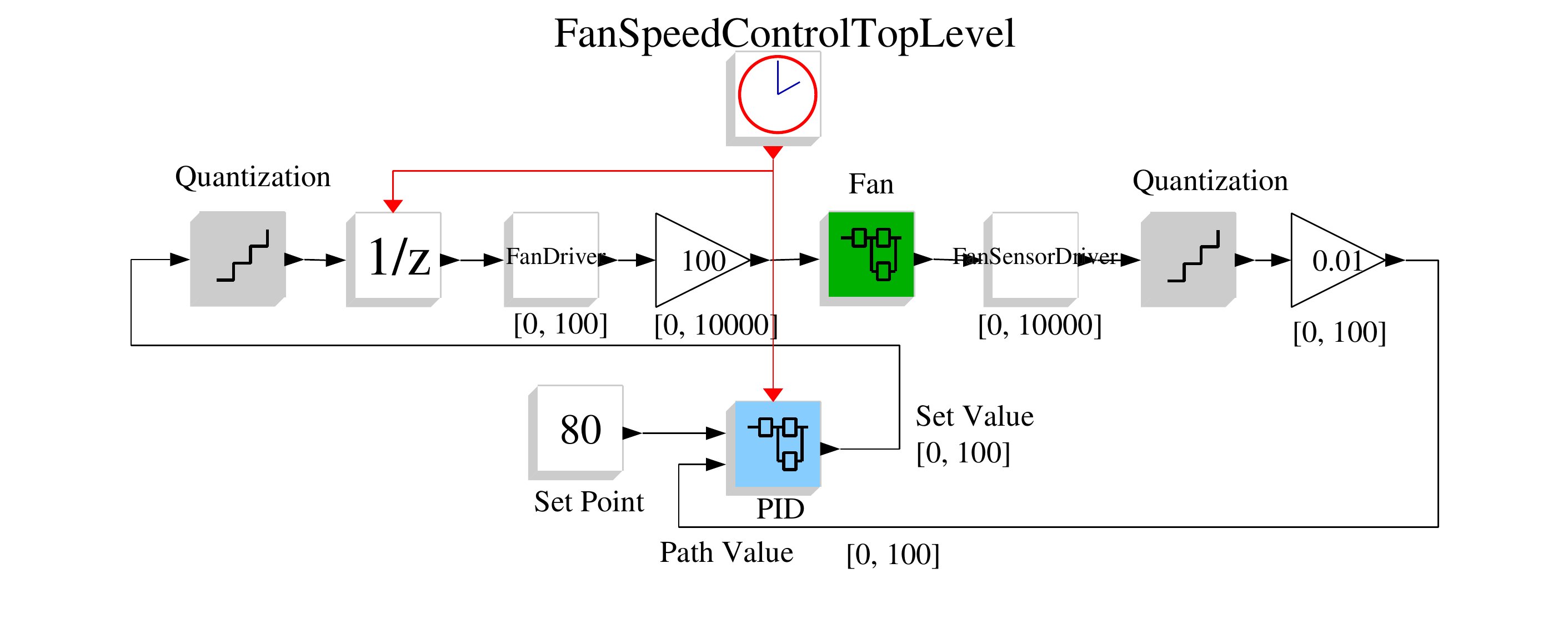}
\caption[Scicos hybrid system model]{Plant and controller modeled in Scicos.}
\label{fig:hybridSystemBeauty}
\end{center}
\end{figure}

Controller and path are modeled as a closed loop and simulated. A top-level model shows the combination of discrete and continuous domain (figure~\ref{fig:hybridSystemBeauty}). The PID controller takes the set point as argument and provides set values between $0$ and $100$. These values are quantized since the modular fan drivers take integers as arguments. The fan driver outputs PWM duty-cycle values that control the fan between $0$ and $10\,000\:RPM$ (this is depicted by the following gain block). Now, in the continuous domain, calculated in Scicos with double values, the fan speed is measured with a fan sensor driver. The resulting $RPM$ values are quantized since the modularly designed drivers provide only integer values. These values are scaled down by the factor $100$ and forwarded to the modularly designed PID control block.

The system is simulated with the parameters $K_P,\:K_I,\:K_S$ tuned according to the rules of Chien, Hrones and Reswick without overshoot (figure~\ref{fig:parameters}).

\begin{figure}[!htbp]
\[
\begin{array}{r c c r c}
T_a & & \ldots \text{effective compenstation time} &=& 330\: ms  \\
T_u & & \ldots \text{deadtime} &=& 78\: ms \\
K_S & & \ldots \text{gain} &=& 0.925\: \frac{\%}{\%}\\
K_P &=& \frac{0.6 \cdot T_a}{K_S\cdot Tu} &=& 2.74\: \frac{\%}{\%}\\
K_I &=& \frac{K_P}{T_n} &=& 0.00832\: \frac{\%}{\%}\\
K_D &=& K_P\cdot T_v &=& 107.02703\: \frac{\%}{\%}\\
T_n &=& T_a &=& 330\: ms\\
T_v &=& 0.5 \cdot T_u &=& 39\: ms\\
S_P & & \ldots \text{set-point} &=& 80 \frac{\%}{\%} \\
T_S & & \ldots \text{sampling time} &=& 5\: ms \\
\end{array}
\]
\caption{PID control algorithm, parameters.}
\label{fig:parameters}
\end{figure}

\begin{figure}[!htbp]
\begin{center}
\includegraphics[width=0.75\textwidth,angle=0]{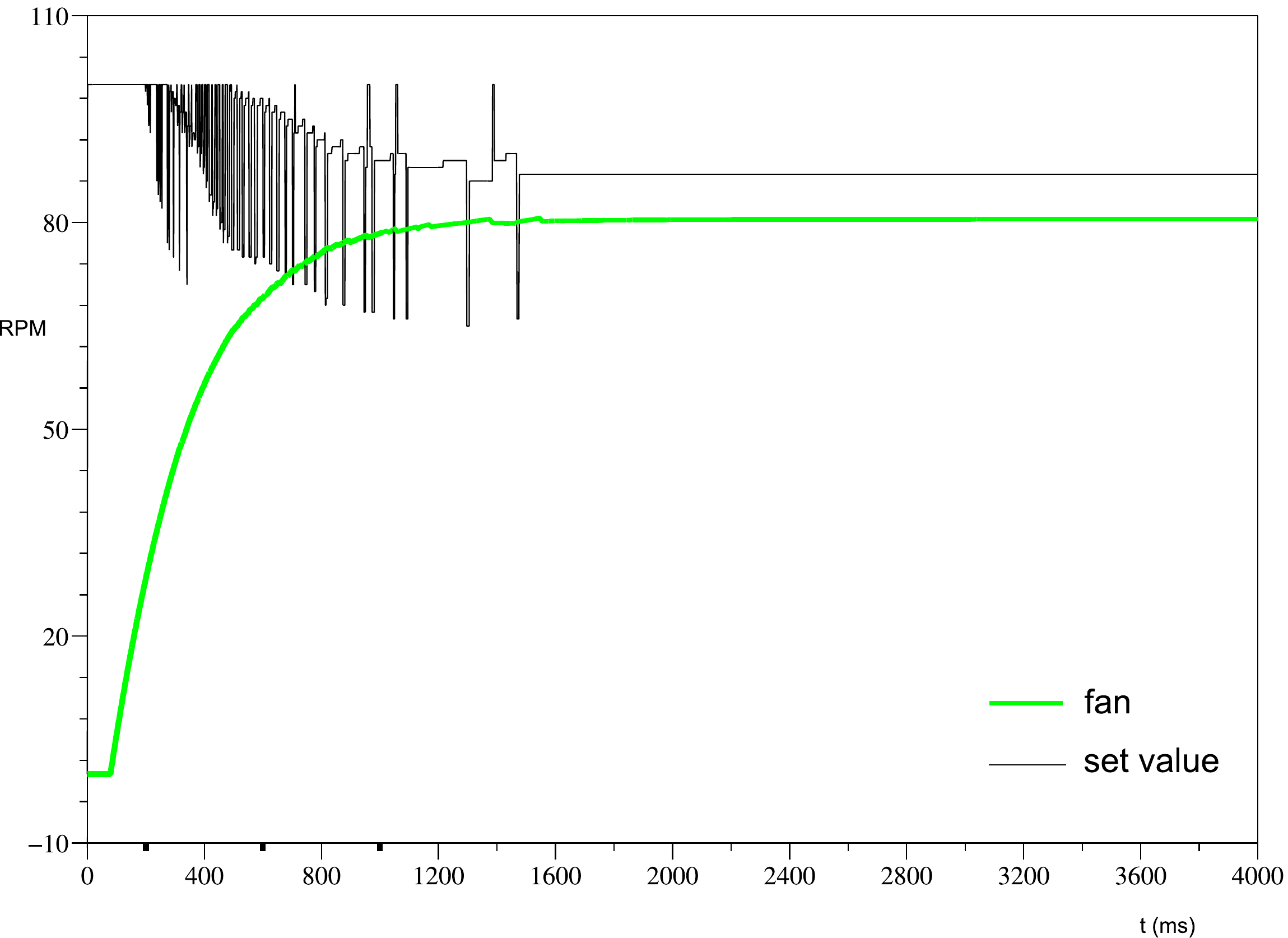}
\caption[Scicos hybrid system simulation]{Plant and controller simulated in Scicos, Input/Output of the fan (continuous domain).}
\label{fig:scicosSimulationHybridFanBeauty}
\end{center}
\end{figure}

The simulation results, input/output values for the fan PID controller are depicted in diagram~\ref{fig:scicosSimulationHybridFanBeauty}). The differential weight factor is responsible for the abrupt changes in the set value.

\paragraph{PID Algorithm - Temporal Design and Automatic Implementation}$~~$\\

\begin{figure}[!htbp]
\begin{center}
	\includegraphics[width=0.55\textwidth]{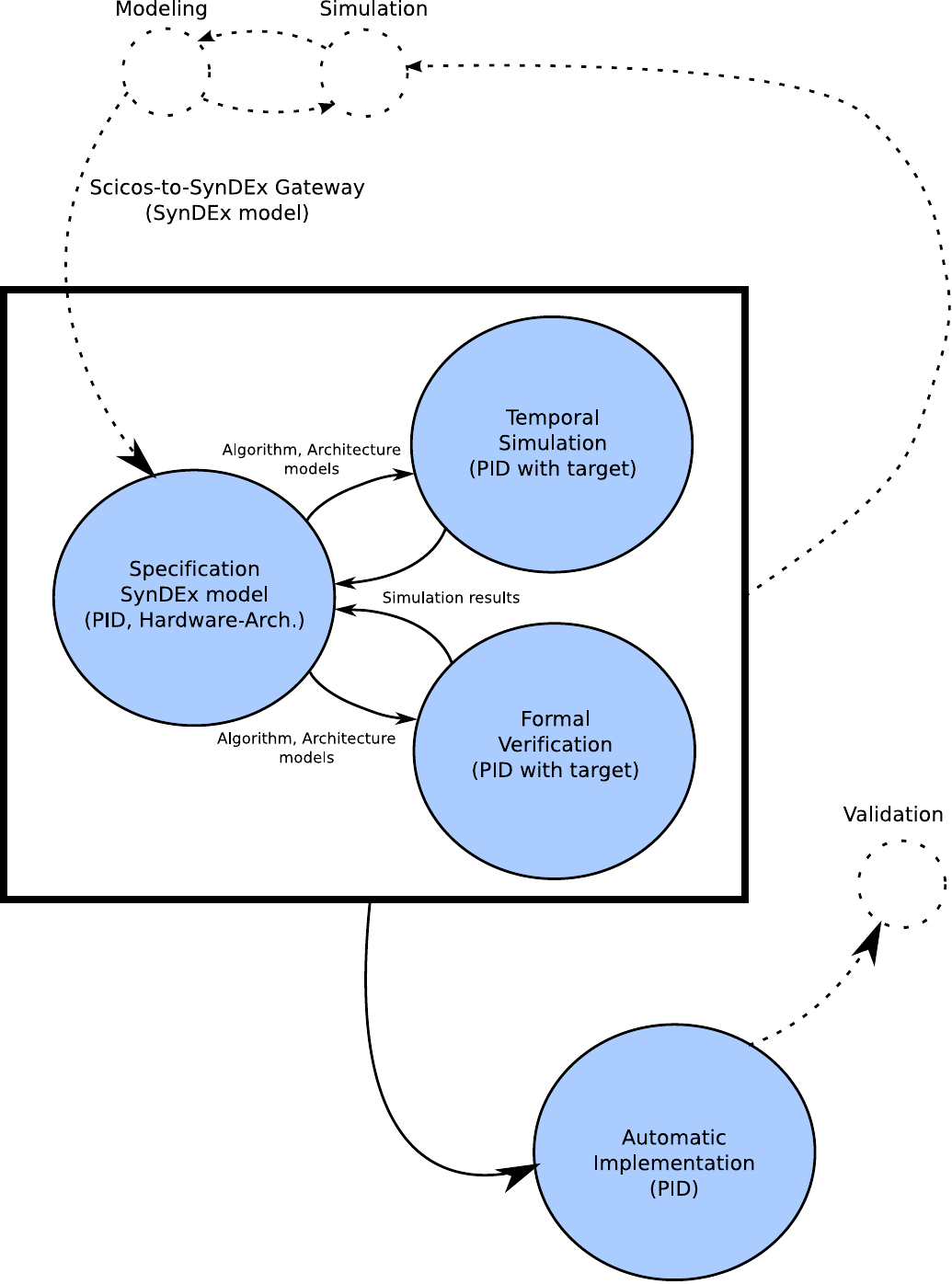}
   \caption[PID-Example. Partial development process.]{Partial software development process, PID-Example. The Scicos model is transformed into a SynDEx model acting as system specification followed by temporal simulations, formal verification, and automated code generation.}
   \label{fig:Scicos:LifeCycleMonoProcessorExampleTemporal}
\end{center}
\end{figure}

After the Scicos simulation results were assumed being right, the Scicos model is transformed to a SynDEx model via the Scicos-To-SynDEx gateway (a model-to-model transformation). SynDEx is then used to design the temporal requirements (the PID control algorithm is triggered every $5\: ms$), model the hardware architecture (and execution environment), simulate these together with the PID algorithm model and then automatically generate a macro code representation of the model (see figure~\ref{fig:Scicos:LifeCycleMonoProcessorExampleTemporal}). After that, the macro code model is expanded into an AVR-GCC compliant C code by the usage of the GNU M4 macro processor\footnote{\url{http://www.gnu.org/software/m4}} in conjunction with the target-dependent macro expansion files, which are consisting of rule-sets for expanding the macro code to C.

The SynDEx set of models for this example consists basically of two main models, the \Keyword{hardware architecture} defining a model of the target hardware, and the \Keyword{algorithm} defining a model of the PID algorithm in connection with the fan drivers.

\begin{figure}[!htbp]
\begin{center}
	\includegraphics[angle=0,width=0.5\textwidth]{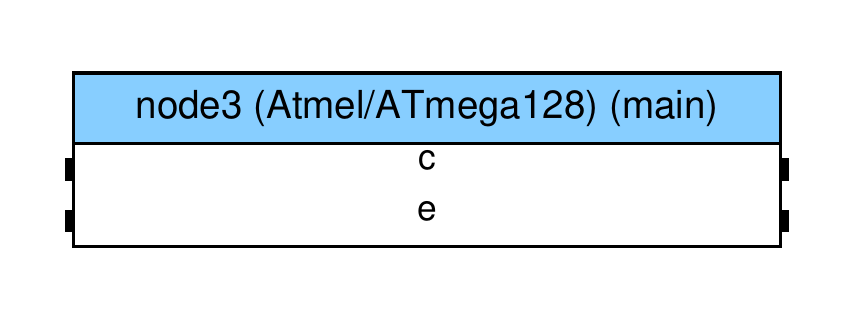}
   \caption{Node3, SynDEx architecture model.}
   \label{fig:SynDEx:Node3Arch}
\end{center}
\end{figure}

\subparagraph{Designing a SynDEx hardware architecture model.} SynDEx requires a target hardware architecture being described by a model, for this purpose an ATmega128 microcontroller SynDEx block is constructed. The ATmega128 block is part of a customized Atmel SynDEx-library. Since a microcontroller on the target hardware can be connected to the bus and enable a communication line to the other nodes on the ESE-board, a communication gate (c) is added to the model. A node could also be connected to an external PC device via a Serial-To-USB circuit, therefore an additional communication gate (e) was added. Since this is a monoprocessor example, the gates will not be used here. The model is depicted in figure~\ref{fig:SynDEx:Node3Arch} and the corresponding SynDEx code is listed here (file Atmel.sdx):\\

\begin{lstlisting}[label=SyndexNodeArch,caption=ESE-Board - ATmega128 Node - SynDEx-Architecture]
%auto-ignore
syndex_version : "7.0.0"
application description : ""

# Libraries

# Algorithms

def operator ATmega128 :
gate COMA c;
gate COMB e;
code_phases:  loopseq initseq endseq;

# Main Algorithm / Main Architecture

# Extra durations

# Software components

# Constraints
\end{lstlisting}

\begin{figure}[!htbp]
\begin{center}
	\includegraphics[angle=0,width=0.5\textwidth]{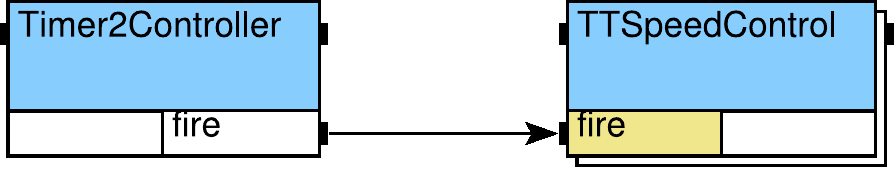}
   \caption[PID-Example. Top-level SynDEx model]{PID-Example. Top-Level SynDEx model.}
   \label{fig:SynDEx:FanSpeedControlTop}
\end{center}
\end{figure}

\subparagraph{Design and parametrization of the SynDEx algorithm model.} The top-level SynDEx algorithm model is depicted in figure~\ref{fig:SynDEx:FanSpeedControlTop}. The figure depicts a block for the control of the timer2 unit of the ATmega128 microcontroller. This block takes parameters that determine when to fire a timer event and activates the following top-level hierarchy block. Latter contains the PID controller algorithm connected to the fan drivers. One reason for choosing this solution was the lack of sufficient computation power of the used development system: If the period of a task is much higher than that of other tasks, the resulting scheduling table will be too large to display, too large to be sufficiently readable by the developer, and additionally tricky scheduling algorithms need to be added to the code generation part if such a scheduling needs to be implemented in the product. This solution, using the timer2 controller block and the following top-level time-triggered (confer to \cite{TUW-137912}) block burns time every micro-period for the sake of scheduling. The price for improved scheduling, readability and an implementation without high effort scheduling code, is thus spending periodic time. A second reason, of no lesser importance, is the fact, that a constant time instance for a periodic task is needed. If there was no timer, there would be a, more or less random, shift of the execution of the time-critical blocks earlier in time, because all time blocks in SynDEx are considered to be atomic and WCET - in fact, tasks may be finished before their specified WCET, pulling the execution of the periodic task forward in time.

\begin{figure}[!htbp]
\begin{center}
	\includegraphics[angle=0,width=0.75\textwidth]{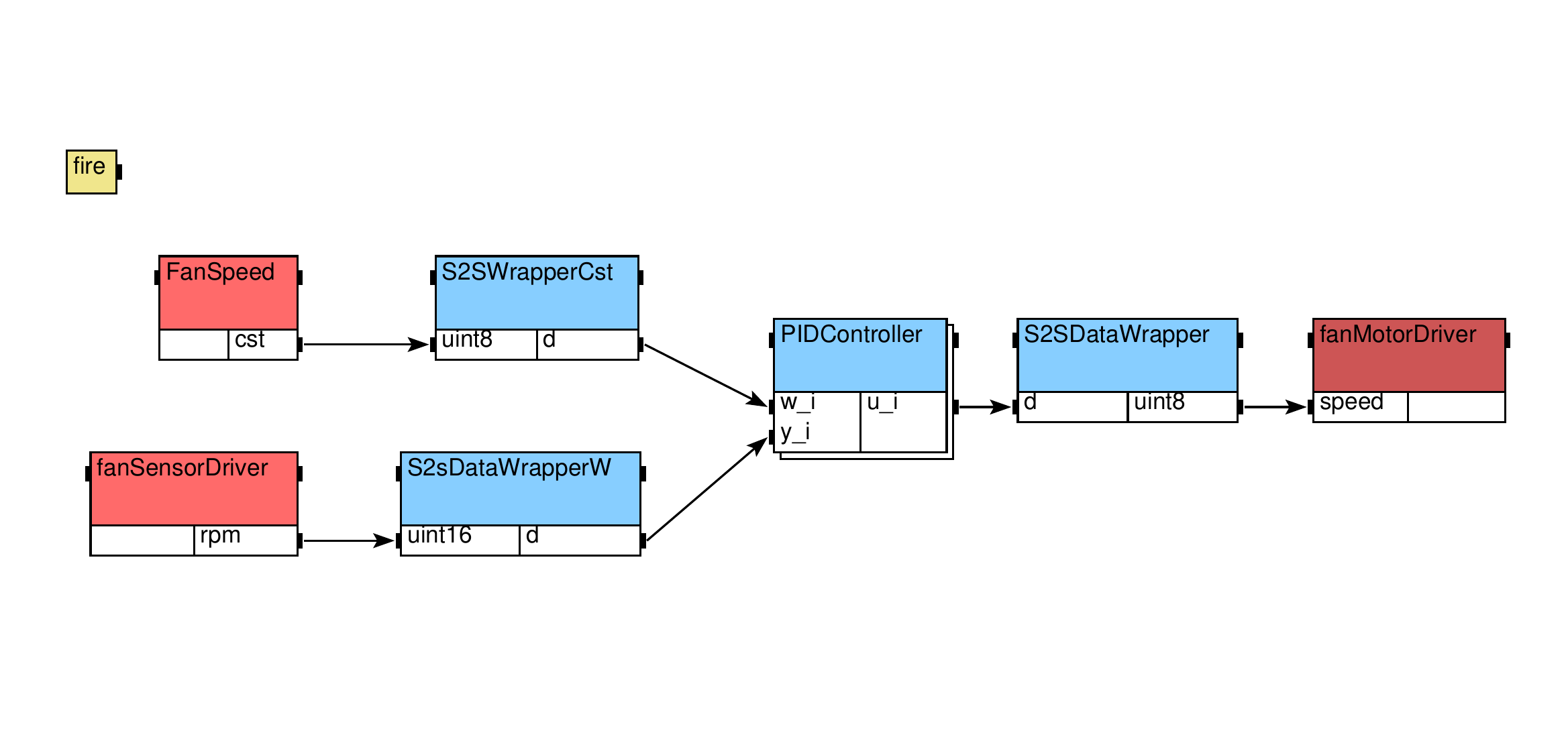}
   \caption[PID-Example. Interfaces]{PID-Example. SynDEx model, Scicos-transformed PID algorithm interfacing fan sensor and fan drivers.}
   \label{fig:SynDEx:FanSpeedControlTTAlgorithm}
\end{center}
\end{figure}

The timer2 (time-triggered) block activates the fan sensor block, the PID controller block and the fan driver (see figure~\ref{fig:SynDEx:FanSpeedControlTTAlgorithm}). The PID block was converted from the Scicos model into the SynDEx model and therefore its data types are double. Fan driver and fan sensor driver blocks are of the types uint8 and uint16 leading to the need of an intermediate interface layer wrapping the types: S2SWrapper-blocks convert double to uint8, uint8 to double and uint16 to double data types, which could mean a loss of information, but since the Scicos model was implemented by using quantization blocks (thus interfacing the real fans with integer values), these conversions make sense. 

\begin{figure}[!htbp]
\begin{center}
	\includegraphics[angle=0,width=1.\textwidth]{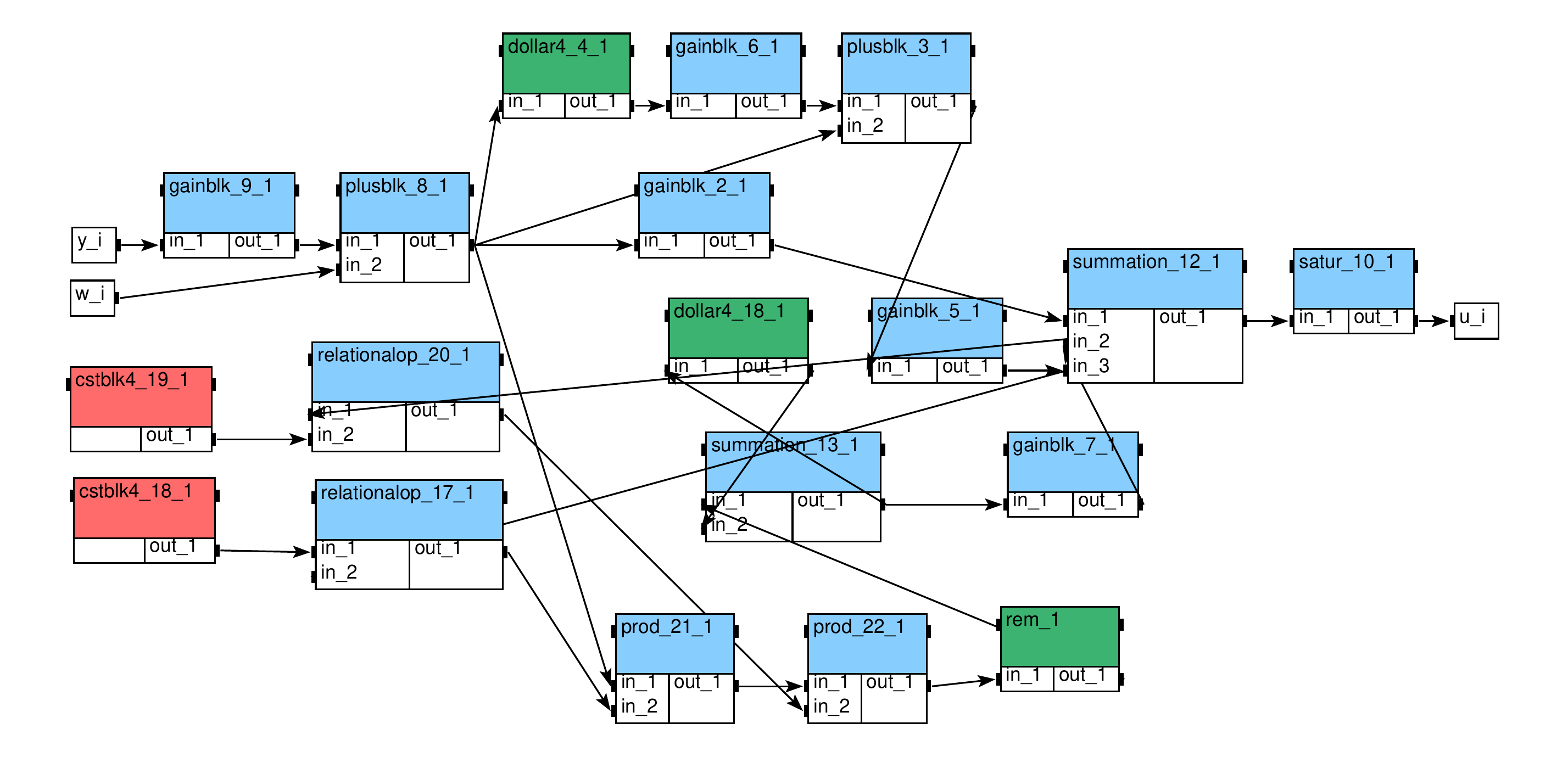}
   \caption[PID-Example. PID SynDEx model]{PID-Example. SynDEx model of the PID controller transformed from the Scicos PID model.}
   \label{fig:SynDEx:FanSpeedControlPID}
\end{center}
\end{figure}

Finally, on the bottom level of the hierarchy, the PID algorithm, which was converted from a Scicos model into a SynDEx model, is shown in figure~\ref{fig:SynDEx:FanSpeedControlPID}. The algorithm takes the set point $w_i$ and the set value $y_i$ as inputs and outputs the path value $u_i$. Note that all the blocks contained in this PID controller are bearing the data structure overhead from the Scicos model: The ability to simulate and change, automatically transform models between Scicos and SynDEx introduces additional data memory costs on the target platform.\\

\paragraph{PID Algorithm - Scheduling of the SynDEx model}$~~$\\

SynDEx takes the algorithm model and schedules the blocks by using a greedy algorithm under provided time constraints. The entire algorithm is supposed to be triggered about every $5\:ms$ (even a Sampling Time of $8.5\:ms$ was calculated, however the real-world properties make the fine-tuning of a PID almost always necessary) and this is accomplished basically in two steps:\\
\begin{itemize}
\item \textbf{Define the periods of each task} (similar to the Liu and Layland task model).\\
The ATmega128 is clocked with a frequency of $14.745\,6\:\textit{MHz}\:$, that means $1\:ms\:$ takes 14745.6 CPU cycles, $5\:ms$ take $73\,728$ CPU cycles. The question raised now is how exactly the period must be chosen to meet the requirements, since this is a single periodic example. A period chosen of $74\:STU\:$ will not interfere with the schedulability of other tasks (since there are no other tasks which periods would have to be n-times the period of this task). To keep it simple, the trigger time of the algorithm is modeled with $80\:STU=5.42\:ms$, however the STUs are chosen, they will mostly imply an inaccuracy between model and real world (depends also on the clock of the MCU).
\item \textbf{Define the durations of each task.}\\
To keep the model human-readable, a standardized, constant duration for each block is chosen: $1\,000\: cycles$ of the ATmega128 shall represent $1\:STU$ (SynDEx Time Unit in the model - this unit is made up for a better nomenclature). All the blocks in this example are executed a lot faster than that time (WCET), leading into an investment of resources into safety.
\end{itemize}

\begin{figure}[!htbp]
\begin{center}
	\includegraphics[angle=0,width=0.95\textwidth]{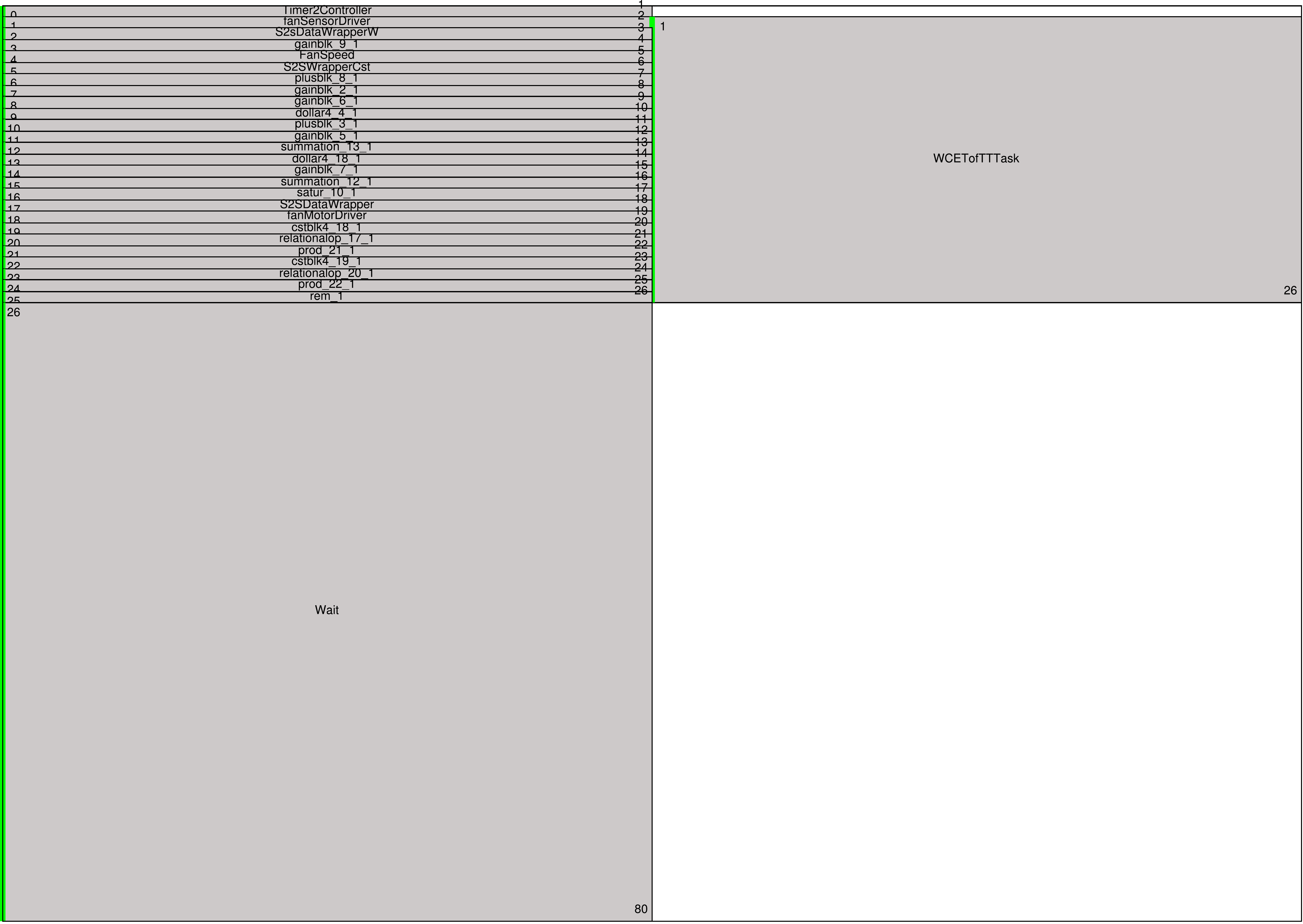}
   \caption[PID-Example. SynDEx scheduling after algorithm adequation]{PID-Example. SynDEx scheduling after algorithm adequation.}
   \label{fig:SynDEx:SchedulingMono}
\end{center}
\end{figure}

After SynDEx runs through the algorithm using the AAA methodology, a resulting scheduling table can be investigated (see figure~\ref{fig:SynDEx:SchedulingMono}). On the left hand side on the processor schedule, there is the timed controller running at the start of each CPU-period, followed by the main algorithm. After that, a wait statement is inserted to fill up the intended period of $80\:STU$. The right hand side is a time reservation slot which will be executed if the timer controller has not fired. If a task has to be executed every $3\:s$ and another one every $5\:ms$, the costs for the calculation of the scheduling will be enormous, resulting in an obfuscated model (provided the calculating time of the model is smaller than the developers patience/time available). Therefore, we reserve time (wait the WCET of the whole algorithm) whenever the timer does not fire in this CPU period.    

\paragraph{PID Algorithm - Macro Code and Executive Code Generation}$~~$\\

At this step the development process is at a stage where it is possible to generate a target language-independent macro code file for each involved node. Using target language-dependent macro definitions these macro code files are then translated into avr-gcc compliant C code which afterwards can finally be compiled, downloaded and executed on the hardware.

A look at the generated macro code explains what is needed to realize these last steps (node3.m4, listing~\ref{lst:SynDExFanSpeedControlNode3}). In this file, after allocating global memory for variables, the \Keyword{main\_} function tag is immediately followed by an \Keyword{proc\_init\_} tag. Latter tag is used to switch a flag in the macro code - this way a macro is then expanded to its corresponding section in the target-dependent definition file, thus the C code for the initialization for the block (if there is any) is inserted. An example of a macro expansion definition for the pid\_gainblock is listed in the appendix (see listing~\ref{lst:SynDEx:Macros:PIDGain}): The generated C code depends on the code phase in the node3.m4 algorithm file. After all initializations of the blocks are handled, the main loop of the algorithm expands to an infinite loop (by using the definition in a processor (ATmega128) specific language macro definition file, see listing~\ref{lst:SynDEx:Macros:Atmel}). In this main loop all the blocks are expanded by their corresponding main loop phase. After finishing the main loop, blocks are expanded according to an end phase. For example, the latter can be used to free resources. The resulting C code is listed in the appendix \ref{lst:SynDEx:FanSpeedControlCCode} which can then be compiled by using the generated makefiles (see listing~\ref{lst:SynDEx:FanSpeedControlGNUMakefile} and listing~\ref{lst:SynDEx:FanSpeedControlMK}). The executable was downloaded onto the ESE-Board and the fan run approximately at the set speed.\\

\subparagraph{A note about data types.} It is possible to automatically map the data types to compiler conforming ones by adding rules to the target language expansion file (e.g. ATmega128.m4x). For example, the type \textit{int} is mapped to \textit{int16\_t} without a redesign of the model. This is useful to support the use of different compilers or libraries, and adds value to the re-usability of models.

The interface data types of the modeled SynDEx blocks are defined by an identifier and the data type size (bytes):
\begin{verbatim}
  typedef_(`uint16',2)
\end{verbatim}
New data types can be defined:
\begin{verbatim}
  typedef_(`uint16_t', 2)
\end{verbatim}
The new data types can be mapped (note that also those which are not mapped to a new type need a mapping definition):
\begin{verbatim}
  define(`uint16_map',`uint16_t')
\end{verbatim}
SynDEx will create all data types with a "\_map" appendix when the "basicAlloc\_" definition is edited as follows:
\begin{verbatim}
  define(`basicAlloc_',`_($1_type_()_map() $1[$1_size_];)')
\end{verbatim}

\paragraph{PID Algorithm Example - Validation}$~~$\\

Simulations of the hybrid system, temporal design and algorithm optimization were done. The question at this point in the development process, is if the models and simulations are a good representation of the life behavior. The algorithm was deployed onto the target architecture and some monitoring code inserted. The velocity data of the fan was sent via UART to node0, which itself forwards all the communication data via USB to the development PC. The collection of velocity data on the target did not interfere with the PID code running on it, since the UART baud rate and communication calls were designed carefully with the help of the SynDEx scheduling model: The \Keyword{Wait} time interval in the scheduling model was used for communication purposes.

\begin{figure}[!htbp]
\begin{center}
	\includegraphics[width=1.00\textwidth]{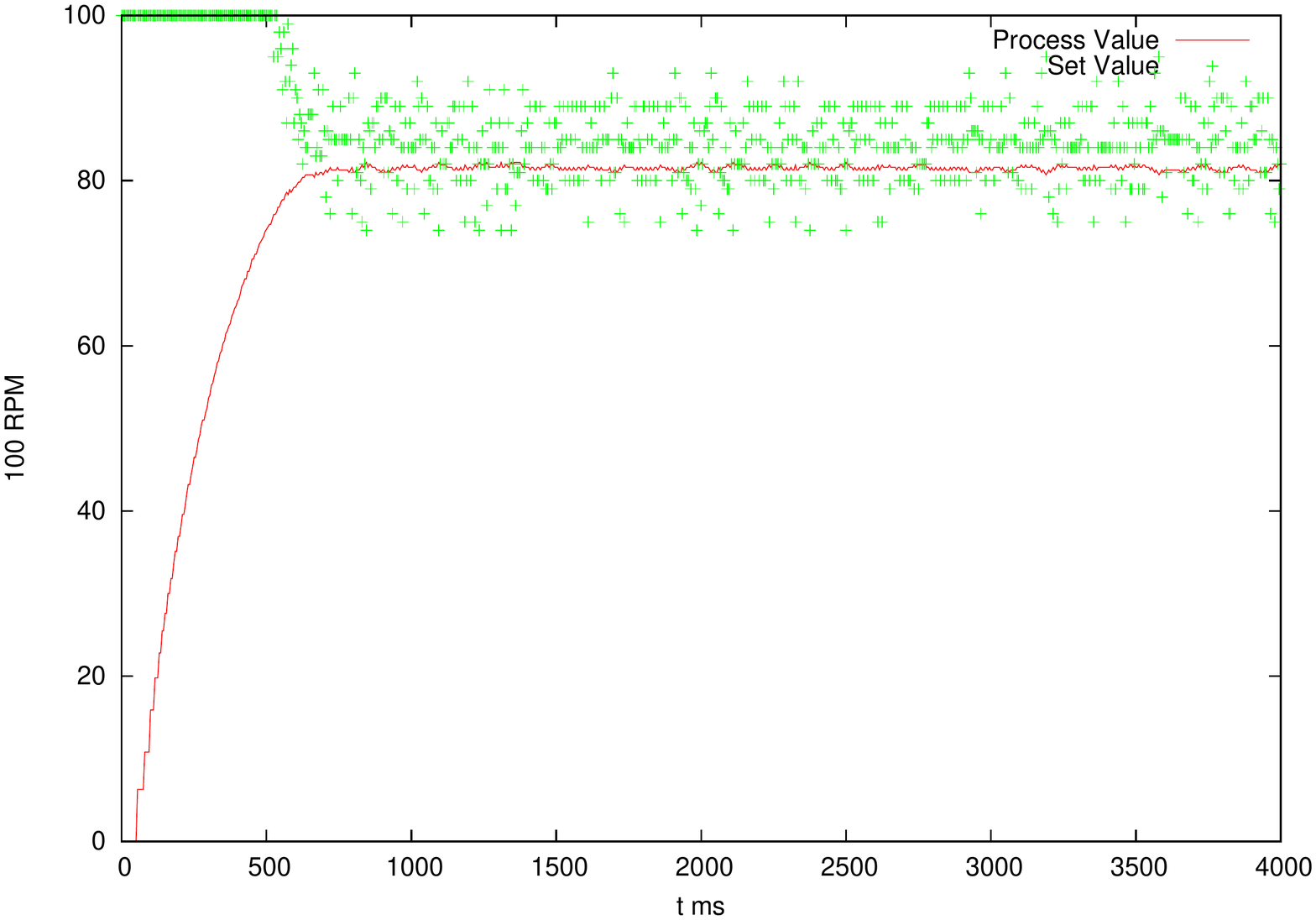}
   \caption[PID-Example. Live data]{PID-Example. Live data diagram of the PID controller and the fan.}
   \label{fig:results:fan0To80}
\end{center}
\end{figure}

Figure~\ref{fig:results:fan0To80} shows the real behavior of the PID controlling the fan. The fan was running approximately at a speed of $80\:\%$ for some seconds. The life behavior differs from the simulated behavior - there might be several reasons for that, however, they are no object of further investigations right here. A PI controller might be more suitable for this situation.

\begin{figure}[!htbp]
\begin{center}
	\includegraphics[width=0.75\textwidth]{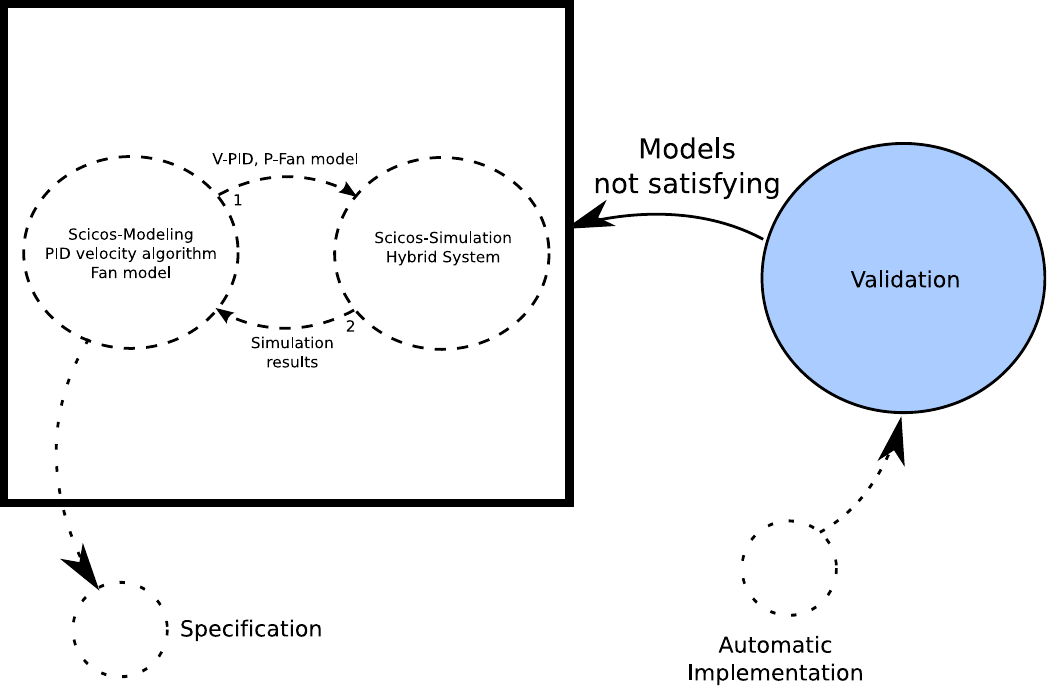}
   \caption[PID-Example. Model redesign.]{PID-Example. Model redesign and new parameters in case of unsatisfying results.}
   \label{fig:results:verification}
\end{center}
\end{figure}

Comparing these live data (figure~\ref{fig:results:fan0To80}) to the Scicos simulation results (figure~\ref{fig:scicosSimulationHybridFanBeauty}) shows different behavior of model and real world. Because of the simple plant model, as well as its roughly calculated parameters, the live data differs. The high fluctuations are explained by the $K_D$ factor and the small fan speed spikes provided by the sensor. If the model was required to be more accurate, a re-modeling with Scicos/SynDEx would be necessary (figure~ \ref{fig:results:verification}). This example escapes the software development cycle with Scicos/SynDEx at this point. 

	\section[Example Multiprocessor]{Example Multiprocessor - Data Observation and Communication with SynDEx}\label{sec:exampleMulti}

\begin{figure}[!htbp]
\begin{center}
	\includegraphics[width=0.75\textwidth]{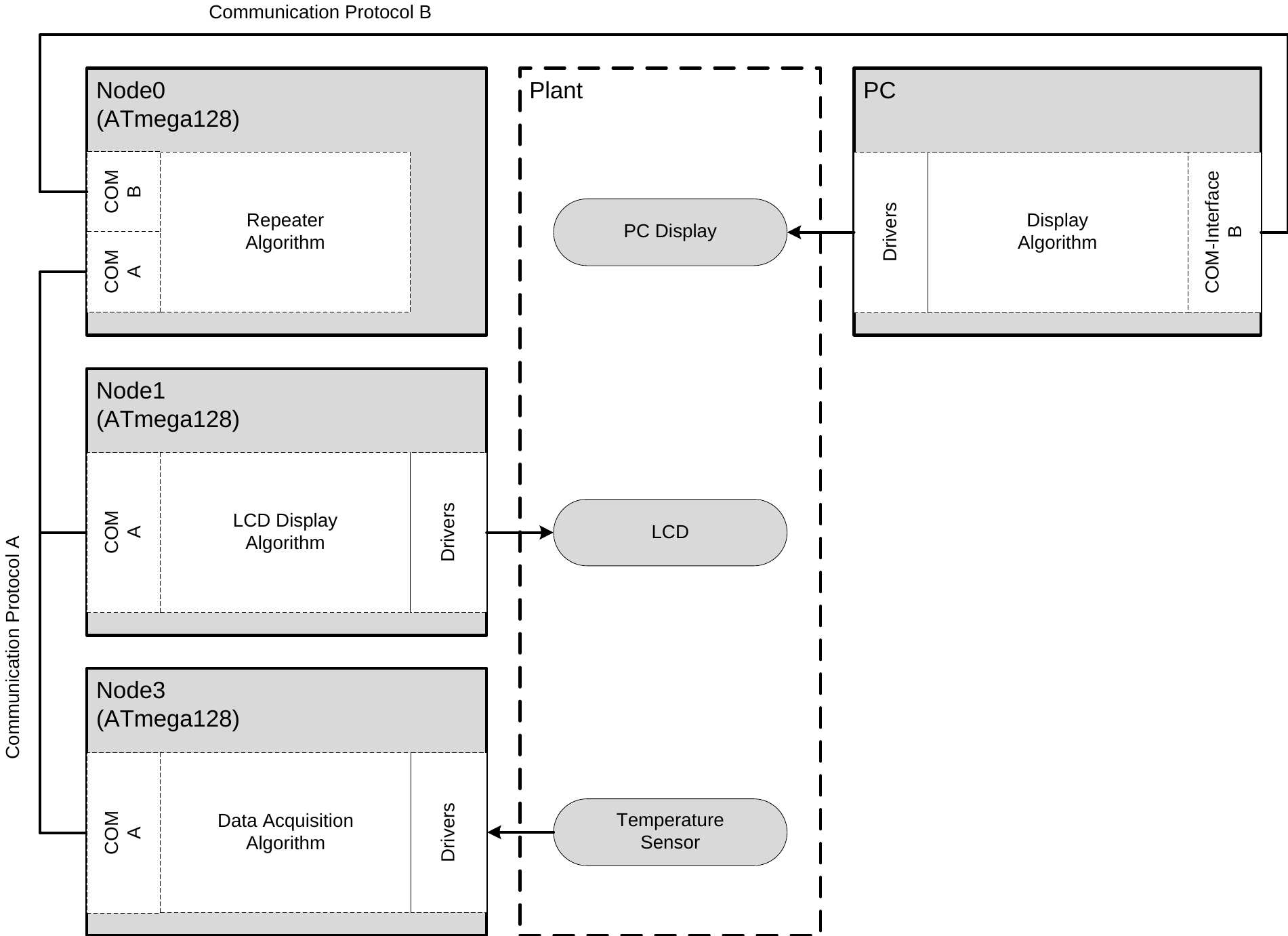}
   \caption{Multiprocessor-Example. Distribution and display of temperature information.}
   \label{fig:ExampleMulti:Principle}
\end{center}
\end{figure}

The aim of this example is to model a multiprocessor application in SynDEx with the ESE-Board as target hardware. Temperature values are periodically acquired at node3. These temperature values will be sent to node1 and displayed on its LCD peripheral. Furthermore, the temperature data is sent via node0 (acting as a repeater) to the development PC and displayed there (figure~\ref{fig:ExampleMulti:Principle}).

Expected results of this example are statements about the feasibility and scope of using SynDEx models on a microcontroller multiprocessor architecture.

\subsection[Hardware Architecture]{Hardware Architecture (Execution Environment)}\label{sec:exampleMulti:hw_arch}

The execution environment consists of several ESE-Board components:
\begin{itemize}
\item \textbf{Microcontroller device:} Node0 - Receive temperature data and forward it to the Desktop PC.
\item \textbf{Microcontroller device:} Node1 - Receive temperature data and control LCD peripheral.
\item \textbf{Microcontroller device:} Node3 - Sample temperature data and send it to Node0 and Node1.
\item \textbf{Microcontroller peripheral} Node1: The liquid crystal display for presenting temperature values.
\item \textbf{Microcontroller peripheral} Node3: The temperature sensor for data sampling.
\item \textbf{Desktop computer device:} PC - display temperature data received.
\end{itemize}

\subsection[Software Architecture]{Software Architecture}\label{sec:exampleMulti:sw_arch}

The principle of this example is to realize an application where information is gathered periodically and distributed within a pre-determined timespan to other targets. In terms of an industrial application, this example could be described like a car-application: The break pedal state is checked regularly and its state sent to the brake system within one millisecond.

\paragraph{Requirements}$~~$\\

Temperature values are measured periodically with $20\:STU$ and displayed by the LCD display on Node1. Furthermore the measured data is sent to the PC at the same time period. The temperature data measured has to be sent to the LCD within a time of $15\:STU$.

\paragraph{SynDEx - Hardware architecture model}$~~$\\

\begin{figure}[!htbp]
\begin{center}
	\includegraphics[angle=0,width=0.75\textwidth]{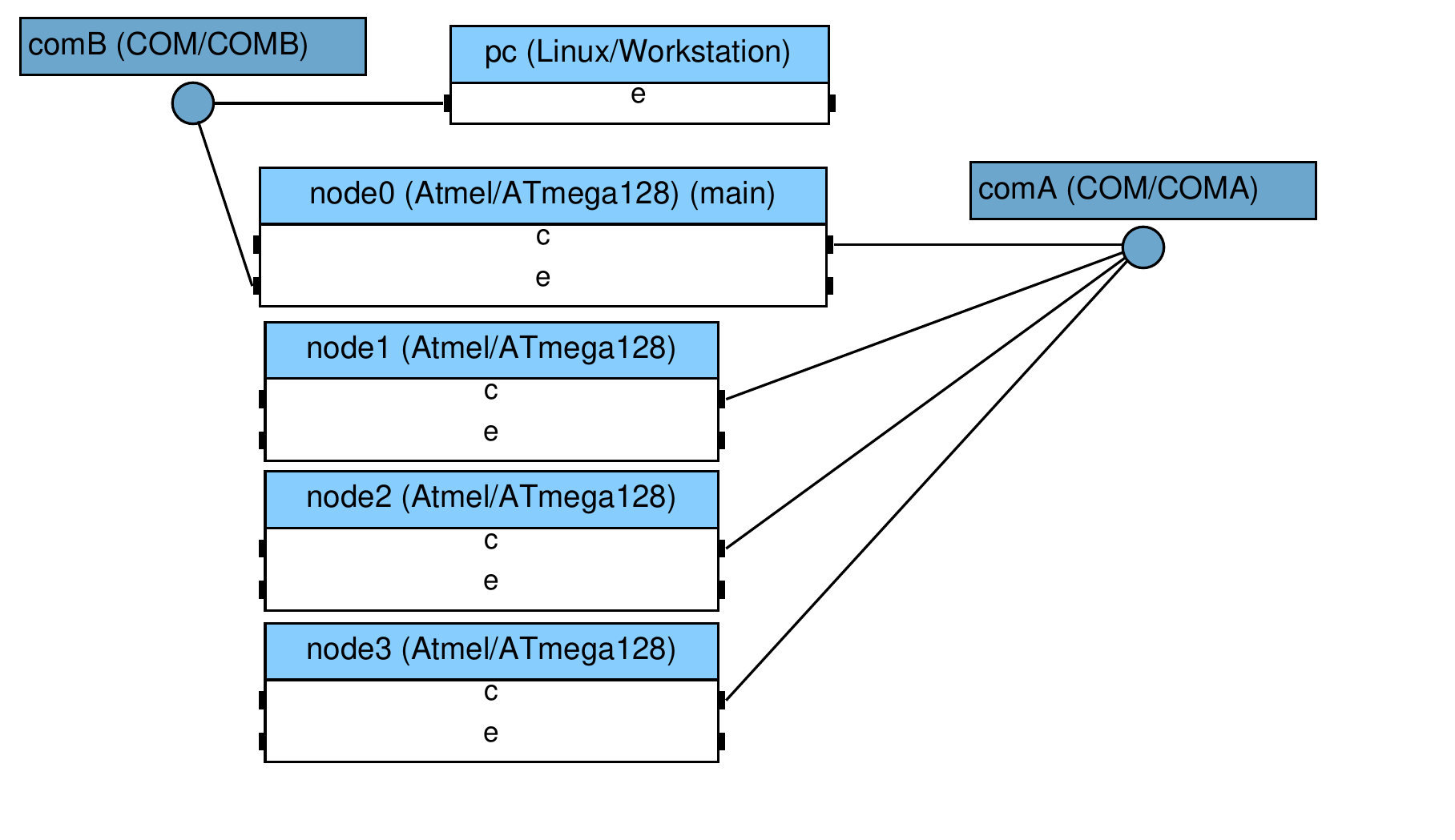}
   \caption[Multiprocessor-Example. SynDEx architecture model]{SynDEx architecture model.}
   \label{fig:ExampleMulti:SynDExHWArch}
\end{center}
\end{figure}

The hardware architecture model designed for this example includes four microcontroller nodes (node0, node1, node2, node3) connected by communication medium "comA", and one desktop PC connected to node0 by communication medium "comB" (figure~\ref{fig:ExampleMulti:SynDExHWArch}). Every communication medium is defined as SAM multipoint without broadcast.

\paragraph{SynDEx - Algorithm model}$~~$\\

\begin{figure}[!htbp]
\begin{center}
	\includegraphics[angle=0,width=0.75\textwidth]{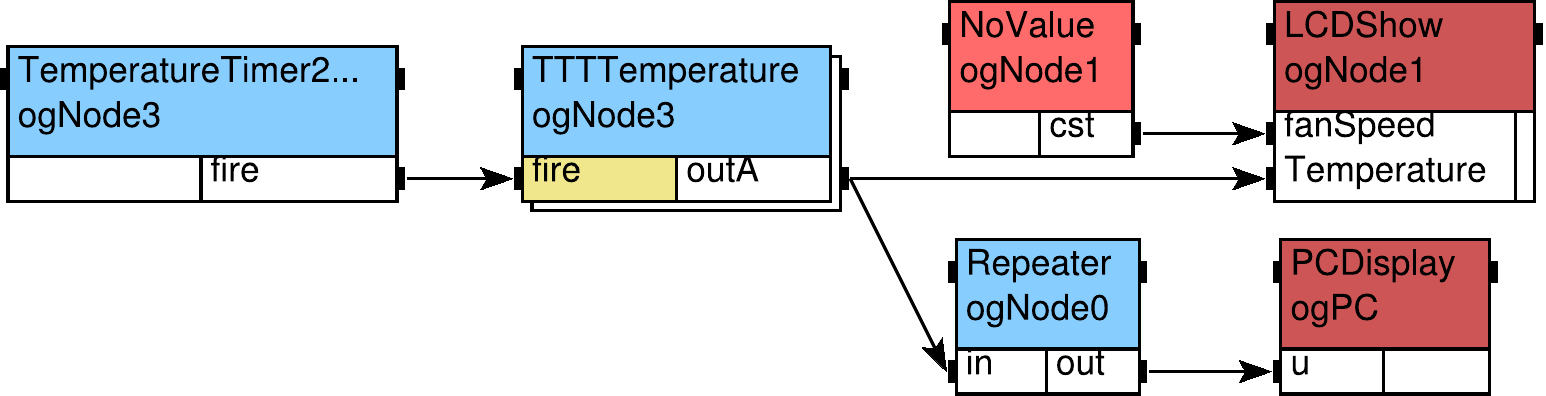}
   \caption[Multiprocessor-Example. A SynDEx algorithm model with four operators]{A SynDEx algorithm model with four operators.}
   \label{fig:ExampleMulti:SynDExAlgorithmModel}
\end{center}
\end{figure}

The algorithm model (figure~\ref{fig:ExampleMulti:SynDExAlgorithmModel}) consists of several interacting blocks. Located on Node3, a timer block (\Keyword{TemperatureTimer2Controller}) periodically triggers the temperature acquisition blocks (\Keyword{TTTTemperature}; returns 0 when not fired, otherwise the temperature) is the first instance in the algorithm. After the temperature data is gathered, it is sent to the \Keyword{Repeater} block on Node0 which forwards it to the \Keyword{PCDisplay} block. The data returned from the \Keyword{TTTTemperature} block is passed to the \Keyword{LCDShow} block located on Node1. \Keyword{NoSpeedValue} is supplied to the \Keyword{LCDShow} block if there are no data available.

\begin{figure}[!htbp]
\begin{center}
	\includegraphics[angle=0,width=0.75\textwidth]{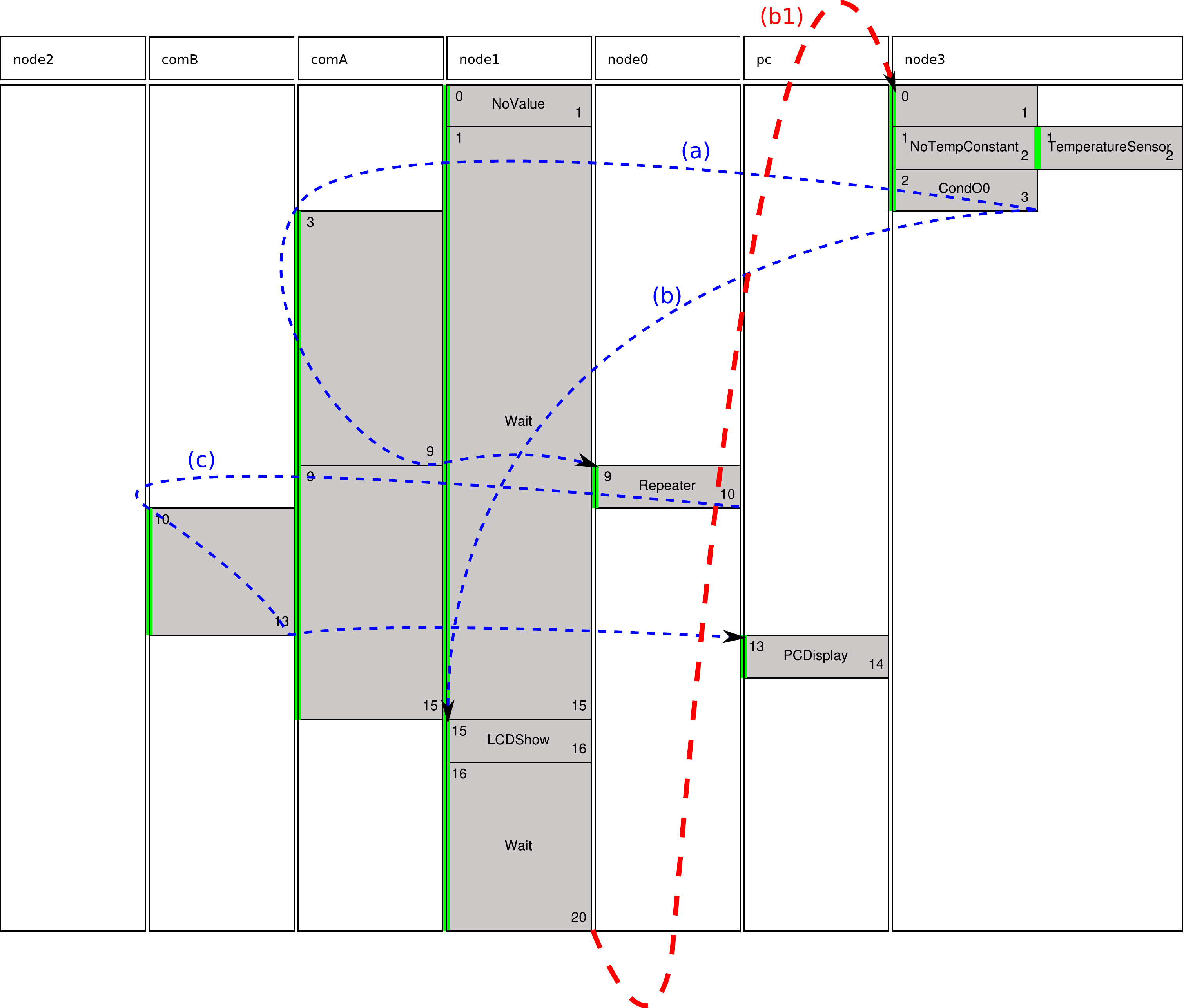}
   \caption[Multiprocessor-Example. SynDEx scheduling.]{SynDEx scheduling for the multiprocessor example.}
   \label{fig:ExampleMulti:SynDExSchedulingMultiResult}
\end{center}
\end{figure}

All blocks in the model have a defined period of $20\:STU$, the durations of the operator-blocks were defined with very high values for the sake of the readability of the model (otherwise the fraction of communication-durations/operator-block-durations might be very high). The resulting SynDEx scheduling of the algorithm can be found in figure~\ref{fig:ExampleMulti:SynDExSchedulingMultiResult} and a brief explanation what happens in this scheduling table is given here:

\begin{itemize}
\item \textbf{node0 (a)}. The \Keyword{Repeater} block is executed after the temperature data from block \Keyword{CondO0} arrived via communication medium \Keyword{comA}.
\item \textbf{node1 (b)}. Initially a no-value block is executed followed by a wait statement until temperature data is received on \Keyword{comA} from \Keyword{CondO0} on node3. With the data available, information is displayed on the \Keyword{LCDShow} block. This node holds also the conceptual trigger for the next inter-repetition synchronization (b1) - it is supposed to hold an additional wait statement to fill out the period of $20\:STU$ (note: in the generated m4 macro code no such statements could be found).
\item \textbf{node2}. This node is not used for this example.
\item \textbf{node3}. The timer controller block triggers the temperature sensor block on time.
\item \textbf{pc (c)}. The \Keyword{PCDisplay} block gets data from the \Keyword{Repeater} block via \Keyword{comB}.
\item \textbf{comA}. The occupation of the communication channel is displayed. First, data is sent from \Keyword{CondO0} (node3) to \Keyword{Repeater} (node0). Second, temperature data from \Keyword{CondO0} (node3) is sent to \Keyword{PCDisplay} (PC).
\item \textbf{comB}. The \Keyword{Repeater} (node0) sends data to the \Keyword{PCDisplay} (PC) using communication protocol B.
\end{itemize}

\paragraph{SynDEx - Macro Code Structure and Communication}$~~$\\

\begin{figure}[!htbp]
\begin{center}
	\includegraphics[angle=0,width=0.75\textwidth]{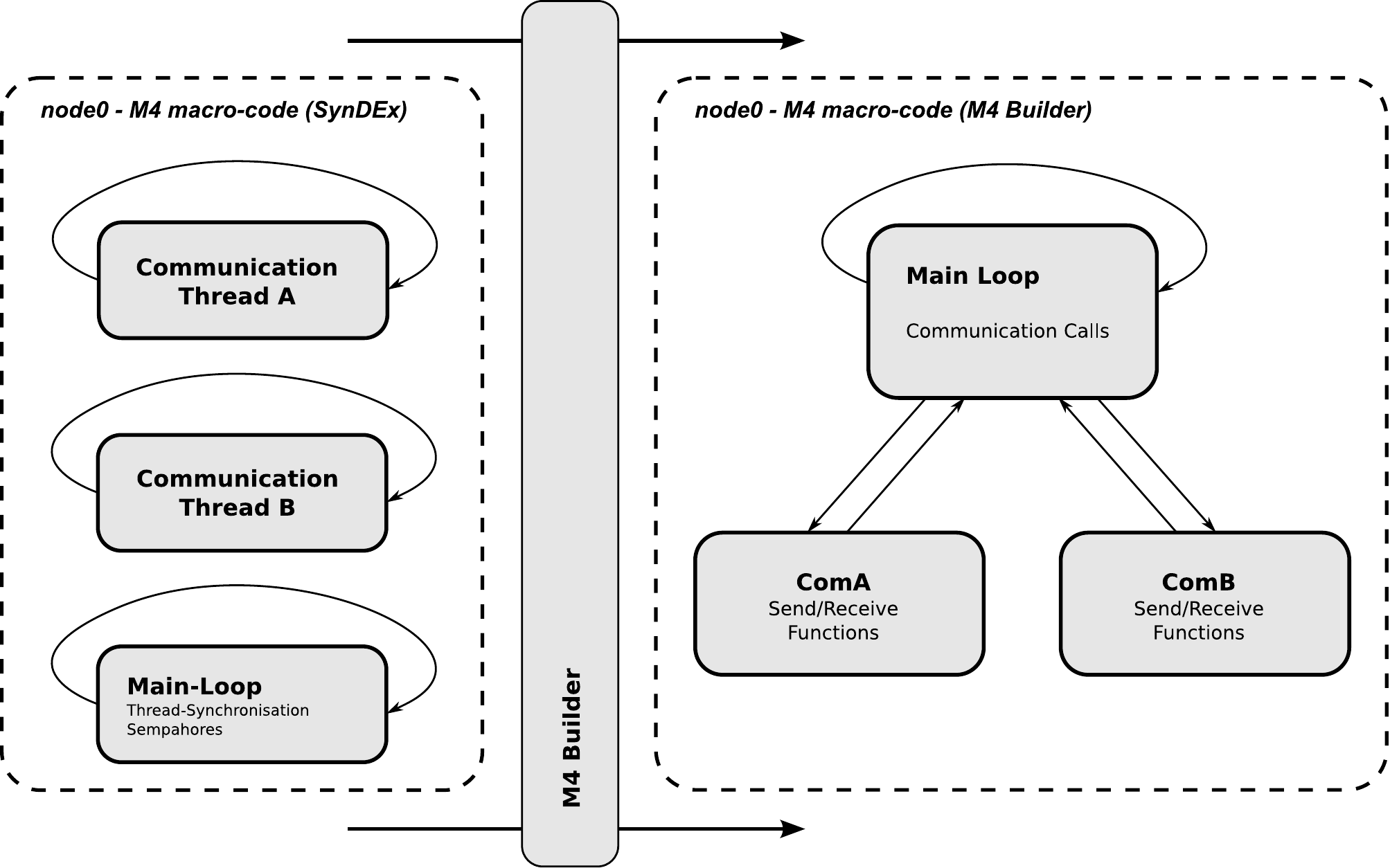}
   \caption{Re-structuring of the SynDEx M4 code.}
   \label{fig:exampleMulti:m4Builder}
\end{center}
\end{figure}

The generated M4 macro code consists of a main loop surrounded by initialize and finalize placeholders for each block. Every communication medium is realized with a separate thread running concurrently to the main loop. The challenge here is to make this code structure fit for a target without supported threads.
Therefore a customized program is built (M4-Builder) to transform these threads to send/receive functions which can be called from within the main-loop blocks (figure~\ref{fig:exampleMulti:m4Builder}). 

Consider listing \ref{lst:exampleMulti:ComBeforeCleaned} which is a portion of the M4 code generated for node0. A thread is assigned to every communication medium: thread\_(COMA,c,node0,node1,node2,node3) for communication medium comA and thread\_(COMB, e, node0, PC) for comB: The parameters defined are the communication medium name, the communication gate (c, e) and the nodes connected by it (e.g. node0 and the PC are connected by comB). Each communication thread holds some semaphores responsible for maintaining the right order of the block execution (Pre0, Suc0, Pre1, Suc1). For example, node0 might only send data to the PC if the repeater block is filled with data.

\lstset{breaklines=true}
\begin{lstlisting}[frame=single,label=lst:exampleMulti:ComBeforeCleaned,caption=SynDEx communication threads]
%auto-ignore
thread_(COMA,c,node0,node1,node2,node3)
  loop_
    Suc1_(_comMultiProcessor_TTTTemperature_CondO0_o_node0_c_empty,,_comMultiProcessor_TTTTemperature_CondO0_o,empty)
    recv_(_comMultiProcessor_TTTTemperature_CondO0_o,ATmega128,node3,node0)
    Pre0_(_comMultiProcessor_TTTTemperature_CondO0_o_node0_c_full,,_comMultiProcessor_TTTTemperature_CondO0_o,full)
  endloop_
endthread_

thread_(COMB,e,node0,pc)
  Pre0_(_comMultiProcessor_Repeater_out_node0_e_empty,,_comMultiProcessor_Repeater_out,empty)
  loop_
    Suc1_(_comMultiProcessor_Repeater_out_node0_e_full,,_comMultiProcessor_Repeater_out,full)
    send_(_comMultiProcessor_Repeater_out,ATmega128,node0,pc)
    Pre0_(_comMultiProcessor_Repeater_out_node0_e_empty,,_comMultiProcessor_Repeater_out,empty)
  endloop_
endthread_

main_
  proc_init_
    spawn_thread_(c)
  spawn_thread_(e)
  eBoardDrivers_eRepeater(_comMultiProcessor_TTTTemperature_CondO0_o,_comMultiProcessor_Repeater_out)
  Pre1_(_comMultiProcessor_TTTTemperature_CondO0_o_node0_c_empty,c,_comMultiProcessor_TTTTemperature_CondO0_o,empty)
  loop_
    Suc0_(_comMultiProcessor_TTTTemperature_CondO0_o_node0_c_full,c,_comMultiProcessor_TTTTemperature_CondO0_o,full)
    Suc0_(_comMultiProcessor_Repeater_out_node0_e_empty,e,_comMultiProcessor_Repeater_out,empty)
    eBoardDrivers_eRepeater(_comMultiProcessor_TTTTemperature_CondO0_o,_comMultiProcessor_Repeater_out)
    Pre1_(_comMultiProcessor_Repeater_out_node0_e_full,e,_comMultiProcessor_Repeater_out,full)
    Pre1_(_comMultiProcessor_TTTTemperature_CondO0_o_node0_c_empty,c,_comMultiProcessor_TTTTemperature_CondO0_o,empty)
  endloop_
  eBoardDrivers_eRepeater(_comMultiProcessor_TTTTemperature_CondO0_o,_comMultiProcessor_Repeater_out)
  wait_endthread_(Semaphore_Thread_c)
  wait_endthread_(Semaphore_Thread_e)
  proc_end_
endmain_

\end{lstlisting}

With the m4Builder the communication threads are transformed to communication calls (see listing~\ref{lst:exampleMulti:ComAfterCleaned} for the transformed node0 macro code) while the blocking behavior (e.g. block until data was received or sent) lies in the responsibility of the communication protocol design. The blocking behavior is necessary to maintain the synchronization within the whole algorithm. Without it, data packets would most probably be missed and the display of the scheduling would differ more from the real-world execution. As seen, every communication medium thread becomes a function with a switch block providing the choice of operation to be done. Semaphores were removed and in the main loop replaced by appropriate communication functions calls. Note that this transformation \textbf{changes the scheduling table} and is only allowed if there are no subsequent operations following on the processors during the communication duration.

\begin{lstlisting}[frame=single,label=lst:exampleMulti:ComAfterCleaned,caption=SynDEx communication functions]
%auto-ignore
#define   _comMultiProcessor_Repeater_out_node0_e_full  0
#define   _comMultiProcessor_TTTTemperature_CondO0_o_node0_c_full  1

alloc_(uint16,_comMultiProcessor_TTTTemperature_CondO0_o,1)
alloc_(uint16,_comMultiProcessor_Repeater_out,1)

void comCOMA(uint8_t comActions){
	switch(comActions){
	case _comMultiProcessor_TTTTemperature_CondO0_o_node0_c_full:
		recv_(_comMultiProcessor_TTTTemperature_CondO0_o,ATmega128,node3,node0)
	break;
	default:
		/* never reached */
	break;
	} /* end Switch */
} /* end comNAME */

void comCOMB(uint8_t comActions){
	switch(comActions){
	case _comMultiProcessor_Repeater_out_node0_e_full:
		send_(_comMultiProcessor_Repeater_out,ATmega128,node0,pc)
	break;
	default:
		/* never reached */
	break;
	} /* end Switch */
} /* end comNAME */

main_
  proc_init_
  eBoardDrivers_eRepeater(_comMultiProcessor_TTTTemperature_CondO0_o,_comMultiProcessor_Repeater_out)
  loop_
    comCOMA(_comMultiProcessor_TTTTemperature_CondO0_o_node0_c_full)
    eBoardDrivers_eRepeater(_comMultiProcessor_TTTTemperature_CondO0_o,_comMultiProcessor_Repeater_out)
    comCOMB(_comMultiProcessor_Repeater_out_node0_e_full)
  endloop_
  eBoardDrivers_eRepeater(_comMultiProcessor_TTTTemperature_CondO0_o,_comMultiProcessor_Repeater_out)
  proc_end_
endmain_

\end{lstlisting}

\paragraph{SynDEx - Communication Protocol Design}$~~$\\

The nature of the macro code allows the realization and the replacement of communication protocols. A protocol which does not violate the visualized scheduling model too much needs certain characteristics. Every send or receive operation must be blocking to maintain the synchronized integrity of the whole algorithm. If it was not blocking, communication would occur randomly and the data integrity would not be given. For example, if node1 is busy with executing the display operations and node 3 sends data to node1 at the same moment, the data would be lost. Another model violation would be a phase shift of the single scheduling between the nodes - that might even be acceptable for some applications.
How the communication protocol works internally, like using handshakes or checksums is free of choice. One thing the protocol does need is an acknowledge mechanism telling the sender that the data was successfully received and it can proceed with its next operation.\\

Each communication transmission must contain an addressing part. Therefore it is possible for the receiving node to verify if the packet is meant to be read from the communication medium. Again, this is necessary because of possible early executions of tasks which could result in two nodes trying to send at the same time which could result in data collisions. Depending if and how bus collisions are handled in this case determines which packet arrives first at the receiver. Imagine two nodes sending temperature values at the same time to the same node and a bus arbitration decides which temperature data will arrive before the other one. If all blocks would have the same real-world behavior as modeled, then addressing would not be necessary leading to a slimmer protocol.

Necessary characteristics of the protocol:
\begin{itemize}
\item Blocking. Send and Receive operations must block to maintain synchronization within the algorithm.
\item Finished transmission notification. Terminate the communication procedure by acknowledging a successfully received transmission - the sender may continue on its schedule.
\item Addressing. Identify receiving and sending nodes, validation of the received data.
\end{itemize}

Next to the mandatory characteristics of the protocol, a good time representation has to be found. Since it has been pointed out that it is hard to maintain a good cognitivity of the model if the execution times of the blocks differ by a magnitude of $10^2$, it might be hard to combine fast executed blocks, which take for example just 50 microcontroller cycles, with a protocol having a baud rate of $57\,600$ bits/sec. An approach to handle this is to reserve more time for every fast block for the price of performance.

	\clearpage
	\section{Results}\label{sec:results}

Several aspects about this conceptually model-driven development method resulted from the examples: How a development process is performed by concrete examples, how much space this particular model-driven approach requires on the ESE-Board target architecture in comparison to hand-written code, statements about the complexity and cognitivity of the macro code structure, and the Scicos/SynDEx models with their differences to real behavior. These results are presented and some statements about modeling with Scicos/SynDEx in combination with a microcontroller-based distributed embedded system are given.

\subsection{Code Size}

The executable code (an ELF file) was analyzed with avr-objdump, avr-size and a memory evaluation library (\Keyword{MemEval})\cite{elm:MemEval} for an approximation of the used dynamic memory: At program start, a bit-pattern is written onto the SRAM between the end of the .bss-section and the stack-pointer. After dynamic memory allocations, the  SRAM area is searched for the largest untouched bit-pattern and its size returned/displayed. The focus of the measurement lies on the overhead produced by Scicos blocks - therefore the heap and stack memory consumptions before the main-loop are of interest. The main-loop consists basically just of Scicos block calls which have a clear structure and are comparable to hand-written code calling components.\\

A pseudo-code clarification how the dynamic memory was estimated is listed in algorithm \ref{alg:results}, while the memory usage is found in table~\ref{fig:results:memory} (note: The listed dynamic memory usage applies not for the dynamic memory consumption of the whole algorithm - it describes only the dynamic memory usage of the introduced Scicos blocks). For comparison, a rough, hand-written PID algorithm which uses modularly designed drivers like in the SynDEx example, could use about $5\,000\: bytes$ Flash and $200\: bytes$ static memory. 

\begin{algorithm}
\caption[Memory Evaluation]{The concept of memory measurement}
\begin{algorithmic}[1]
\STATE Includes, Defines.
\STATE Prototypes, Functions, Globals.
\STATE main()\{
\STATE MemEval $\rightarrow$ Initialize memory.
\STATE Allocate Scicos blocks (includes dynamic memory allocations).
\STATE MemEval $\rightarrow$ Analyze memory.
\STATE MemEval $\rightarrow$ Return largest unused memory area.
\LOOP
\STATE run PID algorithm.
\ENDLOOP
\STATE \}
\end{algorithmic}
\label{alg:results}
\end{algorithm}

\begin{table}[!Bhtp]
\begin{tabular}{| l | c |r|l |}
\hline
Memory type & Memory total $(bytes)$ & \multicolumn{2}{|c|}{Memory used $(bytes)$} \\
\hline
								&			  &\textbf{Scicos/SynDEx} &\textbf{Hand-written} \\
\hline
Flash 							& $131\,072$  & $34\,796$	& $4\,922$ \\
SRAM \footnotesize{(static)}	& $4\,096$	  & $976$		& $226$  \\
SRAM \footnotesize{(dynamic)}	& 			  & $1\,596$ 	& \\
\hline
\end{tabular}
\caption{Memory usage measured in the monoprocessor examples.}
\label{fig:results:memory}
\end{table}
The ELF file was produced by avr-gcc (version 4.2.2) and avr-ld (version 2.17) with the optimization for code size flag (CFLAGS= -Os). The static memory usage is calculated by adding the .data and .bss section sizes (e.g. displayed by an avr-objdump of the ELF file), while the dynamic memory usage is determined by subtracting the unused memory and static memory from the total SRAM capacity.

\begin{eqnarray*}
 Flash &=& .text + .data\\
 &=& 34\,024\: +\: 772\: =\: 34\,796\: bytes.\\
 SRAM(static) &=& .data + .bss\\  
 &=& 772\: +\: 204\: = \: 976\: bytes.\\
 SRAM(dynamic)\: &=& \: SRAM(total)-MemEval(unused\: area)-SRAM(static)\\
 &=& 4\,096-1\,524-976\: = \: 1\,596\: bytes.\\
\end{eqnarray*}

\subsection[Code Structure]{Code Structure - Differences between automatically generated and hand-written code}

In general, hand-written code can have any structure possible. Latter was constructed here in respect to reusability and modular design of the components: Every functionality is designed by a module with appropriate interfaces including at least an init, main, and a end (finalize) function. These modules are used for automatically generated and hand-written applications.\\

\begin{figure}[!htbp]
\begin{center}
	\includegraphics[width=0.90\textwidth]{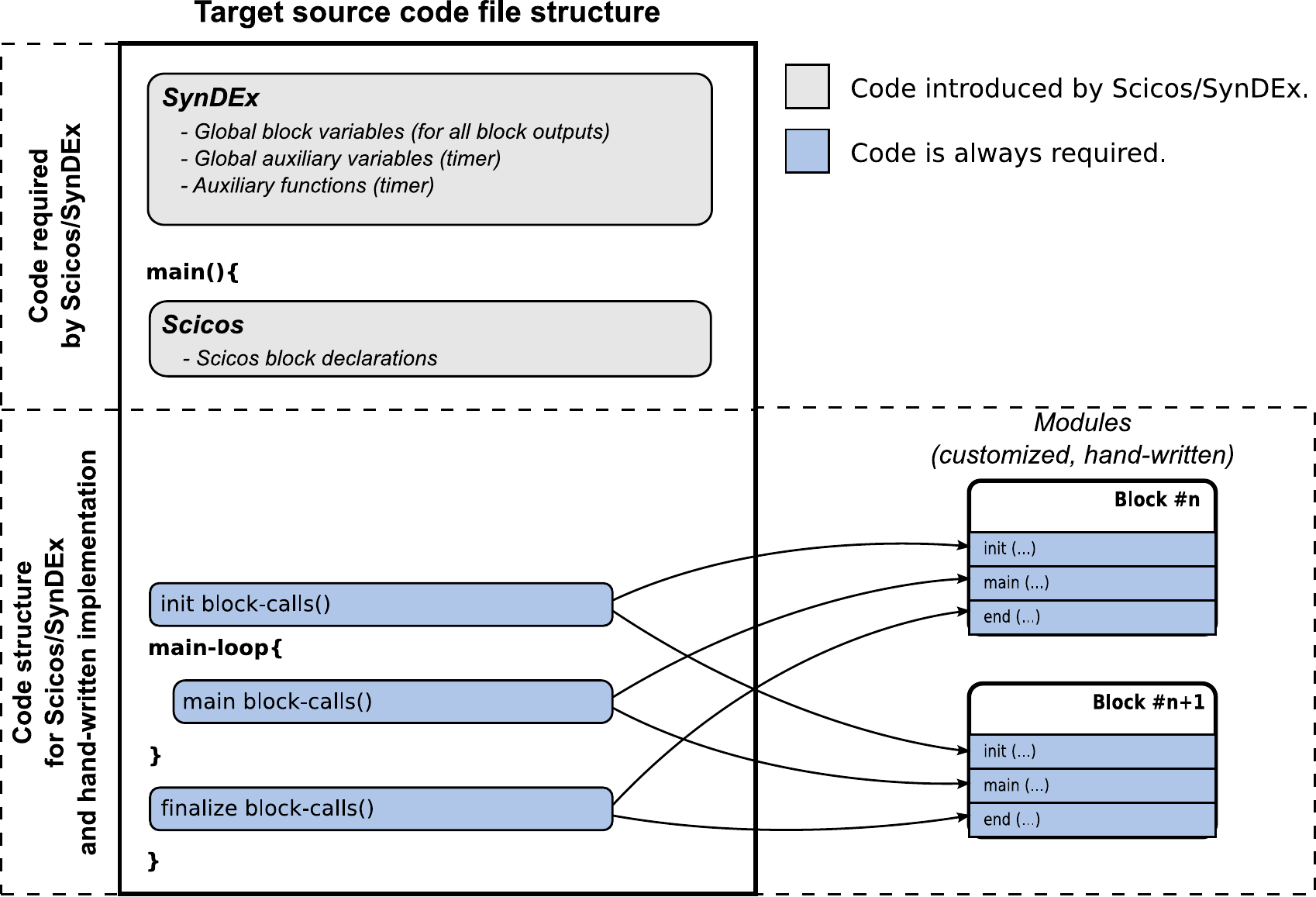}
   \caption{Scicos/SynDEx vs. hand-written code structure.}
   \label{fig:results:codeStructure}
\end{center}
\end{figure}

Three different situations for implementing applications are compared: Scicos/SynDEx, only SynDEx, and a pure hand-written implementation. A design with Scicos/SynDEx results in a scicos\_block structure generated for every designed block (see listing~\ref{lst:scicosblock}). This is necessary to maintain the simulation capability in Scicos. Furthermore SynDEx generates a global variable for each output of the blocks in SynDEx and additionally, customized helper code is needed for designing period-true tasks (e.g. auxiliary code for blocking, timed tasks). This is the case for designing with Scicos/SynDEx and SynDEx only.\\
In the case of hand-written code, no global variables might be necessary in the main module, which they are not in this comparison. In the main-loop of the application, the blocks are first called by their init functions (e.g. resource allocations), then their main functionality is called, followed by a finalization function call (e.g. free resources) - these steps are part of every situation, with/without Scicos/SynDEx or SynDEx (see figure~\ref{fig:results:codeStructure}).

\paragraph{Code complexity and cognitivity}$~~$\\
Scicos and SynDEx do not add any extra complexity compared to hand-written code, but some complexity is caused by needed auxiliary functions. The cyclomatic complexity is the same for the Scicos/SynDEx and SynDEx situation but differs when compared to a hand-written solution by the extra auxiliary functions needed for SynDEx ($cc_{AuxF}$, see table~\ref{fig:results:codeStructureComparison}).
The Scicos PID example model remains human-readable, as well as the SynDEx model. A scheduling table provides a visual representation of all tasks, including their dependencies to each other (successors, predecessors) and this makes it possible to spot unused resources and to optimize the design. Problems occur, if real durations for the tasks are modeled - they are resulting in a scheduling table which is hard to perceive due to optical proportions.

\begin{table}[!Htbp]
\begin{tabular}{| l | c | c | c | c |}
\hline
Development setup & \begin{sideways} Scicos blocks\end{sideways} & \begin{sideways}\parbox{4cm}{SynDEx auxiliary\\ functions/globals}\end{sideways}	& \begin{sideways}\parbox{4cm}{Cyclomatic\\ complexity}\end{sideways}	& \begin{sideways} \parbox{4cm}{Include customized\\ modules (drivers)}\end{sideways}\\
\hline
Scicos/SynDEx	& $\surd$ 	& $\surd$ 			& $cc_{SynDEx}$	& $\surd$ \\
SynDEx 			& $\neg$	& $\surd$			& $cc_{SynDEx}=cc_{HW}+cc_{AuxF}$ & $\surd$ \\
Hand-written	& $\neg$	& $\neg$			& $cc_{HW}$		& $\surd$ \\
\hline
\end{tabular}
\caption{Code structure comparison.}
\label{fig:results:codeStructureComparison}
\end{table}

\subsection{Effort}

The effort for designing a PID algorithm with Scicos/SynDEx is broken down into several key actions (see table~\ref{fig:results:effortTable}). The person hours ($ph$) listed are estimates made during development and are based on following presumptions: The developer is familiar with the working environment, and has expert knowledge about the problem domain. Additionally, the complete tool-chain is prepared and in working condition.\\

\begin{table}[!Htbp]
\begin{tabular}{| c | l | c | c | c |}
\hline
$\#$	& Development tasks									& \begin{sideways}Model-driven\end{sideways}	& \begin{sideways}Hand-written\end{sideways}	& \begin{sideways}Effort $(ph)$\end{sideways}\\
\hline
$1$	& Design Scicos PID model and simulation					& $\surd$		& $\neg$ 		& $~3$ \\
$2$	& Design SynDEx model and temporal design					& $\surd$ 		& $\neg$		& $~3$ \\
$3$	& Implement driver modules (target language code)			& $\surd$ 		& $\surd$		& $~30$ \\
$4$	& Implement Scicos libraries (simulation blocks)			& $\surd$ 		& $\neg$ 		& $~5$ \\
$5$	& Implement SynDEx libraries (code generation)				& $\surd$ 		& $\neg$		& $~10$ \\
$6$	& Implement SynDEx libraries (algorithm \& architecture)	& $\surd$ 		& $\neg$ 		& $~5$ \\
$7$	& Implement PID algorithm by hand							& $\neg$		& $\surd$		& $~2$ \\
\hline

\end{tabular}
\caption{Development effort with an already working tool-chain: model-driven vs. hand-written approach.}
\label{fig:results:effortTable}
\bigskip
\textbf{Clarifications:}
\begin{enumerate}
\item Design and simulation of the hybrid system in Scicos until the results were satisfying.
\item Integration of the PID model in the SynDEx top-model, application of timer blocks for periodic tasks, determination of a proper time resolution, definition of constraints, obtaining and validating the scheduling.
\item These drivers are the same used for the hand-written and model-driven solution. The effort contains testing and modular design of timers, fan speed and sensor drivers.
\item Special blocks modeling real-world items, like the environment fan (envFan) and the environment fan sensor (evnFanSensor), have to be implemented.
\item Implementation of macro expansion files to generate C code (the target language definition): Timers, drivers, auxiliary blocks/functions, type mappings. In this case this is quite a fast to accomplish task since SynDEx is delivered with a C operator target language definition which was adapted for the ATmega128 microcontroller (the resulting target language was not optimized nor complete, and only implemented as needed for the examples).
\item Interface design (formal) for every additional SynDEx block which could not be automatically integrated from Scicos: Timers, type conversions, etc. Design the architecture model library.
\item A solution without Scicos/SynDEx, modular designed drivers are used, the algorithm is written by hand in a source file without intense testing.
\end{enumerate}
\end{table}

Before the Scicos/SynDEx framework can be used in combination with a customized hardware (ESE-Board), several scripts, programs and adaptions of the framework have to be done. The activities in the hereinafter given list require a great manifold of the time required implementing the examples in table~\ref{fig:results:effortTable}. This list does not claim to be complete, several minor workarounds are not included and only issues encountered during the examples are taken into consideration. Note that stating the effort for these steps is difficult for several reasons: a comprehensive documentation for the framework was only partially found, new documentation was published during this work, the requirements for changing the structure of the macro code is dependent on the targets, and many more. \clearpage
\begin{enumerate}
\item \textbf{Installation of the framework.} Install the software framework on a development PC (this point is mentioned because the installation of the framework turned out to be sophisticated where critical information for a working framework was only given in a French description - the appendix of this work includes additional information about that).
\item \textbf{Become familiar with Scicos/SynDEx.} Understand how it works (from the GUI to the underlying scripts), what it can and can not do, search for documentation.
\item \textbf{Implement a m4x builder.} The macro code generated by SynDEx might be not directly useful for the given target hardware, e.g. threads for communication, semaphores and download labels. The m4x builder changes and cleans the structure of the macro code (depicted in figure~\ref{fig:exampleMulti:m4Builder}).
\item \textbf{Implement a m4 builder.} Some strengths of a model-driven approach can only apply if components can be reused in several designs. Combining several Scicos-to-SynDEx transformed designs and integrating them into SynDEx is not possible. A m4 builder for combining several models is most probably be necessary (originally the PID example of this thesis consisted of several Scicos-to-SynDEx transformed components and a m4 builder was provisionally implemented).
\item \textbf{Adapt SynDEx templates.} Some of the generic templates of SynDEx might not be appropriate - therefore they need to be arranged even if it is not recommended by the SynDEx authors (e.g. the Makefile templates).
\end{enumerate}

\subsection{Model versus Reality}

\begin{enumerate}
\item \textbf{Scicos PID model and real fan behavior.} The fan model was designed in Scicos with a PT1 behavior which turned out to be similar (according to the shape of the function) to the real fan behavior after a manual adaption of the PID parameters (compare figures \ref{fig:scicosSimulationHybridFanBeauty} and \ref{fig:results:fan0To80}). One reason contributing to this problem is for sure the large time base chosen ($80\:STU$ instead of $74\:STU$ - even the $74\:STU$ are already a rounded value). Creating a fitting model could turn out to be very sophisticated - extra effort and expert knowledge would be required. Another reason are the data types chosen - differences occur due to the quantification of the plant behavior (the model of the plant behavior is an estimated interpolation of the measured real plant behavior). The fan driver was implemented based on software pulse-width-modulation. A timer is regularly causing interrupts based on the set duty-cycle - no possibilities of modeling such a circumstance were found.

\item \textbf{Scheduling and time representation.} There were two choices for a representation of time in the models.

First, a time base represented by operator cycles brings the advantage of a more precise model, but introduces following disadvantages: It requires more effort to model, the cognitivity of the scheduling table in SynDEx is reduced. Therefore, to scale the model in time, a virtual unit ($STU$ - SynDEx Time Unit) representing $n$ cycles had to be introduced - the price for this is the reservation of additional time-intervals which are not consumed, and the extra differences between model and real-world behavior.
If $1\:STU=1\:cycle$ then the modeled WCET times of blocks would only differ by the clock drift of the real operator.

Second, a natural time base, e.g. microseconds, could be chosen as time base. In this case blocks with a WCET smaller than the time base will be modeled as being too time consuming, which leads to differences between model and reality.
For both cases, a maximal block execution time can be modeled, but an operator might not be able to realize what was required by the application. For example, imagine a task should be executed every $4\:ms$, the time is translated to discrete SynDEx time units, and the algorithm is executed on the target platform. What was modeled might not be feasible on a discrete-time microcontroller target that calculates time based on a clock source and prescalers.

\item \textbf{Periodic tasks.} Durations and periods of tasks can be designed in SynDEx. Just entering the corresponding parameters in an algorithm block results in a task that will be called at least with this period - periods modeled are worst-case. However on the average case, due to the modeled WCET durations of prior called algorithms which terminate earlier than it can be seen in the scheduling, the period of the task will be smaller. This might be desired if a task has to be executed at a given period or better. If true-period tasks are desired, the proposed work-around for timed-execution of tasks is required (details are found in the monoprocessor example). Latter modeling method also introduces differences between the visual presentation of the model and the real-world execution where the timed task has blocking behavior to compensate for the early execution of prior tasks.
\end{enumerate}

\subsection[Systems Design with Scicos/SynDEx]{Scicos/SynDEx and a distributed embedded system with microcontrollers}
During the design of the examples some key differences about what can be modeled with Scicos/SynDEx and what would be needed for designing distributed embedded systems emerged. Of course, there are probably many more requirements - discoveries by the examples are mentioned.

\begin{table}[!htbp]
\footnotesize
\begin{tabular}{|l |c|c|}
\hline
Design requirement				& \parbox{3cm}{Directly supported\\by Scicos/SynDEx}			&	Proposed solution \\
\hline
&&\\
Type mappings					& model re-design required		&   \parbox{6cm}{add rules to target language definition (see PID example, \ref{sec:exampleMono}).}\\
&&\\
Best effort period tasks		& $\surd$						&	- \\
&&\\
True-period tasks				& $\neg$						&	\parbox{6cm}{proposed algorithm design with blocking behavior (section \ref{sec:exampleMono}).}\\
&&\\
Communication (macro code)		& communication threads			&	\parbox{6cm}{transform via m4x-Builder (section \ref{sec:exampleMulti}).}\\
&&\\
\parbox{4.5cm}{Component reusability\\(SynDEx)}	& simple blocks only			&	\parbox{6cm}{hierarchical blocks have to be hard-coded.}\\
								& (no parameters)	&	- \\
&&\\
\parbox{4.5cm}{Component reusability\\(combine transformed\\Scicos models)}	& $\neg$ & \parbox{6cm}{combination by a m4-Builder.} \\
&&\\
Fault tolerance (redundancy)	& samples by broadcast	&	\parbox{6cm}{com-medium modeled as broadcast.}  \\
&&\\
Communication protocol			& theoretically replaceable		&	\parbox{6cm}{blocking, synchronization and addressing (see section \ref{sec:exampleMulti}).}\\
\hline
\end{tabular}
\caption{Scicos/SynDEx on a microcontroller based distributed embedded system.}
\label{fig:results:SSandMC}
\end{table}

Some additional comments clarify the contents of table~\ref{fig:results:SSandMC}:

\textbf{Type mappings.} Blocks are designed with particular input/output types. If they had to be changed afterwards, e.g. include another math-library with different data types, the model would have to be redesigned.

\textbf{Component reusability.} In SynDEx, only simple blocks can be designed and reused in other models. Parameters, e.g. periods are not saved. The nesting of blocks is not supported directly, therefore hierarchical components can only be hard-coded and then be reused in other models/designs. Scicos-to-SynDEx transformed models need to be treated if two or more are required in a SynDEx model. An solution is a customized M4 builder which combines these files through previously inserted section tags.

\textbf{Fault tolerance.} A setup: three different temperature sensors take a synchronized snapshot by a broadcast communication medium. The data is then sent to a microcontroller containing a voter unit.
	\cleardoublepage

	\chapter{Conclusion} \label{sec:conclusion}

In this work, theoretical foundations for modeling embedded systems were presented and a modeling-framework (Scicos/SynDEx) was examined and evaluated. The fitness of this particular model-driven approach for a distributed embedded system with microcontroller targets depends on individual needs and requirements. Required design characteristics, priorities, tool support and many more aspects in systems development are depending very much on the particular problem domains. Thus, it is hardly possible to create a fitness reference covering all thinkable required aspects, however an attempt of creating some valuable statements about advantages and disadvantages of using a model-driven approach in this particular case follows.\\

The combined use of Scicos/SynDEx results in a high demand on dynamic memory introduced by the Scicos block structures. These results (table~\ref{fig:results:memory}) cover only the dynamic memory consumption by the Scicos blocks. A comparison of the overall dynamic memory consumption between Scicos/SynDEx and hand-written code is hard to qualify for the general case, since hand-written solutions can be carried out in many variations. A suggestion for an approximate comparison is to assume that the hand-written solution uses the same libraries (drivers) like the model-driven solution - therefore this part of the used memory should be the same for both solutions. A malicious fact playing against a good comparison of the code size is the nature of compilers - the optimization abilities and differences of several compilers might diffuse results. Thus, the results can only be considered with caution. In the case of developing with Scicos/SynDEx, the static memory consumption of the model-driven approach is significantly higher than that of the hand-written solution, while in the case of using only SynDEx, the memory consumption can be considered as equal to the hand-written solution.\\ 

Next to the code size, the cognitivity of the designed models is of importance - the better the representation of the model the earlier aspects like bad designs or unused resources can be spotted. The code generated by Scicos/SynDEx is not downgrading the readability of the code structure in comparison to hand-written code. Additional structures are inserted by the framework, they are consuming memory, but they do not include obfuscating characteristics, e.g. deep nestings of functions or confusing variable/function names.\\

The development effort of using the Scicos/SynDEx framework is signifficantly higher than that one caused by a hand-written solution, but this might be only valid for the simple one-time applications (which are in fact not that simple, as seen in the examples) made in this thesis. The installation and adaption work necessary for a working, automated framework must not be underestimated - especially the additional scripts and tools for model transformations have to be carefully designed and in the case of designing a critical embedded system these tasks should only be accomplished by experts. Next to establishing a working framework, the design process with a ready-to-use framework requires much more time than the hand-written approach (table \ref{fig:results:effortTable}): $56\:person\:hours$ versus $32\:person\:hours$ (handwritten). Latter results are, of course, not quantitative statements, since these numbers represent a unique case and depend on the skill and preferences of the developer. Advantages of the model-driven approach might show up for designing complicated systems, product families, and when the price for setting up the framework could be neglected and components reused. The communication macro code structure generated by SynDEx in the case of a multiprocessor application opens an unanswered, fundamental question: Why are there communication threads created, if the characteristics of the designed models are atomic and the scheduling is non-preemptive? Threads imply the presence of a scheduler, context switches and additional resource consumption. Even a decoupling of the communication unit from the processor requires coordination resources. These aspects seem to be missing in the models.\\

Building models of real-world entities with a discrete machine causes always differences in the behavior, diminishing these differences results usually in more complex models. Creating an adequate model of the fan behavior did not turn out to be an easy task - the chosen design was not powerful enough to ensure an adequate execution of the simulated models on the target hardware - manual adaptions had to be made for a satisfying PID controller. Since the focus of this thesis was not about model construction, the fan and plant model were kept simple and were redesigned just one time. Expert knowledge and re-modeling of the fan is required to sharpen models. The design process would be improved if manufacturers/vendors included Scicos models within the delivery of the hardware - such models could be integrated in Scicos by the developer - instead of worrying about a customized design and its adequacy (this would also encourage the use of the free Scicos software). The differences between models and real applications seem to be mainly depending on: the chosen representation of time (only a representation of $n\cdot cycles$ was possible, due to representation problems of the scheduling), the real execution time of tasks (they were modeled with WCET), and the introduction of true-period tasks. The modeling of true-period tasks is only possible with work-arounds - adding differences between model and real application.\\

\begin{figure}[!htbp]
\begin{center}
	\includegraphics[width=0.90\textwidth]{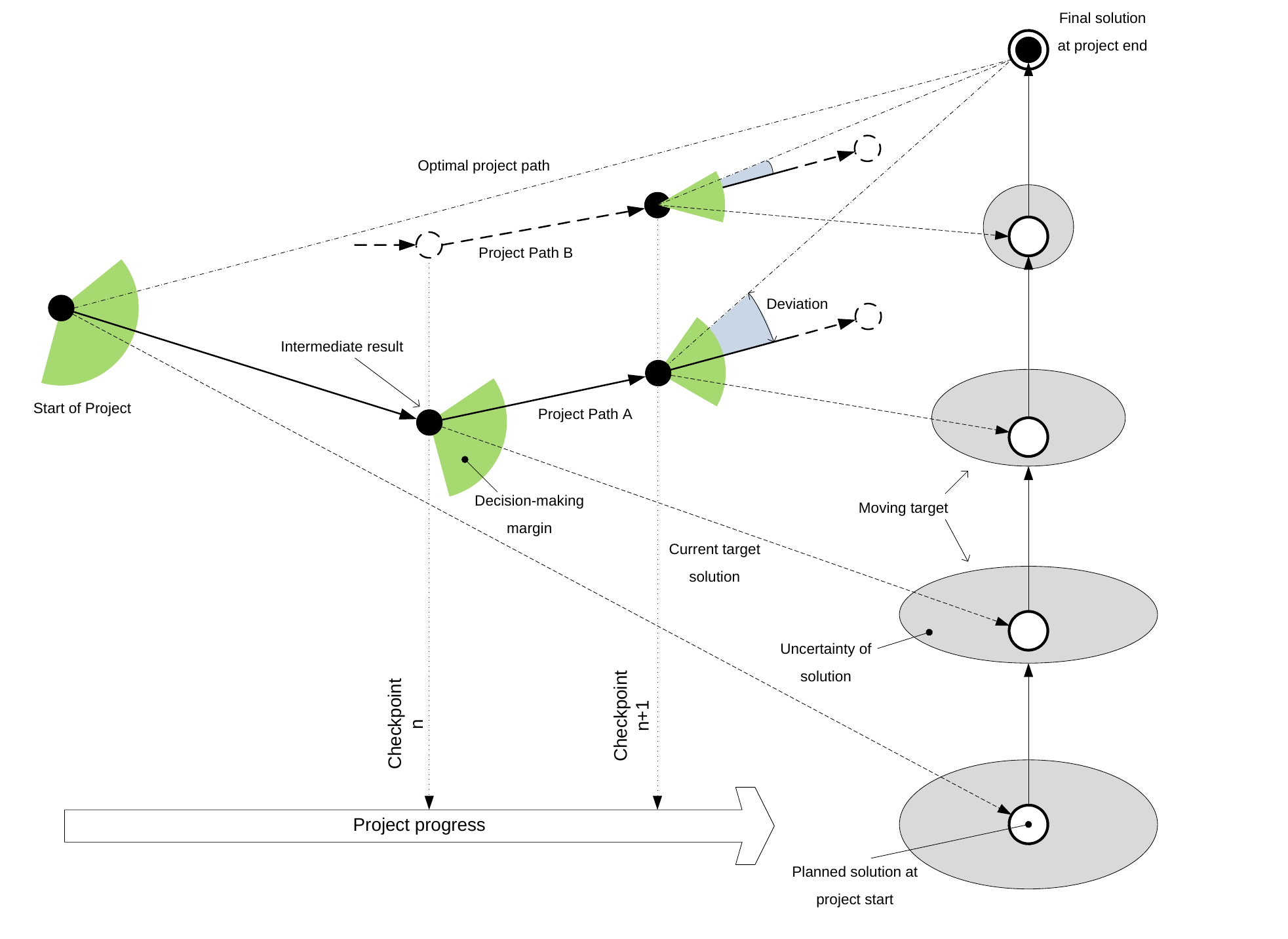}
   \caption[Scicos/SynDEx vs. hand-written solution]{Scicos/SynDEx vs. hand-written solution. Simulation and modeling capabilities (Project Path B) decrease the size of the detour in comparison to hand-written projects (Project Path A). The uncertainty about the solution is reduced earlier, the deviation from the optimal project path diminished.}
   \label{fig:conclusion:projectPaths}
\end{center}
\end{figure}

A major advantage of the model-driven approach (and the idea of Scicos/SynDEx) is the safety of the design concerning deadlocks and resource allocations/optimizations. Such tasks can be error-prone and very complicated if done manually. Another benefit of the framework is that free resources can be easily spotted by the visual representation of the models - this could lead to several benefits as, for example, reduced hardware costs, space for additional features, or just reserved resources for an ongoing evolution of the software. The possibility of model simulation has shown to be of great value during the design of the first PID algorithm (velocity PID). The wrong design was detected early and a redesign started - in the case of a hand-written solution, the parameters of the wrong design could in some cases be refined until a solution for this single version of the application is doable - this could be disastrous if the software requires minor changes, for example if the fan was replaced by another model. This would be followed by confusion and uncertainty about the previously designed solution. Simulation capabilities narrow the path of possible project advancements towards expected solutions earlier than a traditional hand-written approach (figure~\ref{fig:conclusion:projectPaths}). This might lead to reduced costs, development time, and risks.\\

Working with Scicos/SynDEx shows some characteristics of agile development. Models and simulation results are represented visually and system changes, e.g. system enhancements or product evolution, can often be realized faster than with a handwritten solution (which might require additional tests and design checks). Customer collaboration is supported by the models contributing to, or being part of, the specification (this is the idea, but an accurate model is not easy to accomplish, as the examples have shown) - this fact might reduce the need for exhaustive project documentation and reduced contract negotiations.\\

\begin{table}[!Htbp]
\begin{tabular}{| l | c | c |}
\hline
Comparison 			& Model-driven solution				& Hand-written solution \\
\hline
Code size			&	$34\,796\: bytes$				& $4\,922\: bytes$ \\

Memory consumption	& 									& \\
(static)			&	$976\: bytes$					& $226\: bytes$ \\
(dynamic - Scicos)	&	$\approx \frac{100\:bytes}{1\:block}$	& - \\

Code structure		&	extra variables/functions		& - \\
Code cognitivity	&	human-readable					& human-readable \\

Effort				&									& \\
(PID, development)	&	$56\: person\ hours$					& $32\: person\ hours$ \\
(establish tools)	&	high effort						& - \\ 

Simulation			&	early detection of mistakes		& $\neg$ \\
Optimization		&	automatically, safe				& by hand \\ 
Resources (display)	&	visual representation			& source code only \\
Respond to changes	&	lower effort, quick				& higher effort \\
Documentation		&	model as specification			& formless \\
\hline
\end{tabular}
\caption{Fitness of the model-driven approach compared to a hand-written solution.}
\label{fig:conclusion:fitness}
\end{table}

In which scenarios should Scicos/SynDEx be preferred to a hand-written solution? Based on the comparisons made (table \ref{fig:conclusion:fitness}) it should be valid to say that the Scicos/SynDEx framework can unfold its strengths in product families when a basic platform is established and the effort for modeling the components pays off (components are reused). SynDEx, as software, is supposedly meant for rapid prototyping. In the case of the combined use with Scicos on the ESE-Board, this can only be true, if the effort for setting up the tool and the underlying automated build process is neglected. Designing the libraries might still take more time for simple applications than the hand-written approach, but has the advantage of reuse and supports an easy redesign of the application. For instance, if the hardware fan, like the one in the PID example, should be replaced by another model it should be quite comfortable to build the model and then just compare these two models by the simulation results (with the assumption of accurate models). Additionally, models and simulation results might be directly discussed with the customer. Scicos/SynDEx should only be used if there is no shortage of memory capacity on the target platform. SynDEx itself does basically not introduce extra memory costs, however SynDEx does not support hybrid system simulation and modeling of hybrid systems.\\

Designing systems with the framework has shown that it is not only crucial to determine WCETs of tasks, of even the same importance are models which consider the minimal and possible durations of tasks. As seen in this thesis, an early termination of a task leads to gaps in the scheduling model, which might throw over the real behavior of the system. An approach to handle this would be to time-trigger all critical tasks, this would require either a lot of timers or a customized scheduler which can not be modeled with the framework. It might be, that this is not the idea of a synchronized distributed executive, it seems that best-effort applications are lying in the focus of the framework. Modeling time characteristics of a critical, distributed system might be better done with a time-triggered architecture where the problems of early task terminations are diminished (confer to \cite{paukovits:2004}).

The idea of a framework for model-driven development of distributed embedded systems with the support of hybrid system design and resource allocation/optimization sounds promising. Scicos/SynDEx introduces extra memory costs but can pay off in complex systems design. SynDEx itself handles the temporal design and algorithm allocations automatically for distributed systems while basically not requiring extra memory space nor increasing the complexity of the model. The framework is not of a commercial nature, which might be the reason for its immaturity concerning the usability (see the appendix for hints) - the implementation focus lies clearly on applying the underlying concepts of the AAA-Methodology. Implementing a complete model-driven development environment from scratch is an interdisciplinary challenge. Expert knowledge in several fields like model building, formal methods, scheduling, code generation, compiler design, DSL design, dependable system design, communication protocol design, hardware knowledge, software architectures in general, and many more are required. The effort for creating a model-driven development environment can therefore be labeled very demanding. Adapting the Scicos/SynDEx framework requires also a relatively high effort. Thus, it might be reasonable to consider the use of commercial software with already provided hardware/software libraries. The support for the Scicos/SynDEx framework seems to have stopped since no updates for the Scicos-to-SynDEx-Gateway working with the latest Scicos distributions could be found. SynDEx is still under development and with its evolution, based on the fundamental AAA approach for embedded system design, it might turn out to be a very valuable asset in the future.

\section{Outlook}\label{sec:conclusion:outlook}
In this thesis we have seen that the real behavior of systems deviates in different shapes from the modeled systems. First, time and data were abstracted which lead to deviations. Second, tasks were modeled with WCET, but in reality the real task durations can not be predicted. It might be possible to classify the errors caused by discretization and use this information for a better predictability in systems modeling, that means drawing a connection from hybrid systems timings to distributed systems scheduling. Furthermore it would be of interest if there are advantages if several task duration parameters, such as worst-case execution time and best-case execution time, are considered in a scheduling model. In that way the running-away of tasks, as it is the case in single-processor applications in SynDEx, might be prevented. \\

Building a seamless modeling tool-chain with the Scicos/SynDEx framework and the implementation of some examples have shown some potential improvements for the concepts realized in the SynDEx software:
\begin{itemize}
\item The scheduling for communication calls could be integrated into the operator scheduling and thus serving the generation of code for microcontrollers (only one sequencer for the operator scheduling).
\item Support the realization of fixed-period tasks rather than only best-effort tasks.
\item Include code-size indications and constraints which allow a prediction of the code-size for a given hardware target.
\item Support of automatic type-mappings within the graphical tool.
\item Automatic implementation of a small static-scheduler that provides a a way of combining long and short period tasks effectively.
\end{itemize}

Regarding component-based development the effort for design could be reduced by a great deal if manufacturers would supply standardized models for simulation and implementation for their products. This might already be true for some suppliers in combination with certain tool-chains, but still, a commonly accepted format for models is required.
	\cleardoublepage

	\fancyBib
	\newpage
	\phantomsection

	\newpage
	\fancyAcro
	\phantomsection
    \addcontentsline{toc}{chapter}{Acronyms}
    \chapter*{Acronyms}
\begin{acronym}
	\acro{ADC}{Analog Digital Converter\acroextra{. An \acs{ADC} is an \acs{IC} that quantifies analog to digital values.}}
	\acro{AIL}{Architecture Implementation Language\acroextra{. The \acs{AIL} is a description language that allows for an internal representation of the architecture and acts as a connection with tools to simplify the construction, planning, verification, capitalization, and documentation of an architecture.}}
	\acro{AOSTE} {is one of INRIA's research teams focusing on high-level modeling, transformation and analysis and implementation onto embedded platforms}
	\acro{ASIC} {Application-specific integrated circuit}
	\acro{BDD}	{Binary Decision Diagram}
	\acro{CAD}	{Computer Aided Design}
	\acro{CASE}	{Computer Aided Software Engineering}
	\acro{CMA} 	{Centre de Mathematiques Appliquees (Sophia Antipolis, France)}
	\acro{COTS} {Components off the shelf}
	\acro{DC}	{Direct Current}
	\acro{DMA}	{Direct Memory Access}
	\acro{DSL}	{Domain-Specific Language\acroextra{. A \acs{DSL} is a language well suited to describe a specific domain.}}
	\acro{ECU}	{Electronic Control Unit}
	\acro{ES}	{Embedded System}
	\acro{FPGA} {Field Programmable Gate Array}
	\acro{FSM}	{Finite State Machine}
	\acro{HDL} 	{Hardware Description Language}
	\acro{IC}	{Integrated Circuit}
	\acro{IT}	{Information Technology}
	\acro{IDE} 	{Integrated Development Environment}
	\acro{IEC}	{International Electrotechnical Commission}
	\acro{IEEE}	{Institute of Electrical and Electronics Engineers, Inc. \acroextra{\acs{IEEE} is a non-profit organization and the world's leading professional association for the advancement of technology.}}
	\acro{INRIA}{Institut National De Recherche En Informatique Et En Automatique\acroextra{. \acs{INRIA} stands for the French National Institute for Research in Computer Science and Control.}}
	\acro{ISA} 	{Instruction set architecture}
	\acro{ISO}	{International Organization for Standardization}
	\acro{LCD}	{Liquid Crystal Display\acroextra{. A \acs{LCD} is a display based on liquid crystal technology.}}
	\acro{LOC}	{Lines of Code}
	\acro{MARTE}{\acs{UML} Profile for Modeling and Analysis of Real-time and Embedded Systems\acroextra{. \acs{MARTE} is the specification of an \acs{UML} profile adding model-driven development capabilities for real-time and embedded systems to \acs{UML}.}}
	\acro{MDA}	{Model Driven Architecture}
	\acro{MDSD}	{Model-driven Software Development\acroextra{. \acs{MDSD} is a technique for software development with formal models used for automated code generation and documentation.}}
	\acro{MEM} {Micro-Electro-Mechanical System}
	\acro{MOF}	{Meta Object Facility\acroextra{. The \acs{MOF} is the meta-meta-model and core of the \acs{MDA} defined by the \acs{OMG}.}}
	\acro{MTTF} {Mean Time To Failure}
	\acro{NATO}	{North Atlantic Treaty Organization\acroextra{. The \acs{NATO} is an alliance of 26 countries from North America and Europe fulfilling the goals of the North Atlantic Treaty signed on 4 April 1949.}}
	\acro{OMG}	{Object Management Group\acroextra{. The \acs{OMG} is an open consortium for improving interoperability and portability of software systems by defining manufacturer - neutral standards.}}
	\acro{PWM}	{Pulse Width Modulation\acroextra{. \acs{PWM} is a signal modification technique based on the adjustment of ratio between high-/low- time of the signal line.}}
	\acro{ROI}	{Return Of Interest}
	\acro{RPM}	{Rounds per Minute\acroextra{. \acs{RPM} stands for the cycles per minute of a rotating device.}}
	\acro{RTL} 	{Real Time Logic}
	\acro{SIL}	{Safety Integrity Level}
	\acro{SPI}	{Serial Peripheral Interface\acroextra{. The \acs{SPI} bus is a full duplex, synchronous serial data link standard.}}
	\acro{SSM} 	{Safe State Machine}
	\acro{TDMA} {Time Division Multiple Access}
	\acro{UML}	{Unified Modeling Language\acroextra{. The \acs{UML} is a modeling language based on the \ac{OMG}'s \acs{MOF}.}}
	\acro{UNIX} {Unix is a computer operating system originally developed in 1969 by a group of AT\&T employees at Bell Labs.}
	\acro{VDE}	{Verband der Elektrotechnik, Elektronik und Informationstechnik e.V., engl.: Association for Electrical, Electronic \& Information Technologies}
	\acro{VHDL} {stands for VHSIC (Very high speed integrated circuit) hardware description language.}
	\acro{VHSIC}{Very High Speed Integrated Circuit}
	\acro{WCET} {Worst-Case Execution Time}
\end{acronym}

	\cleardoublepage

	\cleardoublepage
	\fancyApp
	\appendix

	\chapter{Notes}
	\section{Scicos/SynDEx - Configuration}

The setup for exercising the examples as well as important Scicos/SynDEx installation notes are discussed as follows in this section.

Scicos is a toolbox of the Scilab package and its development lies in the responsibility of the "Scilab Consortium". Initially it was developed and maintained by INRIA \footnote{INRIA - The French National Institute for Research in Computer Science and Control}. It's distributed freely, including the source code.\\
Scilab - Version: 4.1.1\\

SynDEx has been developed by INRIA in the Rocquencourt Research Unit France by the AOSTE team. It's free for non-commercial use.\\
SynDEx - Version: 7.0.0\\

The avr-gcc compiler packet and tools.\\
avr-gcc - Version: 4.2.2-1\\
binutils - Version: 2.17\\
avr-libc - Version: 1:1.4.7-1\\

A debugging/programming tool interfacing the GDB and JTAG.\\
avarice - Version: 2.6\\

Experiment (ESE-Board) Monitor.\\
JTagZeusProg - Version: 0.2\\

\paragraph{Scilab and Scicos-SynDEx Installation Notes}$~~$\\

First of all, it is important to only use the specified versions of the software. Up to the date this thesis was written, the only setup of software versions listed above seemed to allow Scicos working together with SynDEx via the Scicos-SynDEx interface.

\subparagraph{Scicos}$~~$\\
Necessary files:\\
scilab-4.1.1-src.tar.gz - Only the source version of the scilab-4.1.1 package provides the necessary resources for the Scicos-SynDEx gateway.\\

Compilation notes:\\
For the compilation there is, amongst others, a Fortran compiler necessary - here, the "GNU Fortran - 4.2.4" compiler was used. Additionally some "-dev" packages might also be required: libXext-dev, libxmu-dev, libxmuu-dev, linux-libc-dev, xutils-dev.

\subparagraph{SynDEx}$~~$\\
Necessary files:\\
syndex-7.0.1-linux-i586.tar.gz\\

Extract the contents of "syndex-6.8.5-linux.tgz" and start. "GNU m4" has to be installed to execute the GNUmakefile generated by SynDEx.

\subparagraph{Scicos-SynDEx Gateway}$~~$\\
Necessary files:\\
ScicosToSynDEx.tar\\
s2s\_scicosFiles.tar\\

\begin{figure}[!htbp]
\begin{center}
	\includegraphics[width=\textwidth]{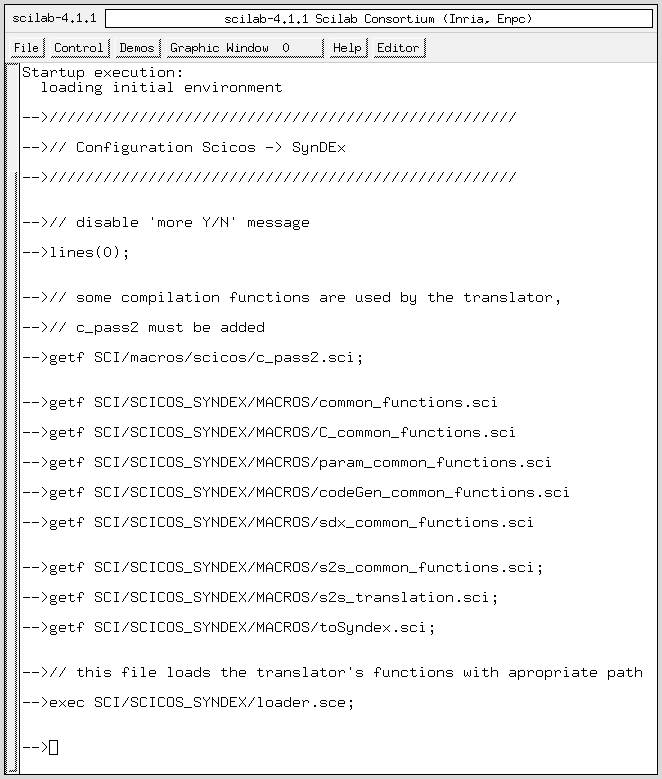}
   \caption{Screenshot of Scilab with the installed SynDEx gateway module.}
   \label{fig:installScicos2SynDEx}
\end{center}
\end{figure}

\begin{figure}[!htbp]
\begin{center}
	\includegraphics[width=\textwidth]{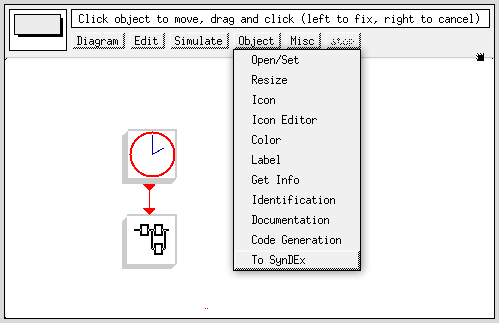}
   \caption{Screenshot of Scicos with the installed SynDEx gateway module.}
   \label{fig:installScicos2SynDExMenuEntry}
\end{center}
\end{figure}

The integration of the Scicos-SynDEx Gateway requires some script file editing and files moved into the right directories:
\begin{itemize}
\item{Create a folder named "SCICOS\_SYNDEX" in the Scilab-install directory.}
\item{Extract "ScicosToSyndex.tar" into the "$<$Scilab-install-dir$>$/SCICOS\_SYNDEX".}
\item{Edit the file "scilab.star" in the $<$Scilab-install-dir$>$: Attach the line "exec SCI/SCICOS\_SYNDEX/dot.scilab" at the end of the file. Scilab will now prompt about the loaded SynDEx module (see figure \ref{fig:installScicos2SynDEx}). With the installation of the Scicos-Syndex Gateway an additional entry inside the graphical interface of Scicos is added: Object$\rightarrow$To SynDEx (see figure \ref{fig:installScicos2SynDExMenuEntry}).}

\item{Extract "s2s\_scicosFiles.tar" into the "$<$SynDEx-install-dir$>$".}
\end{itemize}

	\section{Scicos/SynDEx - Documentation}

\paragraph{Scicos}$~~$\\

\subparagraph{Creating new blocks in Scicos}\label{sec:appendix:scicos:createNewBlocks}$~~$\\

The basic steps of creating blocks for Scicos are documented in this section (see also \cite{Campbell2006}). Each Scicos-Block consists of two parts: an "interfacing function" and a "computational function". The interfacing function handles the interaction with the graphical editor. The computational function defines the behavior of the block during the simulation. Both have to be designed, compiled and "linked" to the Scicos software. Following procedure is suggested:\\

\begin{enumerate}
\item Implement the computational function, e.g. IO\_LEDBar\_8bit.c. In general, the structure of a block's computation function coded in C looks like the following:\\
\begin{verbatim}
  #include "scicos\_block.h"
  #include <math.h>
  void my_block(scicos_block *block, int flag){
  ...
  }
\end{verbatim}
Examples for computational functions can be obtained in the $<$Scilab-installation-directory$>$/routines/scicos (e.g. summation.c). 

\item Implement the interfacing function, e.g. IO\_LEDBar\_8bit.sci.
Examples for interfacing functions can be obtained in the $<$Scilab-installation-directory$>$/macros/scicos\_blocks (e.g. SUMMATION.sci).

\item For the compilation of the files, several Scilab commands have to be executed. The man pages of Scilab should also be checked out\cite{ScilabMan}. To make the job easier, generate a builder file named "builder.sce" with following (example of building the library "ESE" with a 8 bit LED bargraph) content:
\footnotesize
\begin{verbatim}
  comp_fun_lst=['IO_LEDBar_8bit']; // ['foo1','foo2',...]
  c_prog_lst=listfiles('*.c'); // c files in directory
  prog_lst=strsubst(c_prog_lst,'.c','.o');
// generate loader and compile
  ilib_for_link(comp_fun_lst,prog_lst,[],'c','Makelib','loader.sce','ESE');
  genlib('lib_ESE',pwd()); // compile macros and generate lib
\end{verbatim}
\normalsize
Execute this script by starting Scilab, change to the directory where the files are stored (use the command "cd") and execute the script ("exec builder.sce"). Several files are generated: $<$some library files$>$, loader.sce, Makelib (a makefile), lib, and some more.

\item (Optional Step) Create a local scilab script file in the directory of the library files: $<$myScilabScript$>$.scilab. It can be useful for customizing commands for the treatment of the libs and shall at least contain the call of the loader.sce file, e.g.:
\begin{verbatim}
// this file loads the library functions with appropriate path
  exec /home/exa/DA/code/scicos/eBoardEnv/eBoardEnv.scilab;
\end{verbatim}

\item Load the library into Scicos. To make Scicos load the new blocks at start-up the line "exec $<$PATH$>$/$<$MYSCRIPT$>$.scilab" (e.g. "/home/exa/DA/code/scicos/eBoardEnv/eBoardEnv.scilab") shall be appended to the file $<$Scilab-Installation-Directory$>$/scilab.star. If no $<$myScilabScript$>$.scilab is used, just call the loader.sce  directly.
\item The new block is now ready to be used. Start Scicos, in the menu click "edit"$\rightarrow$"Add new block" and enter the name of the block in the popped up dialog field. Now the block can be placed into the Scicos diagram.
\end{enumerate}

\subparagraph{Creating new palettes in Scicos}$~~$\\

Creating a new palette makes the editing process of diagrams more comfortable. To do this, use Scicos to create a new palette.
\begin{enumerate}
\item Create the palette, e.g:
\begin{verbatim}
  create_palette('/home/exa/DA/code/scicos/eBoardEnv') 
\end{verbatim} 
This will create the palette eBoardEnv.cosf in the chosen directory (which must contain all the necessary interfacing functions).
\item Add the new palette to Scicos, e.g:
\footnotesize
\begin{verbatim}
  load('<lib file absolute>');
  scicos_pal($+1,1)='<Name of the Palette>';
  scicos_pal($+1,2)=strcat(<Path to the library files>,'<the palette>.cosf');
\end{verbatim}
\normalsize
For the sake of some working comfort, these lines should be attached to a Scicos script, e.g. to eBoardEnv.scilab:
\begin{verbatim}
  eBoardEnvPath='/home/exa/DA/code/scicos/eBoardEnv';
  eBoardEnvLoader=eBoardEnvPath+'/'+'loader.sce';
  eBoardEnvLib=eBoardEnvPath+'/'+'lib';
  eBoardEnvPal=eBoardEnvPath+'/'+'eBoardEnv.cosf';

// this file loads the library functions with appropriate path
  exec(eBoardEnvLoader);

// load the palette and add it to the Scicos' palette menu
  load(eBoardEnvLib);
  scicos_pal($+1,1)='eBoardEnv';
  scicos_pal($+1,2)=eBoardEnvPal;
\end{verbatim}
\end{enumerate}

\paragraph{Scicos-SynDEx Gateway}$~~$\\

The full documentation, publications and tutorials can be found on the SynDEx homepage \footnote{\url{http://www.syndex.org/scicosSyndexGateway/index.htm}}. For starters, with Scicos and SynDEx, the gateway documentation \cite{Scicos2SynDExGateway} is recommended.\\

\begin{figure}[!htbp]
\begin{center}
	\includegraphics[width=0.75\textwidth]{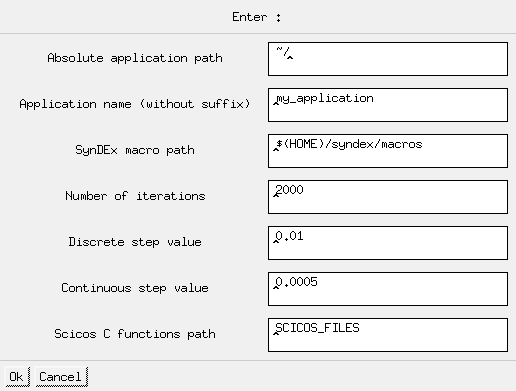}
   \caption[SynDEx parameter-window screenshot]{SynDEx parameter-window screenshot of the Scicos-to-SynDEx gateway.}
   \label{fig:usageScicos2SynDExGateway}
\end{center}
\end{figure}

After the design of an synchronous diagram in Scicos, this diagram can be translated into a SynDEx conform diagram (Note: An asynchronous scicos diagram can't be translated to SynDEx - trying that causes an "error 10000" (something like a clkRoot error)). This is done by making a superblock of the Scicos diagram and then translate it to SynDEx by choosing following entries in the menu: Object$\rightarrow$To SynDEx. This will start the translation process with the pop-up of a parameter window (see figure~\ref{fig:usageScicos2SynDExGateway}).

The parameters are explained as described on the SynDEx homepage:\\
\begin{itemize}
\item{Absolute application path: the path where the SynDEx files will be generated.}
\item{Application name (without suffix): the application name.}
\item{SynDEx macro path: the path where the SynDEx macros are located (usually the path to SynDEx + "/macros").}
\item{Number of iterations: the number of steps the final application has to process.}
\item{Discrete step: the discrete step value (useful when continuous blocks are present).}
\item{Continuous step: the integration step value (useful when continuous blocs are present).}
\item{Scicos C functions path: the path where the Scicos C functions are located.}
\end{itemize}

\begin{figure}[!htbp]
\begin{center}
	\includegraphics[width=0.75\textwidth]{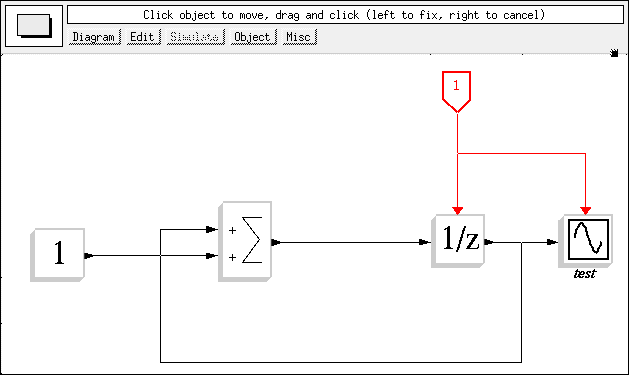}
   \caption{Demo - diagram used for the demonstration of the Scicos2SynDExGateway.}
   \label{fig:testScicos2SynDExGateway}
\end{center}
\end{figure}

\begin{figure}[!htbp]
\begin{center}
	\includegraphics[width=0.75\textwidth]{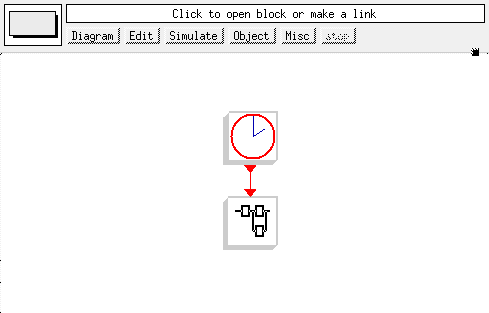}
   \caption{Demo - diagram transformed to a Scicos Superblock.}
   \label{fig:testScicos2SynDExGatewaySuperblock}
\end{center}
\end{figure}

If the Scicos-SynDEx gateway has successfully translated the diagram, the generated files were stored in the $<$Absolute Application path$>$. Following files were created using a simple test-diagram (\ref{fig:testScicos2SynDExGateway}, Scicos Superblock: \ref{fig:testScicos2SynDExGatewaySuperblock}):
\begin{itemize}
\item{demo.sdx - Contains the transformed graph which is readable by SynDEx.}
\item{demo.m4x - Structures and functions for the mechanism of the Scicos-SynDEx macros.}
\item{demo.m4  - Contains architecture properties, for example rootOperator and processor-architecture.}
\item{demo.m4m - Contains the definition of the "rootOperator\_hostname" property.}
\item{RootOperator.m4x - Defines the root operator (for example the microcontroller-application).In a single processor application, like in this demonstration application, the demo.m4m is included inside this file.}
\item{GNUMakefile - ""}
\end{itemize}

\begin{figure}[!htbp]
\begin{center}
	\includegraphics[width=0.75\textwidth]{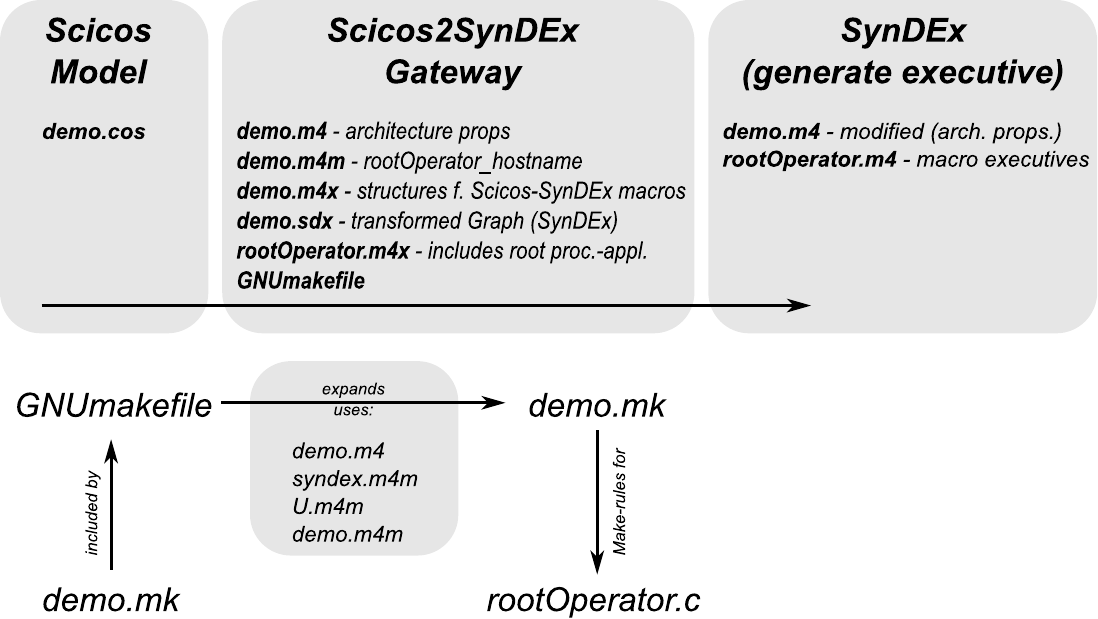}
   \caption{Scicos2SynDEx - Generated artifacts.}
   \label{fig:s2sFileChaos}
\end{center}
\end{figure}

The relationship of the tools and generated artifacts is given in figure~\ref{fig:s2sFileChaos}.

\paragraph{SynDEx}$~~$\\

Notes for the first tries with SynDEx are found in the Scicos-SynDEx gateway documentation\cite{Scicos2SynDExGateway}. This paragraph describes how to use SynDEx to generate code for a mono-processor setup which was already described in french \cite[section 1.4.1]{Scicos2SynDExGateway}, however some additional notes are required for a successful usage.\\

\begin{figure}[!htbp]
\begin{center}
	\includegraphics[width=0.75\textwidth]{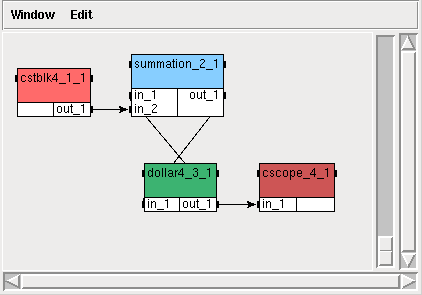}
   \caption{SynDEx - Screenshot, the test diagram.}
   \label{fig:testSynDEx}
\end{center}
\end{figure}

\subparagraph{Section 1.4.4, Pts. 1-3}$~~$\\
When SynDEx is started via command-line, it is mandatory to pass the -libs parameter, otherwise SynDEx will not find even the libraries inside its own distribution: "syndex -libs $<$SynDEx-install-dir$>$/libs". Now it should be possible to open a *.sdx file within SynDEx. For a first start of SynDEx, a test diagram translated by the Scicos-SynDEx Gateway was used (see figure~\ref{fig:testScicos2SynDExGateway}). After starting SynDEx the demonstration file "demo.sdx" was opened using the context menu (see figure~\ref{fig:testSynDEx}).

\begin{figure}[!htbp]
\begin{center}
	\includegraphics[width=0.75\textwidth]{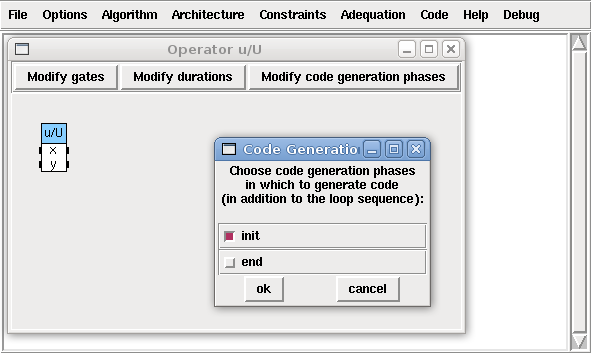}
   \caption{SynDEx - Screenshot, choosing the target architecture.}
   \label{fig:testSynDExOperator}
\end{center}
\end{figure}

Generating code requires, first of all, the target-architecture to be chosen. For test purposes the "U" processor type (within the libs of the distribution) was chosen: "Architecture"$\Rightarrow$"Edit Operator Definition", see figure~\ref{fig:testSynDExOperator}. The automated code generation of SynDEx should generate code for the init and loop phase of the algorithm. The cscope block is not defined in SynDEx, for a successful code generation libraries for this block have to be implemented.

\subparagraph{Section 1.4.4, Pt. 4}$~~$\\
Start the Adequation, for example: "Adequation"$\rightarrow$"No Flatten".\\
Start the code generation by: "Code"$\rightarrow$"Generate Executive". Entering this command will produce an intermediate artifact, not the finished target-architecture code. The file "rootOperator.m4" is created and the file "demo.mk" is changed (in this case a line with "dnl" after the "include..." is inserted, as well as an "endarchitecture\_" at the end of the file; additionally the SynDEx version used for the code generation is inserted).\\
Another note: If the diagram is saved in SynDEx, a demo.sdc file is created - it contains just some miscellaneous version information.

\subparagraph{Section 1.4.4, Pts. 6-7 }$~~$\\
Changing the name of the target machine in demo.m4m is necessary for the "rsh" connecting to the localhost.
The "GNUmakefile" might need some editing, for example to set the right m4 macroprocessor installed on the develop-pc, or to set some path variables. Executing the GNUmakefile will create the demo.mk file, which is included at the end of the GNUmakefile itself. The demo.mk file in this example is used for two things. First, to create (macro-expand) the rootOperator.c file. Second, to compile, link and generate the executive. Note: Anything related to the "rsh" command inside the makefiles can be discarded, since their only purpose is to automate the compilation of the sources on a different target machine.\\
Files created by the GNUmakefile (+ demo.mk) are demo.mk, rootOperator.c, rootOperator.rootOperator.o, rootOperator.\\

\paragraph{Scicos-SynDEx Gateway developer notes on the PID Controller example}$~~$\\

\begin{figure}[!htbp]
\begin{center}
	\includegraphics[width=0.75\textwidth]{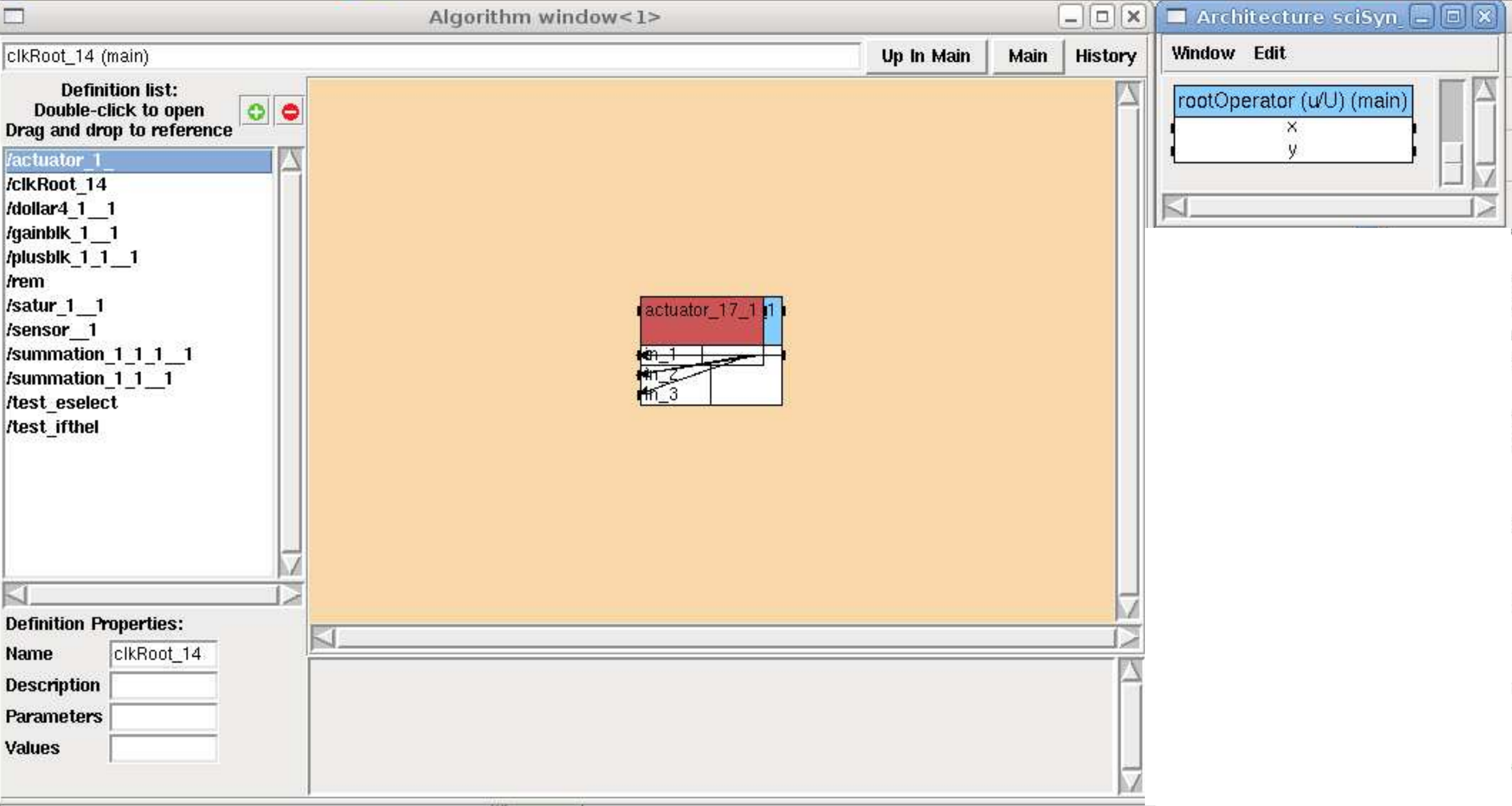}
   \caption{SynDEx - Opening of an example application.}
   \label{fig:syndexDemo3}
\end{center}
\end{figure}

Consider the PID superblock from the mono-processor example (figure~\ref{fig:hybridSystemBeauty} in section \ref{sec:exampleMono}). This block is now converted using the Scicos-to-SynDEx gateway resulting in the following files: FSC.m4, FSC.m4m, FSC.m4x, FSC.sdx, GNUmakefile, rootOperator.m4x. The FSC.sdx file can now be opened in SynDEx (see figure~\ref{fig:syndexDemo3}).

After rearranging the blocks in a human-readable way, following steps will make the algorithm ready to be used in another SynDEx diagram. Only the FSC.sdx file is needed by SynDEx, and the macro-expansion file FSC.m4x is needed for future steps. The FSC.sdx is edited by a text editor (that's more convenient than fighting with the SynDEx bugs in the graphical editor) and all the architecture dependencies and durations removed or commented with "\#". After that the real durations on the target architecture may be inserted (don't forget to include the necessary files if needed). The PID should be able to take two inputs of the type "double": Setpoint (w\_i), Actual Input(y\_i). Additionally the PID will need one "double" output: Actual Output(u\_i). To do this, following lines need to be inserted (the next line right after "def algorithm pidFanController  :"):\\
\begin{verbatim}
 ? double[1] w_i 1;\\
 ? double[1] y_i 2;\\
 ! double[1] u_i 1;\\
\end{verbatim}
Now the file is renamed to "pidFan.sdx" since this algorithm will be a PID block suitable for the fan (RPM - scaling is included in the block, this is necessary for a more comfortable automated edit- and build-process). Additionally, the algorithm is baptized by replacing the "def algorithm clk\_root..." tag by a name of choice. Now either copy the new algorithm in the "$<$SYNDEXHOME$>$/libs" or make a symbolic link to it in that directory. The PID controller block is now ready to be used in SynDEx.\\
The next step proposed is to create a new project folder. If there are some home-made development scripts, then copy these files in the project folder (that could be m4x-builder scripts, a customized GNUMakefile or something else).
Start a new SynDEx project. The new library is found in the SynDEx menu: "File"$\rightarrow$"Included Libraries"$\rightarrow$"pidFan". Start the design process with the use of the new PID algorithm. In the PID example all the blocks have to be called with equal periods. To do this, click on EVERY single block and enter the period. Note: If at least one is forgotten and the algorithm adequation called, a "division by zero" error message occurs: In this case close all files and SynDEx, reopen it (it is good NOT to save after entering periods, because the file will be malformed, not readable, and a start from scratch is required) and do everything again from the last saved point (hopefully before any periods were entered); all blocks previously shifted from one hierarchy level into another must be re-edited as well as all the lost block parameters... Double/Tripple/Multi-check all periods before adequation (don't forget to handle all switch-blocks branches)!\\
After the design of the algorithm, chose the target architecture (e.g. the home-made "eBoard"$\rightarrow$"monoProcNode3") in the "Architecture" menu. Now it should be possible to successfully call the algorithm adequation, obtain the scheduling and generate macro-executive code. Note: Even though it seems that everything was done in the right way, a new message might occur when starting the adequation: "Uncaught exception: File "algorithm/transform.ml", line 287, characters 7-13: Assertion failed" - Why, where, what is wrong? SynDEx won't tell. After some tests it was figured out that a connection the new I/O ports within the block (they were hidden behind a block) had been forgotten - well, let's start from the last saved point again and enter all periods, move hierarchical blocks and re-enter parameters in the previously inter-hierarchy-level moved blocks). The adequation might require some extra durations for some blocks to be entered and don't bother(!?) if the adequation seems to stop half way, or seems not to start or finish. With a diminished trust in adequation results, keep trying some times and it might start / finish or ask for further block durations.\\
If the scheduling is satisfying, macro code can be generated: "Code"$\rightarrow$"Generate Executive(s)". This will generate the files FSCMonoProcessorStage03.m4 and node3.m4, both containing macro code.\\
The next step is the macro code into C code expansion phase. Therefore the GNUmakefile has to be adjusted properly to your needs. Note: Don't forget to touch at least an empty $<$ApplicationName$>$.m4m file - SynDEx does not provide conclusive error messages!\\
In order to successfully expand to C code, self-made m4 definitions files are needed (for this example macro expansion files for the ESE-board, task-timings and the ATmega128 were written). To automatically generate customized Makefiles a customized syndex.m4m is needed. With the customized GNUmakefile in a Unix shell type: "make expand". That will generate the customized Makefile for node3 (FSCMonoProcessor.mk). "make node3.c" will finally generate C code, but things have to be considered. The macro-expansion file (.m4x) generated by the Scicos-to-SynDEx gateway is needed since it holds macro-expansion rules for most of the needed blocks (remember to remove the default sensor, actuator and maybe other example .c includes in this file - otherwise they will be included in the generated C code file). Replace the m4-include calls with C includes, e.g.:
Replace...\\
\begin{verbatim}
  include(/home/exa/DA/syndex-6.8.5/SCICOS_FILES/s2s_common.c)\\
  include(/home/exa/DA/syndex-6.8.5/SCICOS_FILES/gainblk.c)\\
  include(/home/exa/DA/syndex-6.8.5/SCICOS_FILES/plusblk.c)\\
  include(/home/exa/DA/syndex-6.8.5/SCICOS_FILES/dollar4.c)\\
  include(/home/exa/DA/syndex-6.8.5/SCICOS_FILES/satur.c)\\
  include(/home/exa/DA/syndex-6.8.5/SCICOS_FILES/summation.c)\\
\end{verbatim}
with...\\
\begin{verbatim}
  #include "s2s_common.h"\\
  #include "gainblk.h"\\
  #include "plusblk.h"\\
  #include "dollar4.h"\\
  #include "satur.h"\\
  #include "summation.h".\\
\end{verbatim}

Note: Before the macros can be expanded, they might need a rename in a way that already written .m4x macro expansion rule definitions match (use scripts for that, e.g. monoclean.sh). Another note: Be careful when including more than one Scicos-to-SynDEx generated file since some declarations (parameters) will mix up (for this sense I used a self-made m4x-builder java-program).
Yet another note: Remove the line 
\begin{verbatim}
  "enum flag_type {
  flag_init = 4, flag_updateOutputs = 1,
  flag_updateZstates = 2, flag_updateSstates = 0,
  flag_reinit = 6, flag_end = 5};"
\end{verbatim} and the line 
\begin{verbatim}
  "#include <stdio.h>"
\end{verbatim} from the pidFan.sdx file.
Additionally, a remark about memory blocks: Dollar blocks have to be defined with double in this example (otherwise mismatching data-types): e.g. replace alloc\_(int,dollar4\_4\_1\_buf,1) with alloc\_(double,dollar4\_4\_1\_buf,1).
About the number of iterations in the main-loop: In the pidFan.m4x file the number of iterations in the main-loop can be changed to forever by removing the "NBITERATIONS" definition. When doing this, maybe also remove the other comment macros:
\begin{verbatim}
  dnl define(`NOTRACEDEF')
  dnl define(`NBITERATIONS',``1000'')
  dnl define(`dnldnl',``// '')
  dnl define(`    # ',``// '')
\end{verbatim}

Now it should be possible to compile it using avr-gcc: "make".

	\chapter{Listings}
	\section{SynDEx PID-Example, Node3.m4}
\begin{lstlisting}[frame=single,label=lst:SynDExFanSpeedControlNode3,caption=Node3 - SynDEx macro code]
%auto-ignore
include(syndex.m4x)dnl
sinclude(TaskTiming.m4x)dnl
sinclude(eBoard.m4x)dnl
dnl
processor_(ATmega128,node3,fscFinal6b,
SynDEx-7.0.0 (C) INRIA 2001-2009, 2010-10-03 03:02:02)

alloc_(int,Timer2Controller_fire,1)
alloc_(uint16,fanSensorDriver_rpm,1)
alloc_(double,S2sDataWrapperW_d,1)
alloc_(double,gainblk_9_1_out_1,1)
alloc_(uint8,FanSpeed_cst,1)
======================= Snip =================================
...
======================= Snip =================================
alias_(dollar4_4_1_out_1,dollar4_4_1_buf,0,1)
alias_(dollar4_18_1_out_1,dollar4_18_1_buf,0,1)
alias_(rem_1_out_1,rem_1_buf,0,1)

main_
  proc_init_
  HWTimer2Controller(1000,25,Timer2Controller_fire)
  eFanSensorDriver__1(fanSensorDriver_rpm)
  TTTaskReservation(1)
  eUint16ToDouble(fanSensorDriver_rpm,S2sDataWrapperW_d)
  gainblk_1__1(11,4,0,0,0,1,0,0,0,1,5,0,0,0,0,1,1,S2sDataWrapperW_d,gainblk_9_1_out_1)
  eConstantUint8(50,FanSpeed_cst)
  eUint8ToDouble(FanSpeed_cst,S2SWrapperCst_d)
  plusblk_1_1__1(12,2,0,0,0,0,0,0,0,1,5,0,0,0,0,2,1,gainblk_9_1_out_1,S2SWrapperCst_d,plusblk_8_1_out_1)
  gainblk_1__1(16,4,0,0,0,1,0,0,0,0,1,0,0,0,0,1,1,plusblk_8_1_out_1,gainblk_2_1_out_1)
======================= Snip =================================
...
======================= Snip =================================
  loop_
    HWTimer2Controller(1000,25,Timer2Controller_fire)
    switch_(Timer2Controller_fire)
      case_(0)
        TTTaskReservation(1)
      endcase_
      case_(1)
        eFanSensorDriver__1(fanSensorDriver_rpm)
        eUint16ToDouble(fanSensorDriver_rpm,S2sDataWrapperW_d)
        gainblk_1__1(11,4,0,0,0,1,0,0,0,1,5,0,0,0,0,1,1,S2sDataWrapperW_d,gainblk_9_1_out_1)
======================= Snip =================================
...
======================= Snip =================================
      endcase_
    endswitch_
    wait_(54)
  endloop_
  HWTimer2Controller(1000,25,Timer2Controller_fire)
  eFanSensorDriver__1(fanSensorDriver_rpm)
  TTTaskReservation(1)
  eUint16ToDouble(fanSensorDriver_rpm,S2sDataWrapperW_d)
  gainblk_1__1(11,4,0,0,0,1,0,0,0,1,5,0,0,0,0,1,1,S2sDataWrapperW_d,gainblk_9_1_out_1)
  eConstantUint8(50,FanSpeed_cst)
  eUint8ToDouble(FanSpeed_cst,S2SWrapperCst_d)
======================= Snip =================================
...
======================= Snip =================================
  proc_end_
endmain_

endprocessor_
\end{lstlisting}
	\section{SynDEx PID-Example, Scicos-Syndex Gain Block}
\begin{lstlisting}[frame=single,label=lst:SynDEx:Macros:PIDGain,caption=SynDEx Gain Block macro code]
%auto-ignore
define(`gainblk_1__1',`ifelse(
MGC,`MGC',``'',
MGC,`INIT',``/* init phase for block gainblk_1__1  */
pidBlocks[$1-1].type = $2;
pidBlocks[$1-1].ztyp = $3;
pidBlocks[$1-1].ng = $4;
pidBlocks[$1-1].nz = $5;
pidBlocks[$1-1].nrpar = $6;
pidBlocks[$1-1].nipar = $7;
pidBlocks[$1-1].nevout = $8;
pidBlocks[$1-1].nmode = $9;
pidBlocks[$1-1].z = &(pidz[$10]);
pidBlocks[$1-1].rpar = &pidRPAR1[$11];
pidBlocks[$1-1].ipar = &pidIPAR1[$12];
pidBlocks[$1-1].x = &(pidxt[$13]);
pidBlocks[$1-1].xd = &(pidxtd[$14]);
pidBlocks[$1-1].nx = $15;
pidBlocks[$1-1].nin = $16;
pidBlocks[$1-1].nout = $17;
pidBlocks[$1-1].nevprt = 0;
if ((pidBlocks[$1-1].evout = calloc(pidBlocks[$1-1].nevout,sizeof(double)))== NULL ) return 0;
if ((pidBlocks[$1-1].insz = malloc(sizeof(int) * pidBlocks[$1-1].nin))== NULL ) return 0;
if ((pidBlocks[$1-1].inptr = malloc(sizeof(double*) * pidBlocks[$1-1].nin))== NULL ) return 0;
if ((pidBlocks[$1-1].outsz = malloc(sizeof(int) * pidBlocks[$1-1].nout))== NULL ) return 0;
if ((pidBlocks[$1-1].outptr = malloc(sizeof(double*) * pidBlocks[$1-1].nout))== NULL ) return 0;
pidBlocks[$1-1].work=NULL;
pidBlocks[$1-1].inptr[0] = $18;
pidBlocks[$1-1].insz[0] = '$18_size_`;
pidBlocks[$1-1].outptr[0] = $19;
pidBlocks[$1-1].outsz[0] = '$19_size_`;
gainblk(&pidBlocks[$1-1], flag_init);
gainblk(&pidBlocks[$1-1], flag_reinit);
'',
MGC,`LOOP',``/* loop phase for block gainblk_1__1  */
pidBlocks[$1-1].inptr[0] = $18;
pidBlocks[$1-1].outptr[0] = $19;
gainblk(&pidBlocks[$1-1], flag_updateOutputs);
gainblk(&pidBlocks[$1-1], flag_updateZstates);
'',
MGC,`END',``/* end phase for block gainblk_1__1  */
pidBlocks[$1-1].inptr[0] = $18;
pidBlocks[$1-1].outptr[0] = $19;
gainblk(&pidBlocks[$1-1], flag_end);
'')')

\end{lstlisting}
	\section{Atmel ATmega128 macro expansion definitions file}
\begin{lstlisting}[frame=single,label=lst:SynDEx:Macros:Atmel,caption=SynDEx ATmega128 macro expansion definition file]
%auto-ignore
dnl ATmega128.
dnl Purpose: Macro Expansion File for the Atmel's ATmega128 with data type mapping.

======================= Snip =================================
...
======================= Snip =================================

#define(`lang_',`C')

############
# DATA TYPES

# --------
# typedef_(name, size) defines a new type with its size in address units:
# interfacing data-types:
typedef_(`bool',  1) # MUST be defined
typedef_(`double', 4)
typedef_(`int', 2)
typedef_(`uint8',1)
typedef_(`uint16',2)

# mapped target data-types:
typedef_(`uint8_t', 1)
typedef_(`uint16_t', 2)
typedef_(`int8_t', 1)
typedef_(`int16_t', 2)

# MAPPING
# --------
## also types which aren't mapped have to be defined:
define(`double_map',`double')
define(`bool_map',`bool')

## the types to be mapped to different ones:
define(`int_map',`int16_t')
define(`uint8_map',`uint8_t')
define(`uint16_map',`uint16_t')
define(`int8_map',`int8_t')
define(`int16_map',`int16_t')

###################
## MEMORY ALLOCATION

## -----------
## basicAlloc_(label, memoryBank)
#define(`basicAlloc_',`_($1_type_ $1[$1_size_];)')
define(`basicAlloc_',`_($1_type_()_map() $1[$1_size_];)')
 
# -----------
## basicAlias_(newLabel,oldLabel[,offset=0])
define(`basicAlias_', `define(`$1_base_', $2_base_)dnl for `basicCopy_'
`#define' $1 ifelse($3,,`$2',`($2+$3)')')


####################
## CONTROL STRUCTURES

## --------
## basicIf_(buffer,value, tagforElseOrEndif)
define(`basicIf_', `_(if ($1[0]==$2) { /* $3 */)')

======================= Snip =================================
...
======================= Snip =================================

divert
divert`'dnl ---------------------- End of file -------------------------
 
\end{lstlisting}
	\section{SynDEx PID-Example, final C code}
\begin{lstlisting}[frame=single,label=lst:SynDEx:FanSpeedControlCCode,caption=SynDEx PID-Example C code]
%auto-ignore
/* =====> eBoard.m4x */

#include "eFanDriver.h"
#include "eFanSensorDriver.h"
======================= Snip =================================
...
======================= Snip =================================

/*****************************************************************************/
/***** ADDITIONAL GLOBALS ****************************************************/
/*****************************************************************************/

uint8_t HWTimer0Fire;
uint8_t HWTimer1Fire;
uint8_t HWTimer2Fire;
uint8_t HWTimer3Fire;

/*****************************************************************************/
/***** ADDITIONAL COMPUTATIONAL FUNCTIONS ************************************/
/*****************************************************************************/

/* Called when Timer reached destination Time (n * OC-I) */
void HWTimer2Handler(void){
	HWTimer2Fire = 1;
}

/* Reset the Fire-global after 1 repetition */
uint8_t HWTimer2ResetFire(uint8_t fire){

	static uint8_t fireMem;

if(fire==1&&fireMem==1){
		HWTimer2Fire=0;
		fireMem=0;
	}
	if(fire==1&&fireMem==0){

		fireMem=1;
	return 1;	
	}
	if(fire==0){
	return 0;
	}

	/* never reached */
	return 0;
}

======================= Snip =================================
...
======================= Snip =================================


/* SynDEx-7.0.0 (C) INRIA 2001-2009, 2010-10-03 03:02:02,
   application fscFinal6b, processor node3 type=ATmega128 */
/* =====> ATmega128.m4x */

/* `alloc_(int,Timer2Controller_fire,1)' */
int16_t Timer2Controller_fire[1];
/* `alloc_(uint16,fanSensorDriver_rpm,1)' */
uint16_t fanSensorDriver_rpm[1];
/* `alloc_(double,S2sDataWrapperW_d,1)' */
double S2sDataWrapperW_d[1];
/* `alloc_(double,gainblk_9_1_out_1,1)' */
double gainblk_9_1_out_1[1];
======================= Snip =================================
...
======================= Snip =================================

/* `main_()' */
int main(int argc, char* argv[]) { /* for link with C runtime boot */
  
======================= Snip =================================
...
======================= Snip =================================

/* init phase for block gainblk_1__1  */
blocks[11-1].type = 4;
blocks[11-1].ztyp = 0;
blocks[11-1].ng = 0;
blocks[11-1].nz = 0;
blocks[11-1].nrpar = 1;
blocks[11-1].nipar = 0;
blocks[11-1].nevout = 0;
blocks[11-1].nmode = 0;
blocks[11-1].z = &(z[1]);
blocks[11-1].rpar = &RPAR1[5];
blocks[11-1].ipar = &IPAR1[0];
blocks[11-1].x = &(xt[0]);
blocks[11-1].xd = &(xtd[0]);
blocks[11-1].nx = 0;
blocks[11-1].nin = 1;
blocks[11-1].nout = 1;
blocks[11-1].nevprt = 0;
if ((blocks[11-1].evout = calloc(blocks[11-1].nevout,sizeof(double)))== NULL ) return 0;
if ((blocks[11-1].insz = malloc(sizeof(int) * blocks[11-1].nin))== NULL ) return 0;
if ((blocks[11-1].inptr = malloc(sizeof(double*) * blocks[11-1].nin))== NULL ) return 0;
if ((blocks[11-1].outsz = malloc(sizeof(int) * blocks[11-1].nout))== NULL ) return 0;
if ((blocks[11-1].outptr = malloc(sizeof(double*) * blocks[11-1].nout))== NULL ) return 0;
blocks[11-1].work=NULL;
blocks[11-1].inptr[0] = S2sDataWrapperW_d;
blocks[11-1].insz[0] = 1;
blocks[11-1].outptr[0] = gainblk_9_1_out_1;
blocks[11-1].outsz[0] = 1;
gainblk(&blocks[11-1], flag_init);
gainblk(&blocks[11-1], flag_reinit);
======================= Snip =================================
...
======================= Snip =================================
  
  /* `loop_()' */
  for(;;){ /* loop_2 */
	{int i;for (i=0;i<1;i++)  rem_1_out_1[i] = prod_22_1_out_1[i];}
         			Timer2Controller_fire[0]=HWTimer2ResetFire(HWTimer2Fire);
    /* `switch_(Timer2Controller_fire)' */
    switch (Timer2Controller_fire[0]) { /* switch_3 */
      /* `case_(0)' */
      case 0 : /* case_4 */
        eNOP(26); //1
      /* `endcase_()' */
        break; /* case_4 */
      /* `case_(1)' */
      case 1 : /* case_5 */
        /* loop phase for block eFanSensorDriver__1  */
							eFanSensorDriver(fanSensorDriver_rpm);
        /* loop phase for block eUint16ToDouble  */
							S2sDataWrapperW_d[0] = (double)fanSensorDriver_rpm[0];
        /* loop phase for block gainblk_1__1  */
blocks[11-1].inptr[0] = S2sDataWrapperW_d;
blocks[11-1].outptr[0] = gainblk_9_1_out_1;
gainblk(&blocks[11-1], flag_updateOutputs);
gainblk(&blocks[11-1], flag_updateZstates);
======================= Snip =================================
...
======================= Snip =================================
	/* `endcase_()' */
        break; /* case_5 */
    /* `endswitch_()' */
      } /* switch_3 */
    eNOP(54);
  /* `endloop_()' */
  } /* end loop_2 */
  
  /* end phase for block gainblk_1__1  */
blocks[11-1].inptr[0] = S2sDataWrapperW_d;
blocks[11-1].outptr[0] = gainblk_9_1_out_1;
gainblk(&blocks[11-1], flag_end);

======================= Snip =================================
...
======================= Snip =================================
/* `endmain_()' */
  return 0;
} /* end of main */

/* `endprocessor_()' */
\end{lstlisting}
	\section{SynDEx PID-Example, GNUMakefile}
\begin{lstlisting}[frame=single,label=lst:SynDEx:FanSpeedControlGNUMakefile,caption=SynDEx PID-Example GNUMakefile]
%auto-ignore
A=fanSpeedControlMonoProcessorStage03
M4=m4

# these paths have to be modified by user
sdx_MACRO_PATH=/home/exa/DA/syndex-7.0.0/macros

eBoard_MACROS_PATH=/home/exa/DA/code/syndex/macros
eBoard_MACROS_ALGORITHMS_PATH=$(eBoard_MACROS_PATH)/algorithms
eBoard_MACROS_ARCHITECTURES_PATH=$(eBoard_MACROS_PATH)/architectures
eBoard_BUILDENV_PATH=/home/exa/DA/code/eBoardBuildEnv
eBoard_DRIVERS_SRC_PATH=$(eBoard_BUILDENV_PATH)/drivers/src/syndex
eBoard_DRIVERS_INC_PATH=$(eBoard_BUILDENV_PATH)/drivers/include/syndex
eBoard_CONFIG_PATH=$(eBoard_BUILDENV_PATH)/config

# Include general properties for the environment
include $(eBoard_BUILDENV_PATH)/Makefile.set

# Algo_Macros_Path: Path of the algorithm specification files (.m4x and .m4m).
# Archi_Macros_Path: Path of the architecture specification files (.m4x and .m4m).
# S2s_Files_Path: User-made Scicos blocks must be copied in here (.c).
# eBoard_Drivers_Src_Path: User-made Scicos/Syndex computational function blocks (.c and .h).

export syndex_Macros_Algorithms_Path=$(sdx_MACRO_PATH)/algo_libraries
export syndex_Macros_Architectures_Path=$(sdx_MACRO_PATH)/archi_libraries

export eBoard_Macros_Algorithms_Path=$(eBoard_MACROS_ALGORITHMS_PATH)
export eBoard_Macros_Architectures_Path=$(eBoard_MACROS_ARCHITECTURES_PATH)
export eBoard_Drivers_Src_Path=$(eBoard_DRIVERS_SRC_PATH)

export S2s_Files_Path=/home/exa/DA/syndex-6.8.5/SCICOS_FILES

export M4PATH=$(syndex_Macros_Algorithms_Path):$(syndex_Macros_Architectures_Path):$(S2s_Files_Path):$(eBoard_Macros_Algorithms_Path):$(eBoard_Macros_Architectures_Path)

VPATH=$(M4PATH)

#CFLAGS = -DDEBUG -lm -Wall -pedantic

.PHONY: all expand cleanall cleanit cleanc

# replace rootOperator.c by $(A).run in order to run the application with a simple make call
#all : node3.c
all : expand

#clean ::
#	$(RM) \#* *~ *.o *.a *.mnt node3 $(A).mk

expand: $(A).mk

node: node3.c

debugPaths:
	@echo "M4PATH="$(M4PATH)

cleanit : cleanc
	$(RM) $(A).mk

cleanall :
	$(RM) node3.m4

cleanc :
	$(RM) node3.c

$(A).mk : $(A).m4 syndex.m4m ATmega128.m4m $(A).m4m
	$(M4) $< >$@


node3.libs = $(eBoard_DRIVERS_SRC_PATH)/cstblk4.o \
 $(eBoard_DRIVERS_SRC_PATH)/eFanDriver.o \
======================= Snip =================================
...
======================= Snip ================================= 

node3.inc = $(eBoard_DRIVERS_INC_PATH)

# the $(A).mk file is generated by the makefile process, do not edit/modify unless you know exactly what you are doing
sinclude $(A).mk
\end{lstlisting}
	\section{SynDEx PID-Example, .mk Makefile}
\begin{lstlisting}[frame=single,label=lst:SynDEx:FanSpeedControlMK,caption=SynDEx PID-Example .mk Makefile]
%auto-ignore
# SynDEx-7.0.0 (C) INRIA 2001-2009, 2009-10-13 03:54:51
# Makefile for application fanSpeedControlMonoProcessorStage03

# $(M4) must be the GNU macroprocessor m4
# $(Archi_Macros_Path) must be the path to the generic *.m4? macro-files
# $(VPATH) is searched by make for dependent files not found in $(PWD)
# VPATH = $(Archi_Macros_Path)

.PHONY: fanSpeedControlMonoProcessorStage03.all fanSpeedControlMonoProcessorStage03.run clean
fanSpeedControlMonoProcessorStage03.run : fanSpeedControlMonoProcessorStage03.all # load and run fanSpeedControlMonoProcessorStage03:
       # (command args = processors in loading order)
	./$(fanSpeedControlMonoProcessorStage03.root)


# processor node3 type=ATmega128:
fanSpeedControlMonoProcessorStage03.all : node3
fanSpeedControlMonoProcessorStage03.root += node3

#START-procr_------------------------------- START OF PROCESSOR MAKEFILE
# ATmega128.m4m
# Author: exa
# Purpose: Macro expansion file for Atmel's ATmega128.
# Description: Generates the architecture specific makefile.

node3.c : node3.m4 syndex.m4x ATmega128.m4x eBoard.m4x
	$(M4) $< >$@

node3.OBJS       = node3.o
node3.SUBOBJS    = $(node3.libs)

# define variables containing the names of the hex & eep file
node3.FLASH_FILE = node3.hex
node3.EEP_FILE   = node3.eep
node3.ELF_FILE   = node3.elf

======================= Snip =================================
...
======================= Snip =================================

procr_SUBDIRS    = $(dir $(node3.SUBOBJS)) # just the dirs "Sub1/ Sub2/ ..."

======================= Snip =================================
...
======================= Snip =================================
# list of all non-file targets
.PHONY: all makesub lc lmc clean mostlyclean distclean install debug


# 1st the main target
all: $(DREQUIRED) makesub node3.hex node3.eep


# include just existing D-files
ifneq "$(strip $(DFILES))" ""
  include $(DFILES)
endif


# declare implicit rules
%.hex:	%.elf
	$(OBJCOPY) -O $(FORMAT) $< $@

%.eep:	%.elf
	-$(OBJCOPY) $(EEPFLAGS) -O $(FORMAT) $< $@

%.elf:	$(node3.OBJS) $(node3.SUBOBJS)
	$(CC) -Wl,-Map=$*.m $(LDFLAGS) -o $@ $(node3.OBJS) $(node3.SUBOBJS) $(LIBOPT)

======================= Snip =================================
...
======================= Snip =================================

%.o:	%.c
	$(CC) -I$(node3.inc) $(CFLAGS) -Wall -c -o $@ $<

%.o:	%.S
	$(CC) -I$(node3.inc) $(CFLAGS) -Wall -c -o $@ $<

%.i:	%.c
	$(CC) -I$(node3.inc) $(CFLAGS) -E -o $@ $<

%.s:	%.c
	$(CC) -I$(node3.inc) $(CFLAGS) -S -o $@ $<

======================= Snip =================================
...
======================= Snip =================================
include $(eBoard_CONFIG_PATH)/node3.deploy
#END-procr_----------------------------------- END OF PROCESSOR MAKEFILE

\end{lstlisting}
	\section{Scicos Block Struct, .h Header}
\begin{lstlisting}[frame=single,label=lst:scicosblock,caption=Scicos Block Structure .h Headerfile]
%auto-ignore
#ifndef __SCICOS_BLOCK_H__
#define __SCICOS_BLOCK_H__

#include <stdint.h>
#include <stdlib.h>

typedef void (*voidg)(void);

typedef struct {
  int16_t nevprt;
  voidg funpt ;
  int16_t type;
  int16_t scsptr;
  int16_t nz;
  double *z;
  int16_t nx;
  double *x;
  double *xd;
  double *res;
  int16_t nin;
  int16_t *insz;
  double **inptr;
  int16_t nout;
  int16_t *outsz;
  double **outptr;
  int16_t nevout;
  double *evout;
  int16_t nrpar;
  double *rpar;
  int16_t nipar;
  int16_t *ipar;
  int16_t ng;
  double *g;
  int16_t ztyp;
  int16_t *jroot;
  char *label;
  void **work;
  int16_t nmode;
  int16_t *mode;
} scicos_block;



/* extern void do_cold_restart(); 
int get_phase_simulation(void){return 1;}
double get_scicos_time();
int16_t get_block_number();
void set_block_error(int16_t);
void set_point16_ter_xproperty(int16_t* point16_ter);

void * scicos_malloc(size_t );
void scicos_free(void *p);
*/

#define max(a,b) ((a) >= (b) ? (a) : (b))
#define min(a,b) ((a) <= (b) ? (a) : (b))

/*
extern int16_t s_copy();
extern int16_t s_cmp();
*/
enum flag_type { flag_init = 4, flag_updateOutputs = 1, flag_updateZstates = 2, flag_updateSstates = 0, flag_reinit = 6, flag_end = 5};

#endif /* __SCICOS_BLOCK_H__ */
\end{lstlisting}
	\section{Textual SynDEx PID algorithm, pidFan.sdx}
\begin{lstlisting}[frame=single,label=lst:syndex:textual,caption=Textual SynDEx algorithm]
%auto-ignore
syndex_version : "7.0.0"
application description : "Algorithm generated by the Scicos To SynDEx translator"

# Libraries
include "Atmel.sdx";
include "TaskTiming.sdx";
include "eBoard.sdx";
include "pid.sdx";
include "eBoardDrivers.sdx";

# Algorithms
def algorithm FanSpeedControlMonoProcessor  :
conditions: true;
references:
 eBoardDrivers/eFanSensorDriver__1 fanSensorDriver @-188,133;
 eBoardDrivers/eFanDriver_1_<0> fanMotorDriver @379,71;
 eBoardDrivers/eDoubleToUint8 S2SDataWrapper @242,71;
 eBoardDrivers/eUint16ToDouble S2sDataWrapperW @-44,133;
 eBoardDrivers/eConstantUint8<50> FanSpeed @-135,38;
 eBoardDrivers/eUint8ToDouble S2SWrapperCst @-7,38;
 TaskTiming/TTTaskContainer TTSpeedControl @193,-39;
 TaskTiming/HWTimer2Controller<1000;25> Timer2Controller @-21,-39;
 pid/pidControllerFinal PID @135,66;
dependences:
 strong_precedence_data fanSensorDriver.rpm -> S2sDataWrapperW.uint16;
 strong_precedence_data S2SDataWrapper.uint8 -> fanMotorDriver.speed;
 strong_precedence_data S2sDataWrapperW.d -> PID.y_i;
 strong_precedence_data FanSpeed.cst -> S2SWrapperCst.uint8;
 strong_precedence_data S2SWrapperCst.d -> PID.w_i;
 strong_precedence_data Timer2Controller.fire -> TTSpeedControl.fire;
 strong_precedence_data PID.u_i -> S2SDataWrapper.d;
code_phases:  loopseq;


# Architectures

# Main Algorithm / Main Architecture
main algorithm FanSpeedControlMonoProcessor ;
main architecture eBoard/monoProcNode3;

# Extra durations

# Operation groups, previously called Software components

# Constraints
\end{lstlisting}
	\cleardoublepage


\addcontentsline{toc}{chapter}{Bibliography}
	
\begin{thebibliography}{CPPSV06}

\bibitem[{Atm}06]{AVR}
{Atmel Corporation}.
\newblock {\em ATmega128}, 2006.
\newblock Rev. 2467N–AVR–03/06. Available online at
  \url{http://www.atmel.com/products/avr/}.

\bibitem[Boe88]{BoehmSpiral88}
Barry Boehm.
\newblock {A Spiral Model of Software Development and Enhancement}.
\newblock {\em Computer}, 21(5):61--72, May 1988.

\bibitem[Bro04]{Brown}
Alan~W. Brown.
\newblock {Model Driven Architecture: Principles and Practice}.
\newblock {\em Software and Systems Modeling}, 3:314--327, 2004.
\newblock 10.1007/s10270-004-0061-2.

\bibitem[CE00]{eisenecker2000}
Krzysztof Czarnecki and Ulrich~W. Eisenecker.
\newblock {\em {Generative Programming: Methods, Tools, and Applications}}.
\newblock ACM Press/Addison-Wesley Publishing Co., New York, NY, USA, 2000.

\bibitem[CNC06]{Campbell2006}
Stephen Campbell, Ramine Nikoukhah, and Jean-Philippe Chancelier.
\newblock {\em {Modeling and Simulation in Scilab/Scicos}}.
\newblock Springer Science+Business Media, Inc., 2006.

\bibitem[Col01]{nla.cat-vn3847107}
Defense Systems~Management College.
\newblock {\em { Systems Engineering Fundamentals : Supplementary Text /
  Prepared by the Defense Acquisition University Press }}.
\newblock The Press, Fort Belvoir, Va. :, 2001.

\bibitem[CPPSV06]{LanguagesToolsForHybridSystemsDesign}
Luca Carloni, Roberto Passerone, Alessandro Pinto, and Alberto
  Sangiovanni-Vincentelli.
\newblock {Languages and Tools for Hybrid Systems Design}.
\newblock {\em Foundations and Trends in Design Automation}, 1(1):1--204, 2006.

\bibitem[EBK03]{TUW-137912}
W.~Elmenreich, G.~Bauer, and H.~Kopetz.
\newblock {The Time-Triggered Paradigm}.
\newblock In {\em Proccedings of the Workshop on Time-Triggered and Real-Time
  Communication Systems}, 2003.

\bibitem[EK08]{ElAmam2008}
Khaled~El Emam and Akif~G{\"u}nes Koru.
\newblock {A Replicated Survey of IT Software Project Failures}.
\newblock {\em IEEE Software}, 25:84--90, 2008.

\bibitem[Elm09]{elmenreich:09}
W.~Elmenreich, editor.
\newblock {\em {Embedded Systems Engineering}}.
\newblock Vienna University of Technology, Austria, Vienna, Austria, 2009.
\newblock ISBN 978-3-902463-08-1.

\bibitem[Elm10]{elm:MemEval}
W.~Elmenreich.
\newblock {Evaluating the static and dynamic memory consumption for AVR
  microcontroller programs}.
\newblock Networking Embedded Systems Blog, October 2010.
\newblock Available at \url{http://netwerkt.wordpress.com/2010/10/06/memeval}.

\bibitem[Fau06]{Scicos2SynDExGateway}
Cyril Faure.
\newblock {Traducteur Scicos/SynDEx - Installation et Utilisation, v2.2.2},
  April 2006.
\newblock INRIA, Project Eclipse.

\bibitem[fRiCSC03]{scilabdownloads}
INRIA The French National~Institute for Research~in Computer~Science and
  Control.
\newblock {Launch of the Scilab Consortium Dedicated to Scientific Computing},
  May 2003.
\newblock Available online at
  \url{http://www.scilab.org/aboutus/pressroom/press_release/pr_20030522}.

\bibitem[GLS99]{codes99}
Thierry Grandpierre, Christophe Lavarenne, and Yves Sorel.
\newblock {Optimized Rapid Prototyping For Real-Time Embedded Heterogeneous
  Multiprocessors}.
\newblock In {\em Proceedings of 7th International Workshop on
  Hardware/Software Co-Design, CODES'99}, pages 74--78, Rome, Italy, May 1999.

\bibitem[Gro01]{UML}
Object~Management Group.
\newblock {\em {Model Driven Architecture (MDA)}}, July 2001.
\newblock Rev. ormsc/2001-07-01. Available online at
  \url{http://www.omg.org/cgi-bin/doc?ormsc/2001-07-01}.

\bibitem[GS03]{greenfield1}
Jack Greenfield and Keith Short.
\newblock {Software Factories: Assembling Applications with Patterns, Models,
  Frameworks and Tools}.
\newblock In {\em OOPSLA '03: Companion of the 18th annual ACM SIGPLAN
  Conference on Object-oriented programming, systems, languages, and
  applications}, pages 16--27, New York, NY, USA, October 2003. ACM.

\bibitem[IEE90]{IEEE90}
IEEE.
\newblock {Standard Glossary of Software Engineering Terminology. IEEE Std
  610.12-1990}.
\newblock Technical report, {IEEE Computer Society Press}, 1990.

\bibitem[IEE00a]{IEEE100-2000}
IEEE.
\newblock {I}{E}{E}{E} 100 {T}he {A}uthoritative {D}ictionary of {I}{E}{E}{E}
  {S}tandards {T}erms {S}eventh {E}dition.
\newblock {\em IEEE Std 100-2000}, 2000.

\bibitem[IEE00b]{IEEE1471}
IEEE.
\newblock {IEEE 1471-2000 Recommended Practice for Architectural Description
  for Software-Intensive Systems}.
\newblock Technical report, IEEE Computer Society, 2000.
\newblock Available online at
  \url{http://ieeexplore.ieee.org/servlet/opac?punumber=7040}.

\bibitem[iGfiSm03]{Iteratec}
iteratec Gesellschaft f{\"u}r~iterative Softwaretechnologien~mbH.
\newblock {Kurzbeschreibung iteratec Vorgehensmodell}, March 2003.
\newblock Available online at
  \url{www.iteratec.de/download/iteratec_Vorgehensmodell.pdf}.

\bibitem[Koe09]{Koessler2009}
Alexander Koessler.
\newblock {A Platform for Teaching and Research on Distributed Real-Time
  Systems}.
\newblock Master's thesis, Technical University of Vienna, Institute for
  Computer Engineering, Treitlstr. 3/2/182-2, 1040 Vienna, Austria, March 2009.

\bibitem[Kop97]{Kopetz1997}
Hermann Kopetz.
\newblock {\em {Real-Time Systems: Design Principles for Distributed Embedded
  Applications}}.
\newblock Kluwer Academic Publishers, Norwell, MA, USA, 1997.

\bibitem[KS04]{esm04}
R\'{e}my Kocik and Yves Sorel.
\newblock {A Methodology to Reduce the Design Lifecycle of Real-Time Embedded
  Control Systems}.
\newblock In {\em Proceedings of European Simulation and Modelling Conference,
  ESM'04}, Paris, France, October 2004.

\bibitem[NAN03]{NAN2003}
Masoud Najafi, Azzedine Azil, and Ramine Nikoukhah.
\newblock {Implementation of Continuous-Time Dynamics in Scicos}.
\newblock In {\em 15TH European Simulation Symposium and Exhibition}, Delft,
  The Netherlands, October 2003.

\bibitem[NR68]{nato1968}
Peter Naur and Brian Randell.
\newblock {Software Engineering: Report of a conference sponsored by the NATO
  Science Committee, Garmisch, Germany, Brussels, Scientific Affairs Division,
  NATO}, October 1968.
\newblock Available online at
  \url{http://homepages.cs.ncl.ac.uk/brian.randell/NATO/}.

\bibitem[OMG03]{MDA}
OMG.
\newblock {MDA Guide Version 1.0.1}.
\newblock Technical Report omg/2003-06-1, {OMG}, June 2003.
\newblock Available online at
  \url{http://www.omg.org/cgi-bin/doc?omg/03-06-01}.

\bibitem[Pau04]{paukovits:2004}
Christian Paukovits.
\newblock {Modellierung und Scheduling von flexiblen, zeitgesteuerten
  Kommunikationsprotokollen}.
\newblock Bachelor's thesis, Technische Universit{\"a}t Wien, Institut f{\"u}r
  Technische Informatik, Treitlstr. 3/3/182-1, 1040 Vienna, Austria, 2004.

\bibitem[PCM01]{AIL2001}
Arjun Panday, Damien Couderc, and Simon Marichalar.
\newblock {AIL: Description of a Global Electronic Architecture at the Vehicle
  Scale}.
\newblock In {\em DATE '01: Proceedings of the Conference on Design, Automation
  and Test in Europe}, page 112, Piscataway, NJ, USA, 2001. IEEE Press.

\bibitem[Roy70]{Royce1970}
Winston~W. Royce.
\newblock {Managing the Development of Large Software Systems: Concepts and
  Techniques}.
\newblock In {\em {Technical Papers of Western Electronic Show and Convention
  (WesCon)}}, 1970.

\bibitem[Sci10]{ScilabMan}
{\em {Scilab Manual}}, April 2010.
\newblock Rev. 5.2.2. Available online at
  \url{http://www.scilab.org/product/man/index.html}.

\bibitem[Sel03]{Selic}
Bran Selic.
\newblock The pragmatics of model-driven development.
\newblock {\em IEEE Software}, 20:19--25, 2003.

\bibitem[SK98]{springer-prozessautomatisierung}
Gerhard-Helge Schildt and Wolfgang Kastner.
\newblock {\em Proze{\ss}automatisierung}.
\newblock Springer, Wien, 1998.
\newblock ISBN 3-211-82999-7.

\bibitem[Som04]{SommSoft2007}
Ian Sommerville.
\newblock {\em {Software Engineering}}.
\newblock Addison-Wesley, Harlow, England, 7 edition, May 2004.
\newblock ISBN 0-321-21026-3.

\bibitem[Sor94]{mpcs94}
Yves Sorel.
\newblock {Massively Parallel Systems with Real Time Constraints, the Algorithm
  Architecture Adequation Methodology}.
\newblock In {\em Proceedings of Conference on Massively Parallel Computing
  Systems, MPCS'94}, Ischia, Italy, May 1994.

\bibitem[SPF07]{Smith07}
Paul~F. Smith, Sameer~M. Prabhu, and Jonathon Friedman.
\newblock {Best Practices for Establishing a Model-Based Design Culture}.
\newblock Technical report, The Mathworks, 2007.

\bibitem[SVEH07]{stahl}
Thomas Stahl, Markus V{\"o}lter, Sven Efftinge, and Arno Haase.
\newblock {\em Modellgetriebene Softwareentwicklung}.
\newblock dpunkt.verlag, May 2007.

\end{thebibliography}
\end{document}